\documentclass[letter,11pt]{article}
\pdfoutput=1 % if your are submitting a pdflatex (i.e. if you have
\usepackage{jheppub} % for details on the use of the package, please
\graphicspath{{figs/}}

\usepackage[T1]{fontenc} % if needed
\usepackage{topcapt}
\usepackage{multirow}

\usepackage[table]{xcolor}
\usepackage{rotating}
\usepackage{hhline}
\usepackage{adjustbox}
\usepackage{boldline}

\usepackage{booktabs} 

\newcolumntype{P}[1]{>{\centering\arraybackslash}p{#1}}
\usepackage{slashed}
\usepackage{float}
\usepackage{array}
\usepackage{empheq}

\definecolor{LightCyan}{rgb}{0.88,1,1}
\usepackage[hypcap=true,list=true]{subcaption}

\newcommand{\beq}{\begin{equation}}
\newcommand{\eeq}{\end{equation}}
\newcommand{\bea}{\begin{eqnarray}}
\newcommand{\eea}{\end{eqnarray}}
\newcommand{\bal}{\begin{align}}
\newcommand{\eal}{\end{align}}
\newcommand{\bit}{\begin{itemize}}
\newcommand{\eit}{\end{itemize}}
\newcommand{\ben}{\begin{enumerate}}
\newcommand{\een}{\end{enumerate}}

\renewcommand{\eqref}[1]{Eq.~(\ref{eq:#1})}

\newcommand{\secref}[1]{Sec.~\ref{sec:#1}}

\newcommand{\appref}[1]{App.~\ref{sec:#1}}
\newcommand{\figref}[1]{Fig.~\ref{fig:#1}}
\newcommand{\figsref}[2]{Figs.~\ref{fig:#1} and \ref{fig:#2}}

\newcommand{\tabref}[1]{Tab.~\ref{tab:#1}}

\renewcommand{\t}{\tilde}

\newcommand{\f}{\frac}

\newcommand{\OO}{\mathcal{O}}

\newcommand{\ab}{\,{\rm ab}^{-1}}
\newcommand{\fb}{\,{\rm fb}^{-1}}

\newcommand{\gev}{{\ \rm GeV}}
\newcommand{\tev}{{\ \rm TeV}}

\newcommand{\met}{\slashed{E}_\text{T}}
\renewcommand{\d}{\text{d}}
\newcommand{\pt}{p_\text{T}}

\title{\boldmath The Leptoquark Hunter's Guide:\\ Large Coupling}

\author[a]{Martin Schmaltz}
\author[a]{Yi-Ming Zhong}

\affiliation[a]{Physics Department, Boston University, Boston,  MA 02215, USA}

\emailAdd{schmaltz@bu.edu}
\emailAdd{ymzhong@bu.edu}

\abstract{Leptoquarks have recently received much attention especially because they may provide an explanation to the $R_{D^{(*)}}$ and $R_{K^{(*)}}$  anomalies in rare $B$ meson decays. In a previous paper we proposed a systematic search strategy for all possible leptoquark flavors by focusing on leptoquark pair production. In this paper, we extend this strategy to large (order unity) leptoquark couplings which offer new search opportunities: single leptoquark production and $t$-channel leptoquark exchange with dilepton final states. We discuss the unique features of the different search channels and show that they cover complementary regions of parameter space. We collect and update all currently available bounds for the different flavor final states from LHC searches and from atomic parity violation measurements. As an application of our analysis, we find that current limits do not exclude the leptoquark explanation of the $B$ physics anomalies but that the high luminosity run of the LHC will reach the most interesting parameter space.}

\begin{document} 
\maketitle
\flushbottom

\section{Introduction and overview}
\label{sec:intro}

\begin{figure}[!htbp]
 \centering
 \renewcommand{\thesubfigure}{PP-1}
             \begin{subfigure}[t]{0.3\textwidth}
        \centering
        \includegraphics[height=0.51\textwidth]{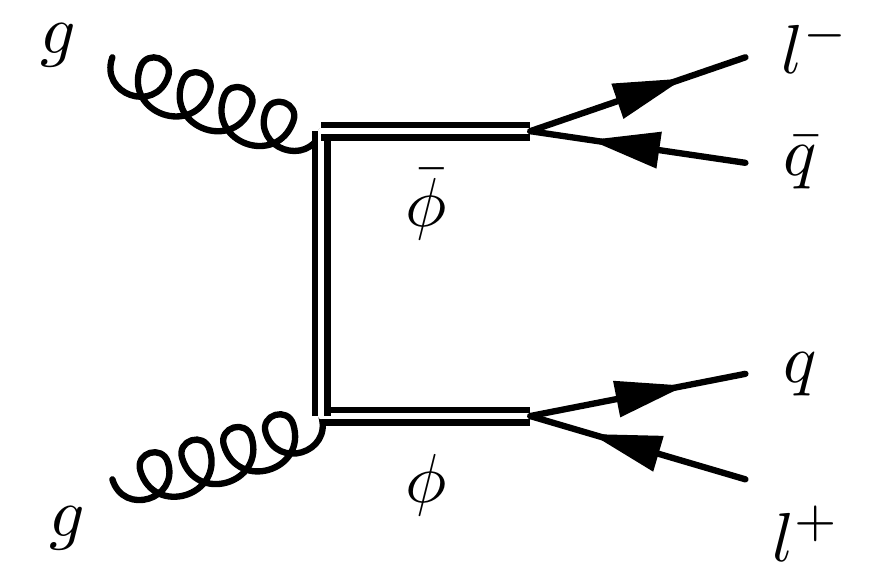}
        \caption{}
         \end{subfigure}%
~
 \renewcommand{\thesubfigure}{PP-2}
        \begin{subfigure}[t]{0.3\textwidth}
        \centering
        \includegraphics[height=0.51\textwidth]{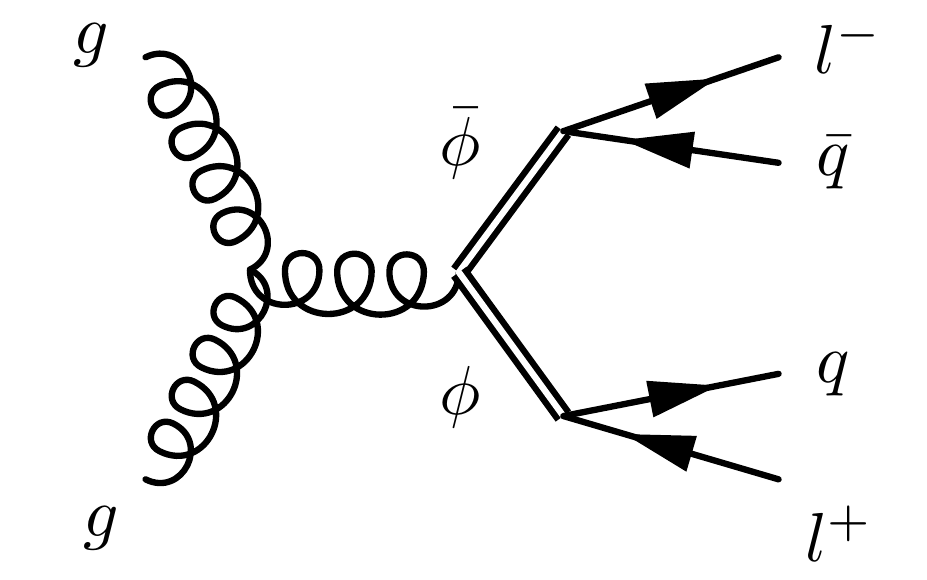}
        \caption{}
         \end{subfigure}%
~         
 \renewcommand{\thesubfigure}{PP-3}
         \begin{subfigure}[t]{0.3\textwidth}
        \centering
        \includegraphics[height=0.51\textwidth]{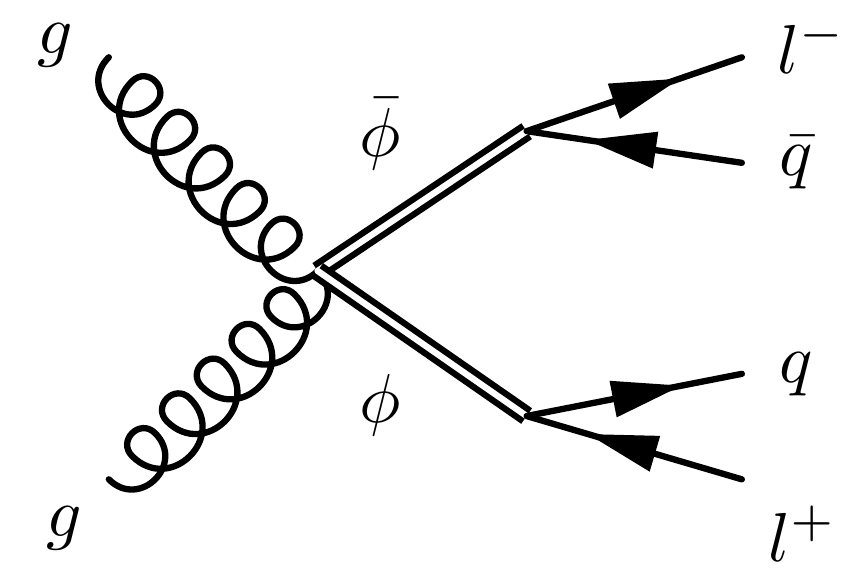}
        \caption{}
    \end{subfigure}
    ~ 
 \renewcommand{\thesubfigure}{PP-4}            
         \begin{subfigure}[t]{0.3\textwidth}
        \centering
        \includegraphics[height=0.51\textwidth]{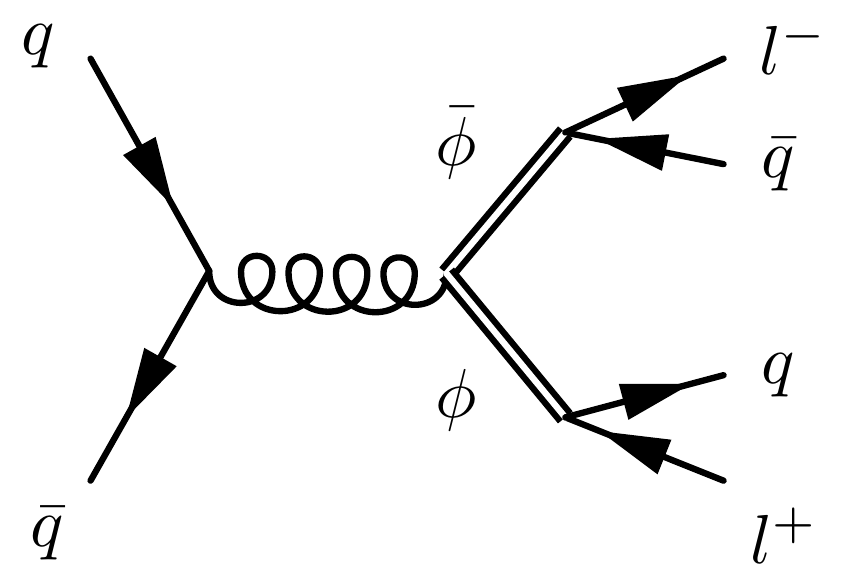}
        \caption{}
    \end{subfigure}
        ~
 \renewcommand{\thesubfigure}{PP-5}                             
         \begin{subfigure}[t]{0.3\textwidth}
        \centering
        \includegraphics[height=0.51\textwidth]{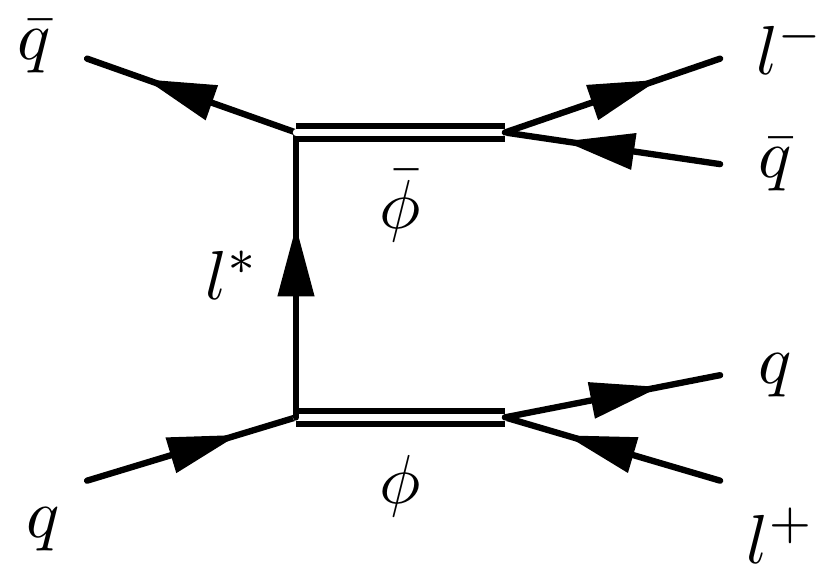}
        \label{fig:pp5}
        \caption{}
    \end{subfigure}
  ~         
 \renewcommand{\thesubfigure}{DY}
         \begin{subfigure}[t]{0.3\textwidth}
        \centering
        \includegraphics[height=0.5\textwidth]{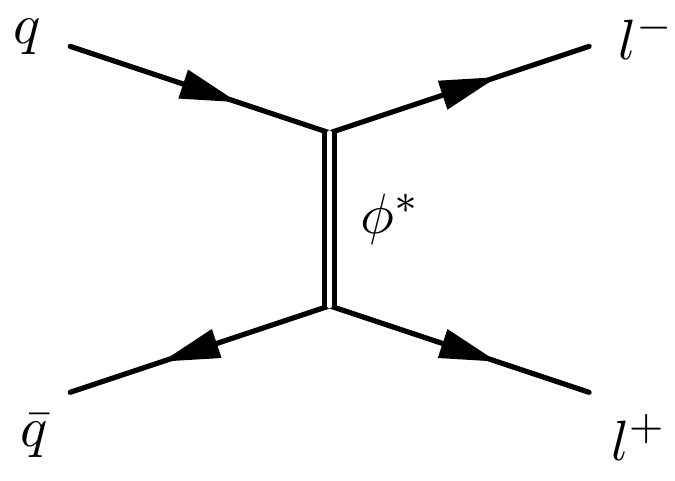}
        \caption{}
    \end{subfigure}   
    ~   
 \renewcommand{\thesubfigure}{SP-1}                             
   \begin{subfigure}[t]{0.3\textwidth}
        \centering
        \includegraphics[height=0.51\textwidth]{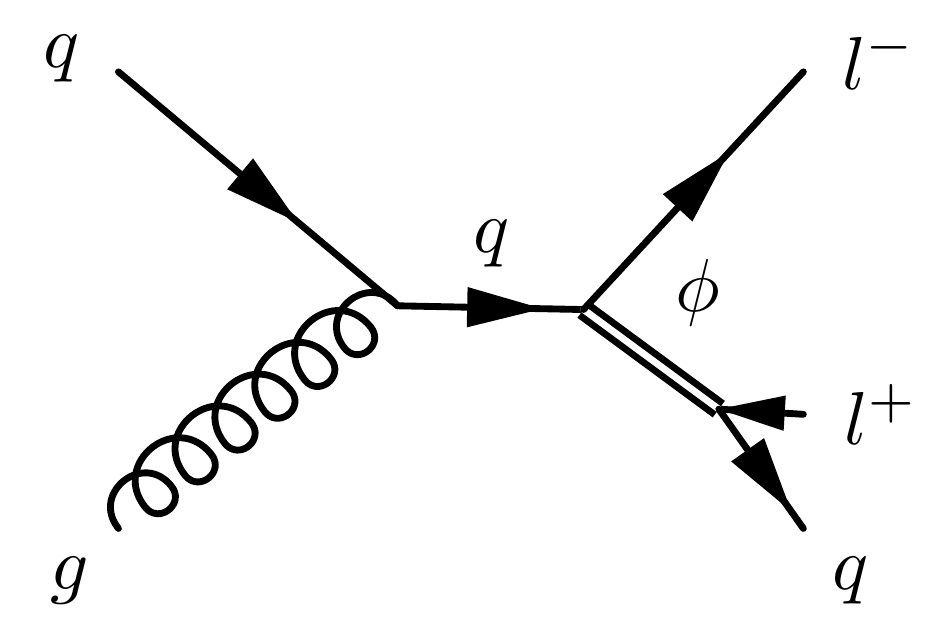}
        \caption{}
         \end{subfigure}%
~
 \renewcommand{\thesubfigure}{SP-2}                             
        \begin{subfigure}[t]{0.3\textwidth}
        \centering
        \includegraphics[height=0.51\textwidth]{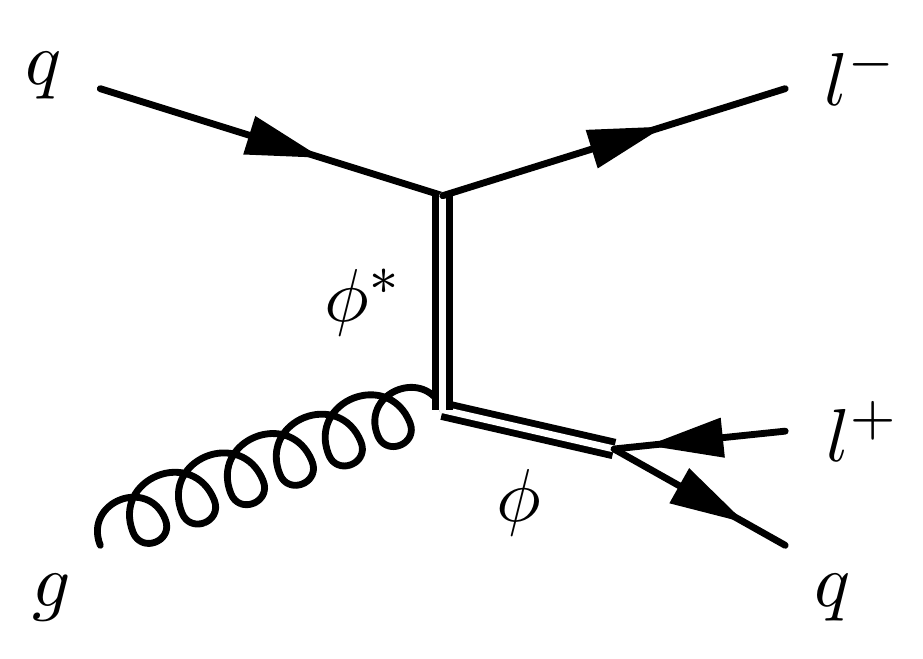}
        \caption{}
         \end{subfigure}%

   \caption{Leading order diagrams for leptoquark pair production $pp \to \phi \bar{\phi} \to (l q) (lq)$ (PP-1 to PP-5), Drell-Yan like dilepton production $pp \to l^+ l^-$ (DY), and single leptoquark production $pp \to l^+ l^- j$: (SP-1 and SP-2). The amplitudes PP-1 to PP-4 are proportional to $g_s^2$ whereas PP-5 and the DY diagram scale as $\lambda^2$. Single production diagrams,  SP-1 and SP-2, are proportional to $g_s \lambda$.}
   \label{fig:mypairdiagrams}
\end{figure}

A leptoquark (LQ) is a beyond-the-Standard-Model particle which can decay to a lepton and a quark. In our previous paper on  leptoquark pair production~\cite{Diaz:2017lit} we emphasized that any of the quark or lepton flavors may be involved in the coupling to leptoquarks. In particular, there is no model-independent reason to require the generation number of the lepton to be identical to that of  the quark.%
\footnote{It is often said that experimental bounds on flavor changing neutral currents  require that leptoquarks couple only to quarks and leptons of the same generation. This is a misunderstanding. To be safe from flavor changing neutral currents, a leptoquark should only have large couplings to a single quark flavor and a single lepton flavor, their generation numbers need not be the same.}  Leptoquark searches  should therefore investigate all possible combinations of the Standard Model (SM) leptons and quarks. We also defined ``minimal leptoquark'' (MLQ) models which include only one leptoquark at a time, coupled to only one SM lepton and one quark field. We focused on searches for minimal leptoquark pair-production at the LHC in~\cite{Diaz:2017lit}. Pair production is the dominant production mode at a proton-proton collider when the leptoquark-lepton-quark coupling, $\lambda$, is much smaller than 1. In this limit the production of leptoquarks only depends on the strong coupling, $\alpha_s$, because the contribution from the $t$-channel exchange of the lepton (diagram PP-5 in~\figref{mypairdiagrams}) is negligible. Therefore the only new physics parameter entering the scalar leptoquark cross section prediction is the leptoquark mass.%
\footnote{This is not true for vector leptoquarks. Vector leptoquark models require a ultraviolet completion with additional states and couplings which introduce model dependence even for leptoquark pair production~\cite{Diaz:2017lit}.} LHC constraints on the cross section can be directly translated into lower bounds on the leptoquark mass after accounting for possible multiplicity factors.

In this paper, we focus on leptoquark searches at large coupling, $\OO(0.1)  \lesssim \lambda \lesssim 3 $, where leptoquark production cross sections at the LHC depend on both the leptoquark mass and the leptoquark coupling. In addition to the pair production, a process with $t$-channel exchange of the leptoquark yielding a Drell-Yan-like dilepton final state, and single-leptoquark production are important for constraining leptoquark models at large coupling. The leading diagrams for the pair production, the Drell-Yan-like dilepton (DY) production, and the single production are shown in~\figref{mypairdiagrams}. The diagrams are valid for scalar leptoquarks, $S$, or vector leptoquarks, $V$, which we collectively denote as $\phi$.

\begin{figure}[!htbp]
   \centering
   \includegraphics[width=0.48\textwidth]{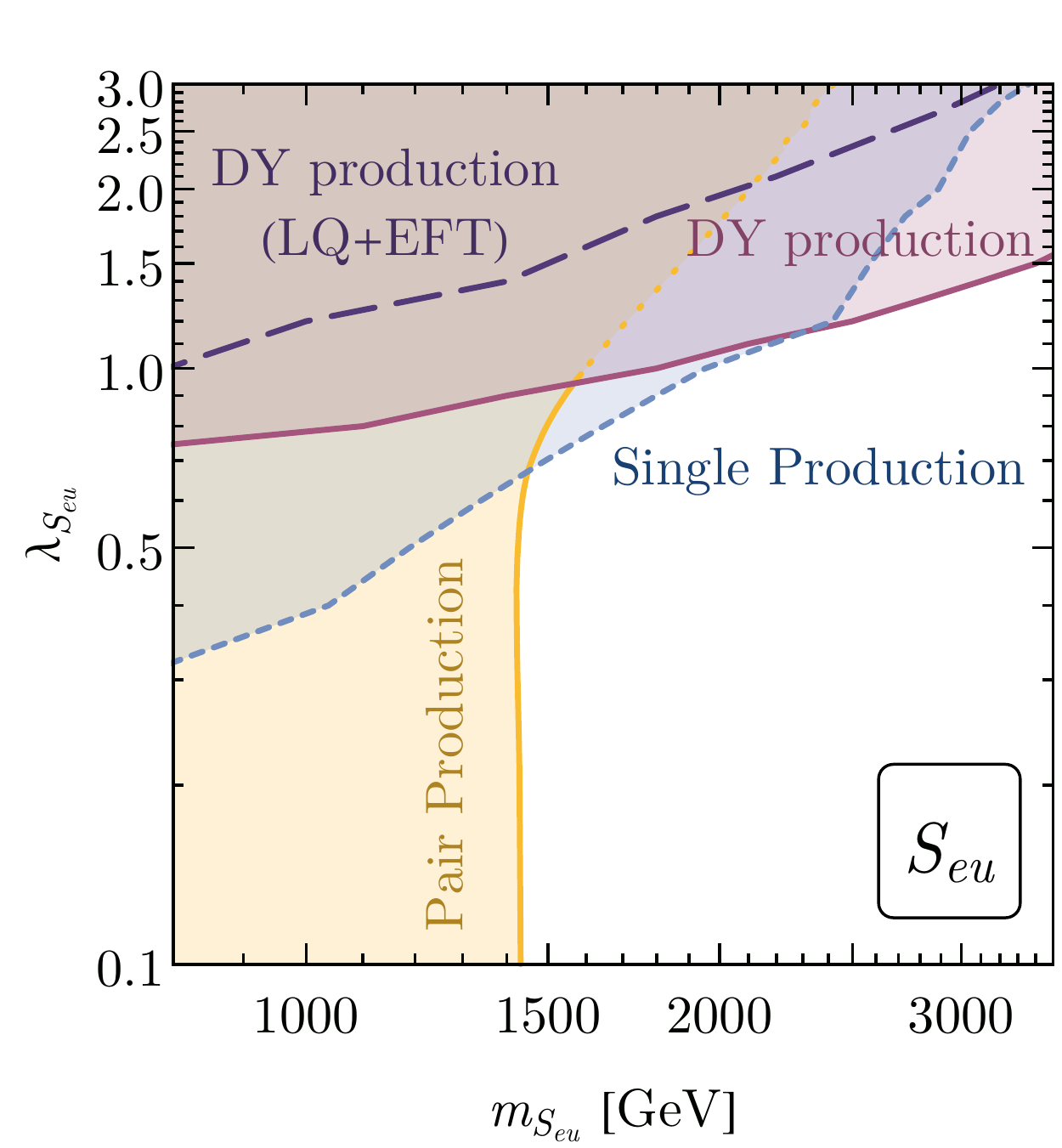}
 \caption{Bounds on scalar minimal leptoquark, $S_{e u}$, with flavor structure $U^c E^c$ from Run 2 pair production (yellow) ~\cite{CMS-PAS-EXO-17-009}, projected Run 2 single production (blue), and Run 2 DY-production (purple)~\cite{Sirunyan:2018exx} at the LHC. For the single production search, no Run 2 analysis has been published yet, and thus our ``bound'' is only a projected limit. 
Note that the pair-production search excludes leptoquarks with low masses while the DY search is very constraining at large coupling $\lambda$. This leaves only a relatively small region of parameter space with intermediate masses and couplings which can be probed by single production. Note however that since the DY search does not involve the reconstruction of a resonance it is more model dependent and can be weakened in the presence of a four-fermion operator which destructively interferes with leptoquark exchange. We indicate the weakened DY bound in the presence of a maximally interfering operator with a dashed purple line. This leptoquark also contributes to atomic parity violation and parity-violating electron scattering experiments; we derive the corresponding bounds in~\appref{others} and include them in~\figref{equark}.}
   \label{fig:muu}
\end{figure}

Constraints on leptoquark models from the three types of searches, the pair production, the DY production, and the single production, exclude regions  in the leptoquark mass versus coupling plane. As an example, we show that the bounds on a scalar leptoquark that couples to electrons and up quarks in~\figref{muu}.  One sees an important result which also applies to most other leptoquark flavors: off-shell leptoquark exchange (i.e. the DY production) together with the pair production provide the most significant bounds on leptoquark parameter space with the pair production excluding low masses and the DY production excluding large couplings. The single production provides sensitivity to a smaller additional region in parameter space near where the bounds from the pair production and the DY production cross. This suggests that the single production is less important in ruling out significant portions of leptoquark model parameter space. However, the fact that the exchanged leptoquark in the DY search is virtual makes the corresponding bound more model-dependent than those from the pair-production and single-production searches. This means that higher dimensional operators generated from larger scales can interfere with the $t$-channel exchange diagram of the leptoquark and change the new physics cross section. In~\figref{muu}, a dashed line indicates the weakened bound one obtains in the presence of such a dimension 6 operator with coefficients chosen to maximize destructive interference with the leptoquark exchange diagram.

In the following, we summarize the most important features of leptoquark searches at large coupling:
\begin{enumerate}

\item Diagrams proportional to the leptoquark coupling can originate from $q g$, $\bar q g$, or $q \bar{q}$ initial states where the initial quark flavors depend on which quark the leptoquark couples to. 
Thus cross sections for production of leptoquarks with different quark-flavors depend on the parton distribution functions (PDF), leading to very different bounds for the different quark flavors even if the final states are experimentally indistinguishable (for example, leptoquarks that couples to $u, d$, $s$ or $c$ quarks produce similar final state jets with very different cross sections).  An extreme case is the top-quark where the PDF in the proton vanishes.

\item
For pair production, and small leptoquark couplings, the cross section is independent of quark flavor and thus only simple 
multiplicity factors from production and decay differentiate between different leptoquark models contributing to the same final state~\cite{Diaz:2017lit}. 
This  allows straightforward comparison of pair production bounds for various leptoquark models (scalar versus vector, coupling to left-handed $SU(2)_{weak}$ doublets versus right-handed singlet fermions). At large leptoquark couplings, this is no longer possible because of the significant flavor-non-universal leptoquark couplings to initial state quarks and separate bounds must be obtained for each different leptoquark flavor.

\item At large coupling single-leptoquark production and processes with off-shell leptoquarks become important~\cite{Wise:2014oea, Bessaa:2014jya, Faroughy:2016osc,Raj:2016aky,Bansal:2018eha,Angelescu:2018tyl}. Off-shell processes include $t$-channel leptoquark exchange producing an dilepton final state and the partially off-shell diagram SP-2. Comparing to the gluon-gluon initiated pair production, these production channels are suppressed by smaller PDFs (except for leptoquarks coupling to $u$ or $d$ valence quarks) but enhanced by the coupling.  

\item For scalar leptoquarks coupling to a single combination of lepton and quark with coupling $\lambda$ the decay width is given by
\beq
\Gamma_S = \f{\lambda^2}{16 \pi } m_S \left(1 - \f{m_q^2}{m_S^2}\right)^2  \approx \f{\lambda^2}{16 \pi} m_S \ ,
\label{eq:scalarwidth}
\eeq
where in the first equality we have displayed the dependence on the quark mass which is significant only for the case of the top-quark. For vector leptoquarks one has
\beq
\Gamma_V = \f{\lambda^2}{24\pi} m_V \left(1 -\f{m_q^2}{m_V^2}\right)^2\left(1 + \f{m_q^2}{2 m_V^2}\right) \approx \f{\lambda^2}{24\pi} m_V \ .
\label{eq:vectorwidth}
\eeq
Note that the decay width is larger than $10\%$ of the leptoquark mass for $\lambda \gtrsim 2.2\,(2.7)$ for the scalar (vector) leptoquark. The width can be further enhanced if a leptoquark couples to more than one lepton-quark combination. Large leptoquark widths are a problem for those leptoquark searches that rely on leptoquark mass peak reconstruction. 

\item 
Since proton PDF's do not include top-quarks, the single production (SP-1, SP-2), the DY production, and the PP-5 diagram for the pair production do not exist for minimal leptoquarks that only couples top-quarks. However, diagrams PP-1 through PP-4 can still produce final states in which one or both of the leptoquarks are off shell. This is important at large leptoquark coupling and when the leptoquark mass is so large that on-shell pair production is not possible. We leave studies of this case to future work.

\item
For the special case of leptoquarks that couple to electrons and up or down quarks, the reach of atomic parity violation (APV) experiments can be competitive to that of collider searches.
We combine the results from APV experiments and parity-violating electron scattering (PVES) experiments and derive the constraints on $\phi_{eu}$ and $\phi_{e d}$ for all possible electroweak quantum numbers, see~\appref{others}.   

\end{enumerate}

Given these differences from the small coupling case, it is clear that LHC searches for leptoquarks with large couplings can be more interesting than the simple pair production case. Such searches are the focus of our paper. We review the existing constraints and overlapping searches, and provide  a systematic  framework for future search efforts.

The structure of our paper is as follows: In~\secref{normalization} we define our convention for the leptoquark coupling normalization, and in ~\secref{productionanddecay}, we discuss  key features of the production and decay of leptoquarks for the cases of pair production, single production and DY production. In~\secref{searches}, we systematically discuss searches for the different possible flavors of minimal leptoquarks. In~\secref{4321} we take a closer look at a specific model with a vector leptoquark which has recently garnered a lot of attention because it nicely explains the $B$-physics anomalies.  We end with a discussion of how one could distinguish between leptoquarks with similar flavor final states but different spins and chiralities in~\secref{distinguish}. We also include a number of appendices on how we perform our cut-and count analyses~\appref{cutandcount}, on constraints from APV and PVES measurements~\appref{others}, on details of the 4321 model~\appref{4321app}, on leptoquarks with different spin and chiral couplings~\appref{alternativecouplings}, and on parton-level leptoquark cross sections~\appref{partonlevel}.

\section{Definition and normalization of the leptoquark coupling}
\label{sec:normalization}

For most of the paper we consider leptoquarks which couple to only one lepton and quark flavor. We also take the leptoquarks to be singlets under $SU(2)_{weak}$. Assuming $SU(2)_{weak}$ invariant couplings we take the leptoquark to couple to the $SU(2)_{weak}$ singlet leptons and quarks (i.e. the right-handed fermions). We use the notation where all singlet fields are represented by left-handed charge conjugate fields ($U^c, D^c, E^c, N^c$).\footnote{Throughout the paper we follow the notation given in App.~A of~\cite{Diaz:2017lit} for leptoquarks and other field contents. Tab.~XIII of~\cite{Diaz:2017lit} shows the nomenclature for MLQ models in the BRW~\cite{Buchmuller:1986zs} and PDG~\cite{Olive:2016xmw} conventions.}  Note that we have also included a light neutrino singlet field, $N$, which could be either the Dirac partner of the neutrino in the $SU(2)_{weak}$ doublet if neutrino masses are Dirac or a heavy sterile neutrino. In~\appref{alternativecouplings} we explore leptoquarks with more general quantum numbers that can also couple to the $SU(2)_{weak}$ fermion doublets.

For example, the scalar leptoquark coupling to singlet electrons and up quarks is given by
\beq
{\mathcal L} \supset \lambda_{eu} S_{eu} (E^c U^c)^* +h.c.\ ,
\eeq 
where the spinor indices of $E^c$ and $U^c$ are contracted anti-symmetrically
$\epsilon^{\alpha\beta} E^c_\alpha U^c_\beta \equiv (E^c)^T i\sigma^2 U^c$. In four-component notation
this is
\beq
{\mathcal L} \supset 
 \lambda_{eu} S_{eu} e_R^T i\sigma^2 u_R  + h.c.
=\lambda_{eu} S_{eu} e^T i\gamma^0\gamma^2 P_R u +h.c. \ ,
\label{eq:rightScoupling}
\eeq 
where we used $e=(e_L,e_R)$ and $u=(u_L,u_R)$ for the Dirac spinors of the fermions so that, for example, $E^c=i \sigma^2 e_R^*$. Note also that the leptoquark $S_{eu}$ couples to a lepton and a quark, i.e. no anti-particles.
\eqref{rightScoupling} also defines our normalization for the coupling $\lambda_{eu}$. 

Similarly, the vector leptoquark coupling to positrons and up quarks is
\beq
{\mathcal L} \supset  \lambda_{eu} V^*_{\mu\,eu} (U^c)^\dag \bar \sigma^\mu E^c
= - \lambda_{eu} V^*_{\mu\,eu}  e_R^\dag \sigma^\mu u_R
=  -\lambda_{eu} V^*_{\mu\,eu}\, \bar e \gamma^\mu P_R u \ ,
\label{eq:rightVcoupling}
\eeq 
plus the hermitian conjugate. We see that in this case the leptoquark couples to an anti-lepton and a quark.

\section{Features in the production and decay for scalar vs. vector leptoquarks}
\label{sec:productionanddecay}

\subsection{Pair production}
In~\cite{Diaz:2017lit},  we compared the cut efficiencies for scalar vs. vector leptoquarks in the limit of small $\lambda$ and found that the relative cut efficiencies are similar with up to $10\%$ differences. Therefore there is no need to optimize cuts differently for scalar and vector leptoquark searches. We do not expect this to change significantly at large $\lambda$ in the parameter space which is still allowed by the pair-production bounds. This is because there the leptoquark mass is rather large ($\gtrsim$ 1 TeV) and the distributions of final state particles are largely determined by the decay kinematics.  

\subsection{Single production}
\label{sec:spcuts}
Here we investigate if the similarity between cut efficiencies for scalar and vector leptoquarks also holds in the case of single-production. We take $\phi_{eu}$ as an example and consider the process $p p \to e^+ e^- j$. More specifically, we use the leptoquark-couplings
$S^*_{eu} E^c U^c$ and $V^*_{\mu\,eu} (U^c)^\dag \bar \sigma^\mu E^c$ for the scalar and vector cases, respectively (together with their conjugates). The relevant Feynman diagrams are SP-1 and SP-2 in~\figref{mypairdiagrams}.

\begin{figure}[t]
   \centering
   \includegraphics[width=0.55\textwidth]{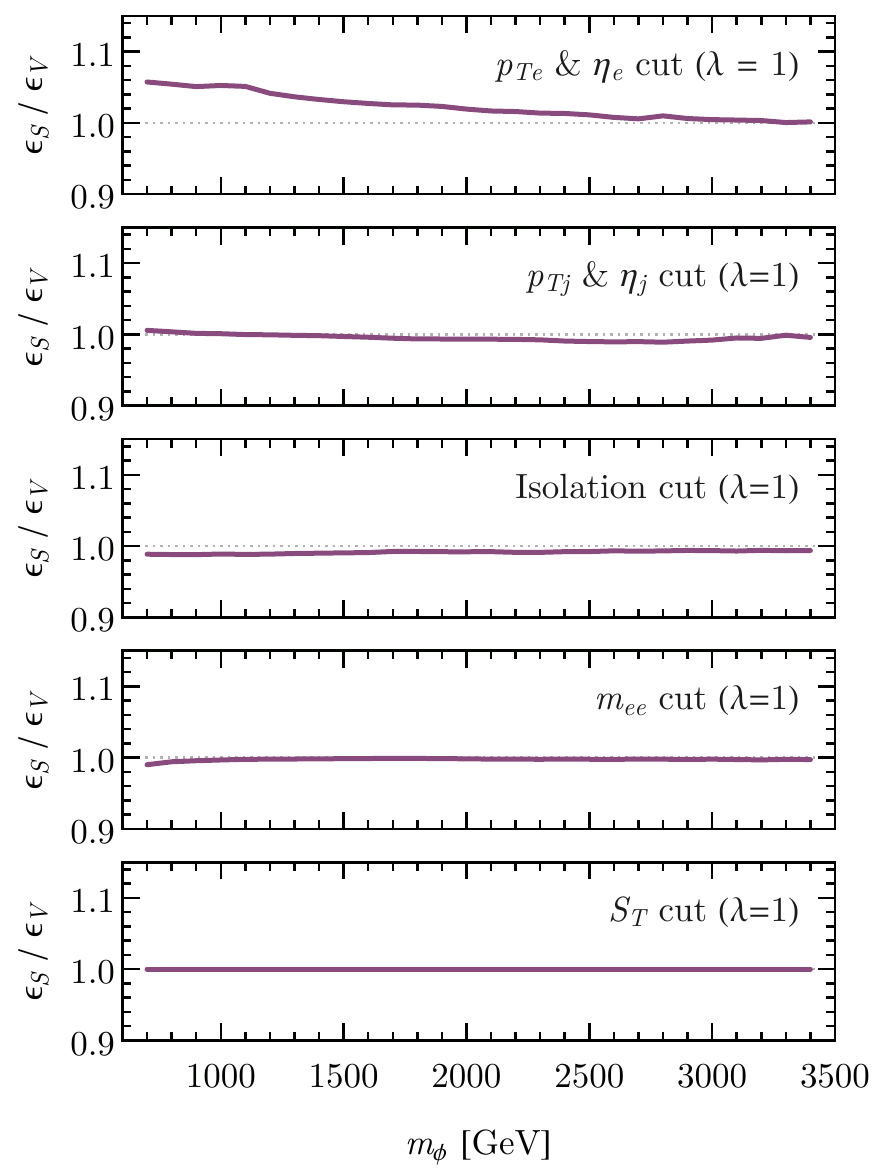} % requires the graphicx package
   \caption{Ratios of relative signal efficiencies for scalar and vector leptoquarks as functions of leptoquark mass $m_\phi$ and for fixed coupling $\lambda=1$. We take $\phi_{eu}$ as an example and investigate the process $pp \to e^+ e^- j$. The panels show the relative efficiencies of a series of selection cuts from the CMS search~\cite{Khachatryan:2015qda}. See text for more details.}
   \label{fig:scalarvsvector}
\end{figure}

We simulate the process for LHC 13 TeV with \texttt{MadGraph 5 aMC@NLO 2.6.0}~\cite{Alwall:2014hca} (\texttt{MG5} for short in below) with UFO-format model files built in \texttt{FeynRules 2.3.27}~\cite{Alloul:2013bka}. We perform a leading-order (LO) matrix-element-level (ME) simulation with CTEQ6L1 PDF and set the factorization and renormalization scales to the leptoquark mass. Basic cuts from the CMS analysis for a leptoquark single-production search~\cite{Khachatryan:2015qda} are adopted and summarized in \tabref{sumarryofcuts} (a comprehensive summary of the basic and higher level cuts are shown in~\tabref{sumarryofcutsfull}). We compute the relative efficiencies, $\epsilon_S$ and $\epsilon_V$, at each level of cuts listed in \tabref{sumarryofcuts} for scalar and vector leptoquark's respectively. We then plot the ratio $\epsilon_S/\epsilon_V$ as a function of $m_\phi$ for $\lambda_\phi = 1$ in~\figref{scalarvsvector}.

\begin{table}[!htbp]
   \centering
   \topcaption{Summary of basic cuts for CMS  single leptoquark production $eej$ search~\cite{Khachatryan:2015qda}. Leptons are labeled in order of decreasing transverse-momentum, $p_\text{T}$.  The full list of selection cuts in~\cite{Khachatryan:2015qda} can be found in~\tabref{sumarryofcutsfull}.}
   \label{tab:sumarryofcuts}
   \begin{tabular}{@{} lc @{}} % Column formatting, @{} suppresses leading/trailing space
      \hline
      Generator & $p_\text{T$e_{1,2}$} > 20$ GeV, $p_\text{T$j$} > 50$ GeV,  $|\eta_{e_{1,2}}|<2.6$, $|\eta_j|<2.6$\\
      \hline
      Trigger      & Two electrons $p_\text{T} > 33$ GeV, $|\eta|<2.4$ \\
      \hline
      Electron $p_\text{T}$ \& $\eta$         & $p_\text{T$e_{1,2}$} > 45$ GeV, $|\eta_{e_{1,2}}|< 1.442$ or  $1.56 < |\eta_{e_{1,2}}|< 2.1$  \\
      Jet $p_\text{T}$ \& $\eta$         & $p_\text{T$j$} > 125 \gev$, $|\eta_{j}|<2.4$ \\
      Isolation    &  $\Delta R_{e e}$ > 0.5, $\Delta R_{e j} > 0.3$ \\
      $m_{ee}$ & $> 110 \gev$\\
      $S_\text{T}$ &  $>  250 \gev$ \\
      \hline
   \end{tabular}

\end{table}

\begin{figure}[t]
   \centering
   \includegraphics[width=0.33\textwidth]{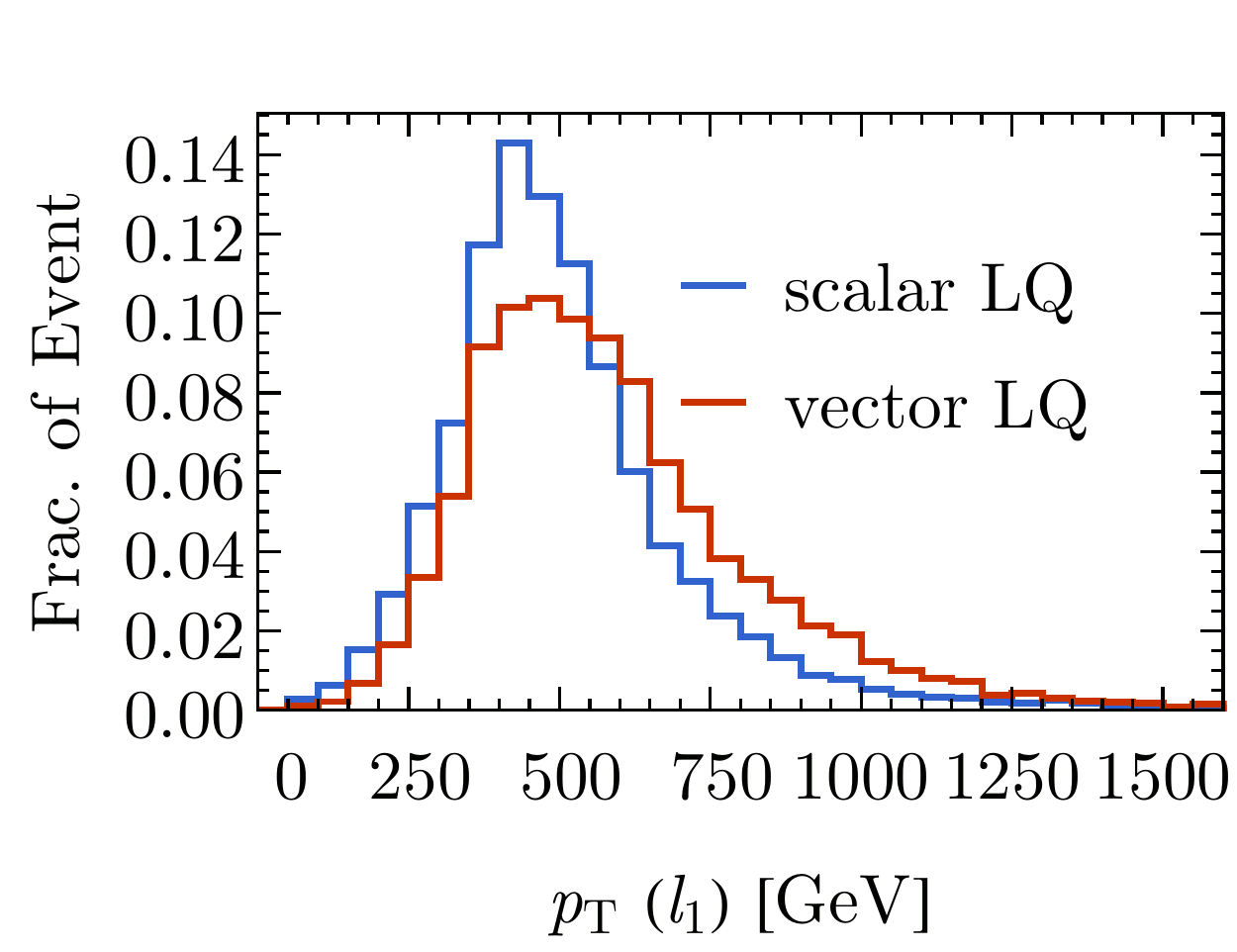}~\includegraphics[width=0.33\textwidth]{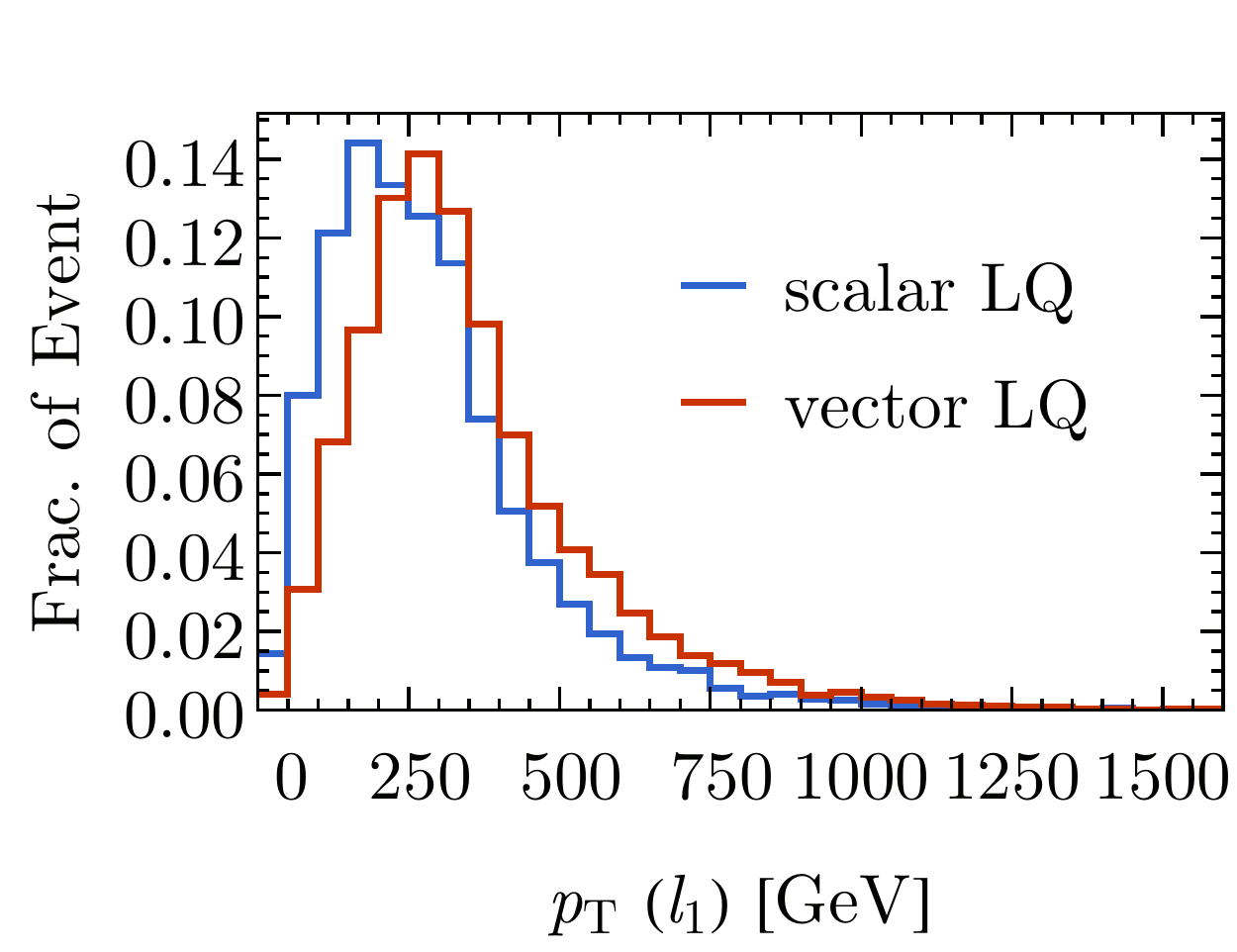}\\
   \includegraphics[width=0.33\textwidth]{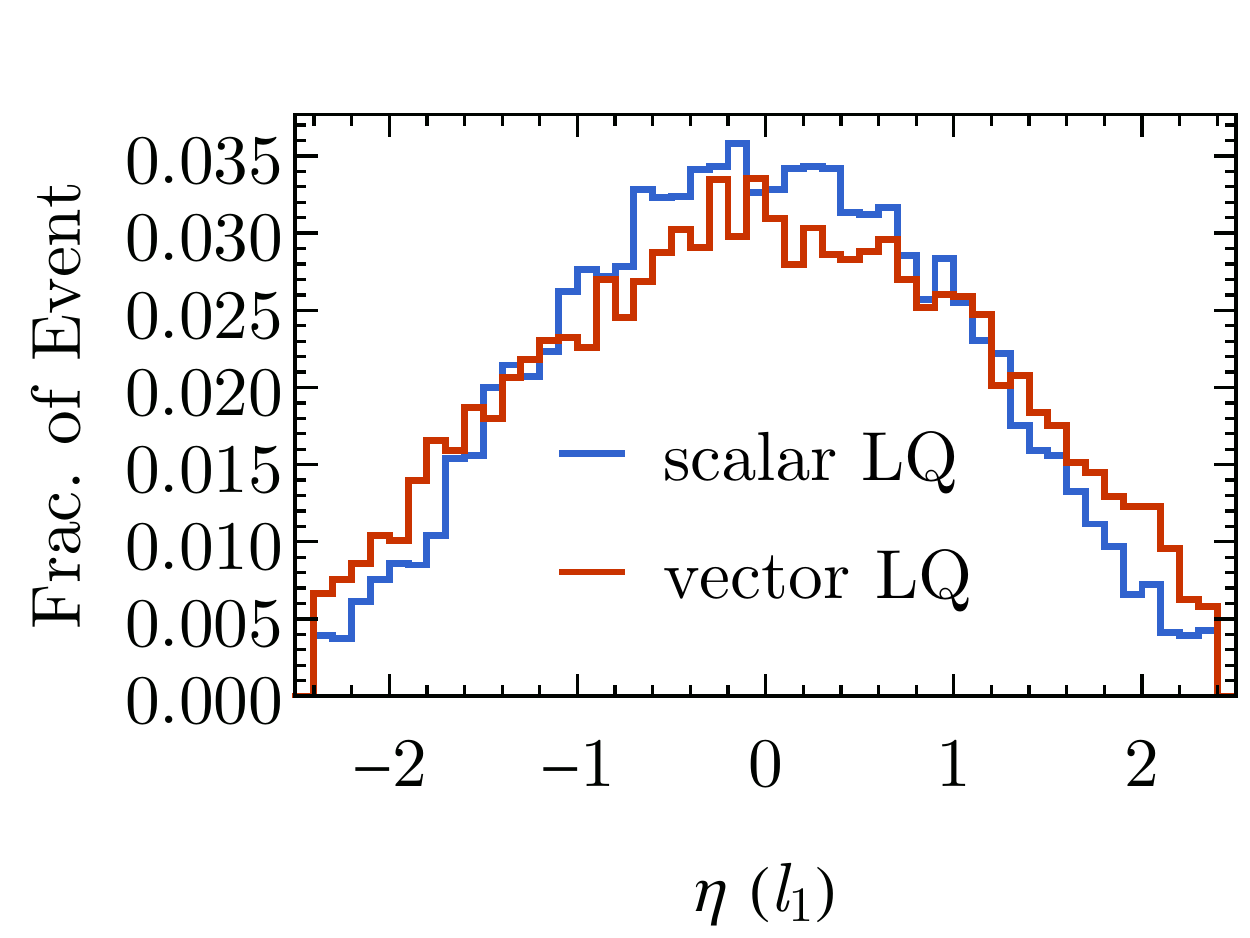}~\includegraphics[width=0.33\textwidth]{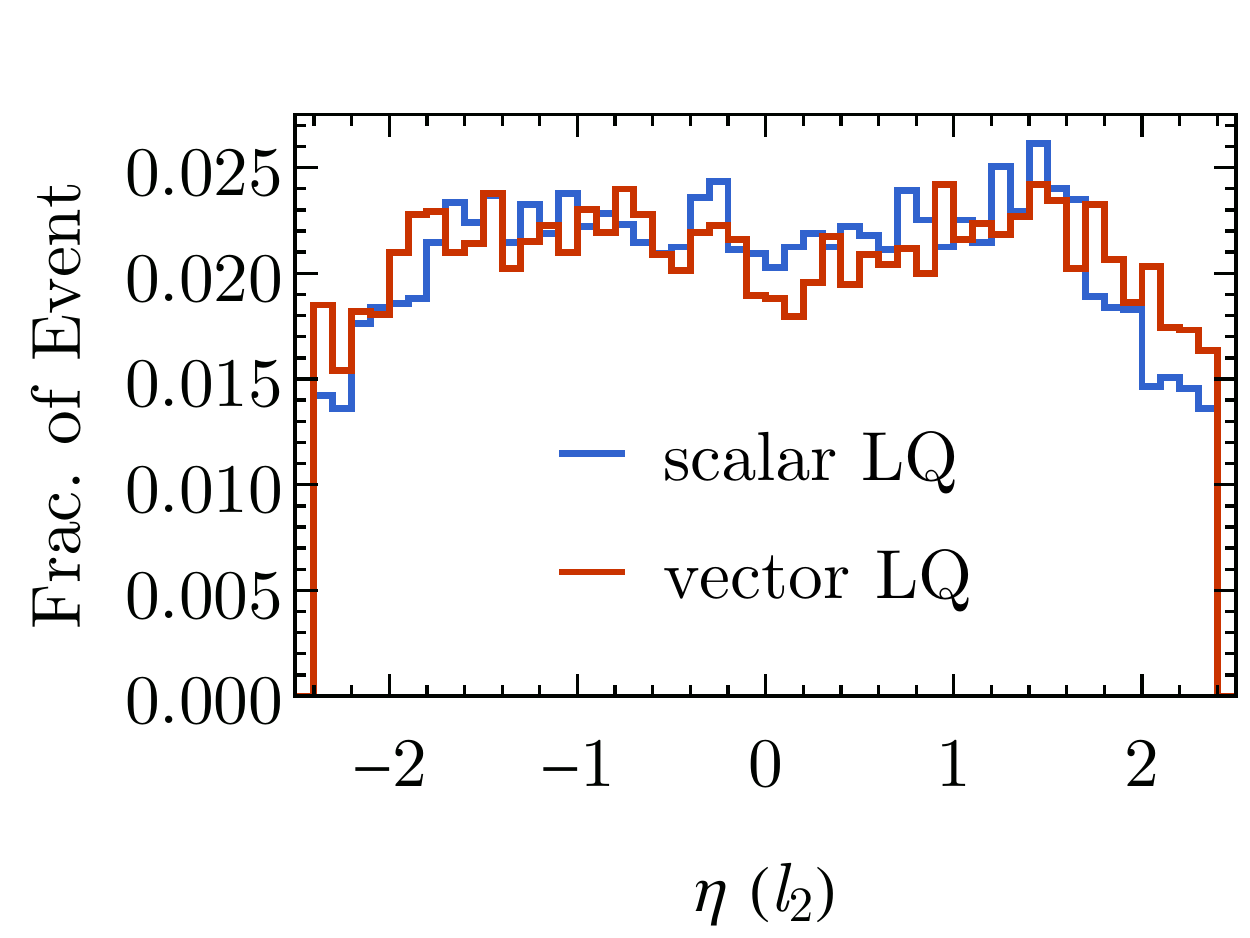} % requires the graphicx package
   \caption{$\pt$ and $\eta$ distribution for the two electrons given $m_{\phi_{eu}} = 1 \tev$ and $\lambda_{\phi_{eu}} = 1$.}
   \label{fig:disA}
\end{figure}

Overall, we see that the cut efficiencies for the vector and scalar cases are very similar. The overall differences in $\epsilon_S$ and $\epsilon_V$ are within 10\%. Looking more closely, one notices that $\epsilon_S/\epsilon_V>1$ for the cut on lepton transverse-momentum $p_{\text{T} e}$ and lepton pseudo-rapidity $\eta_e$ at small leptoquark masses. Thus more leptons fail either the $\pt$ or the $\eta$ cut in the vector leptoquark case. To understand this behavior we plotted the distributions of the leptons for both scalar and leptoquark cases in~\figref{disA}. Comparing the scalar vs. vector $\pt$ of the second lepton (panel 2) one sees that lepton is softer in the case of the scalar. However this difference has only negligible impact on cut efficiencies because all leptons are sufficiently hard to easily pass the $\pt$ cuts. More important is that the second lepton in the vector case has a slightly more forward distribution in $\eta$ (panel 4), it is therefore more likely to fail the $|\eta|<2.1$ cut. This difference explains the slightly lower efficiency of the lepton cuts in the vector case~\figref{scalarvsvector}.

\begin{figure}[t]
   \centering
   \includegraphics[width=0.35\textwidth]{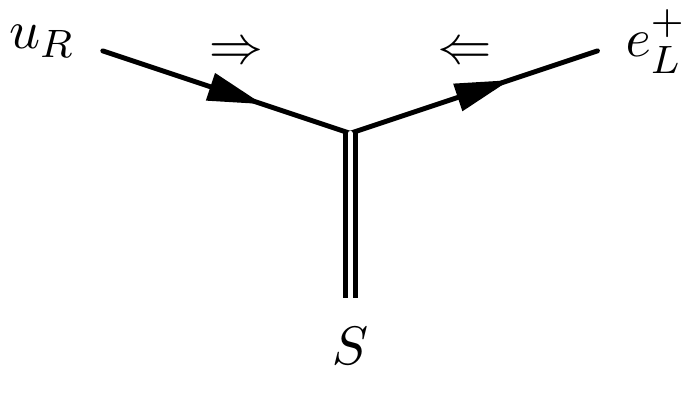}\quad\quad\includegraphics[width=0.35\textwidth]{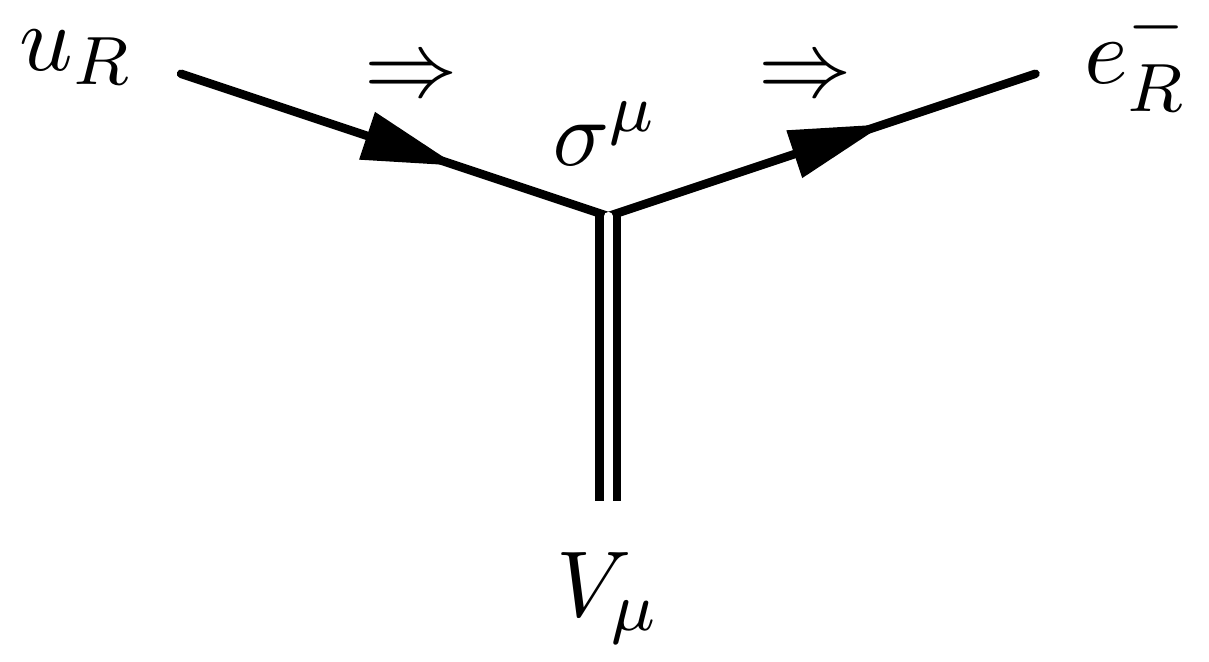} % requires the graphicx package
   \caption{Vertices for $e^-$ in scalar and vector scenarios. $\Leftarrow$ and $\Rightarrow$ indicates the spin directions.}
   \label{fig:spinflip}
\end{figure}

We can understand both behaviors, that for the scalar case the softer of the two leptons is softer but also more central (small $|\eta|$)  than for the vector case by considering the fermion spins. First, note that the lepton from the decay of the leptoquark is typically the harder and more central of the two. Thus for understanding the cut efficiencies we are interested in the other lepton which stems from the $t$-channel-like vertex shown in~\figref{spinflip} where we use $\Rightarrow$ and $\Leftarrow$ to label the spin direction. The interaction for the scalar leptoquark $e_R^T i\sigma^2 u_R$ annihilates a right helicity quark and creates a left-helicity positron, thus the interaction would require a spin flip if the positron were to go in the forward direction. This gives rise to a $(1-\cos \theta)$ dependence  of the cross section for the angle between the incoming quark and outgoing positron. Thus the positron tends to be central or even backwards relative to the direction of the incoming quark. This also explains why it is not as energetic because the overall event is typically boosted in the direction of the incoming quark (which is harder than the gluon). For the vector,  interaction $e_R^\dag \sigma^\mu u_R$ preserves the helicity through the interaction. Therefore the fermion spin overlap gives a $(1+\cos\theta)$ in this case, favoring the forward direction. In addition, for light leptoquarks, there is also a $t$-channel forward singularity further favoring large $\eta$. However, for larger leptoquark masses the $t$-channel enhancement of the forward direction is suppressed  and the efficiency of the $p_\text{T}$ and $\eta$ cuts between scalar and vector leptoquark become more similar.

\subsection{Drell-Yan}

\begin{figure}[t]
   \centering
   \includegraphics[width=0.35\textwidth]{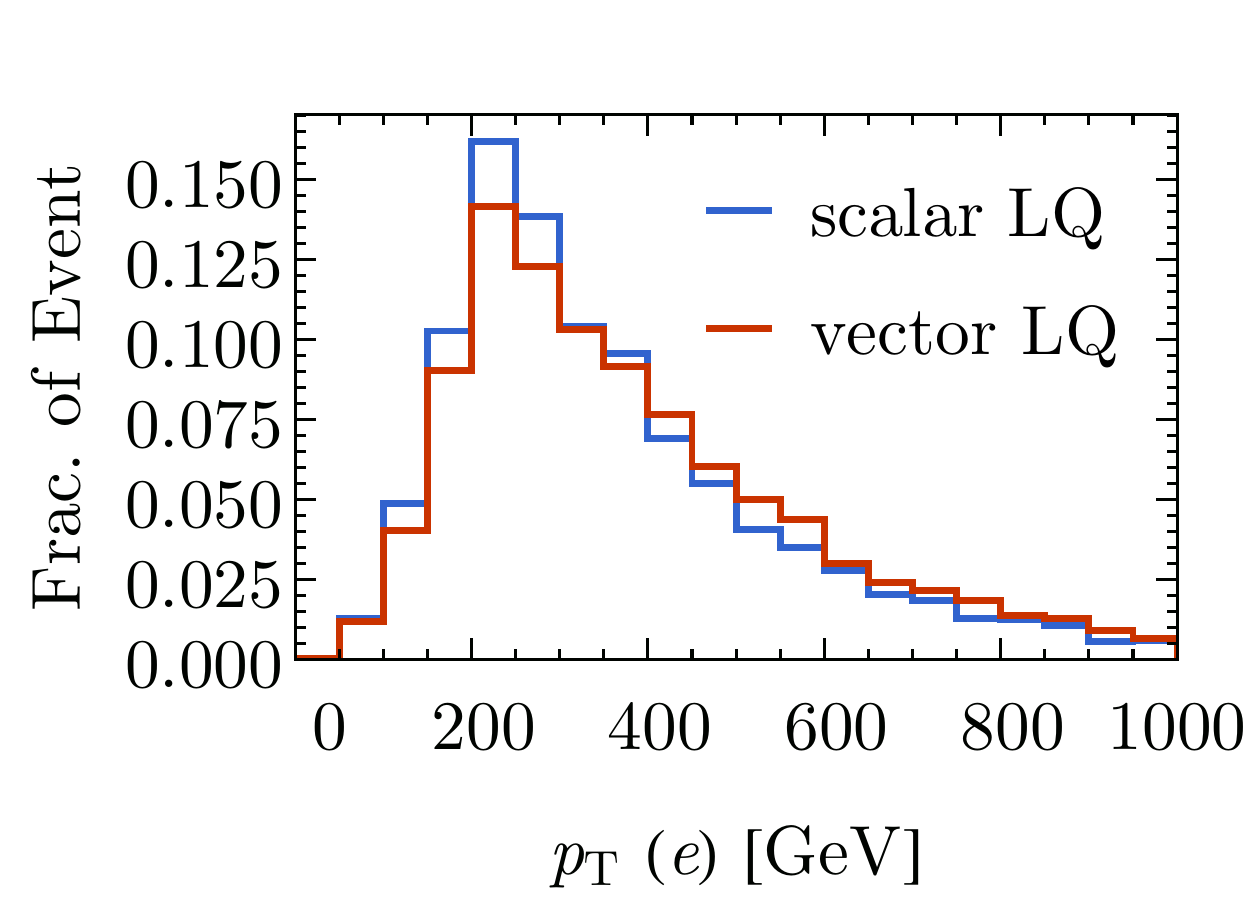}~\includegraphics[width=0.33\textwidth]{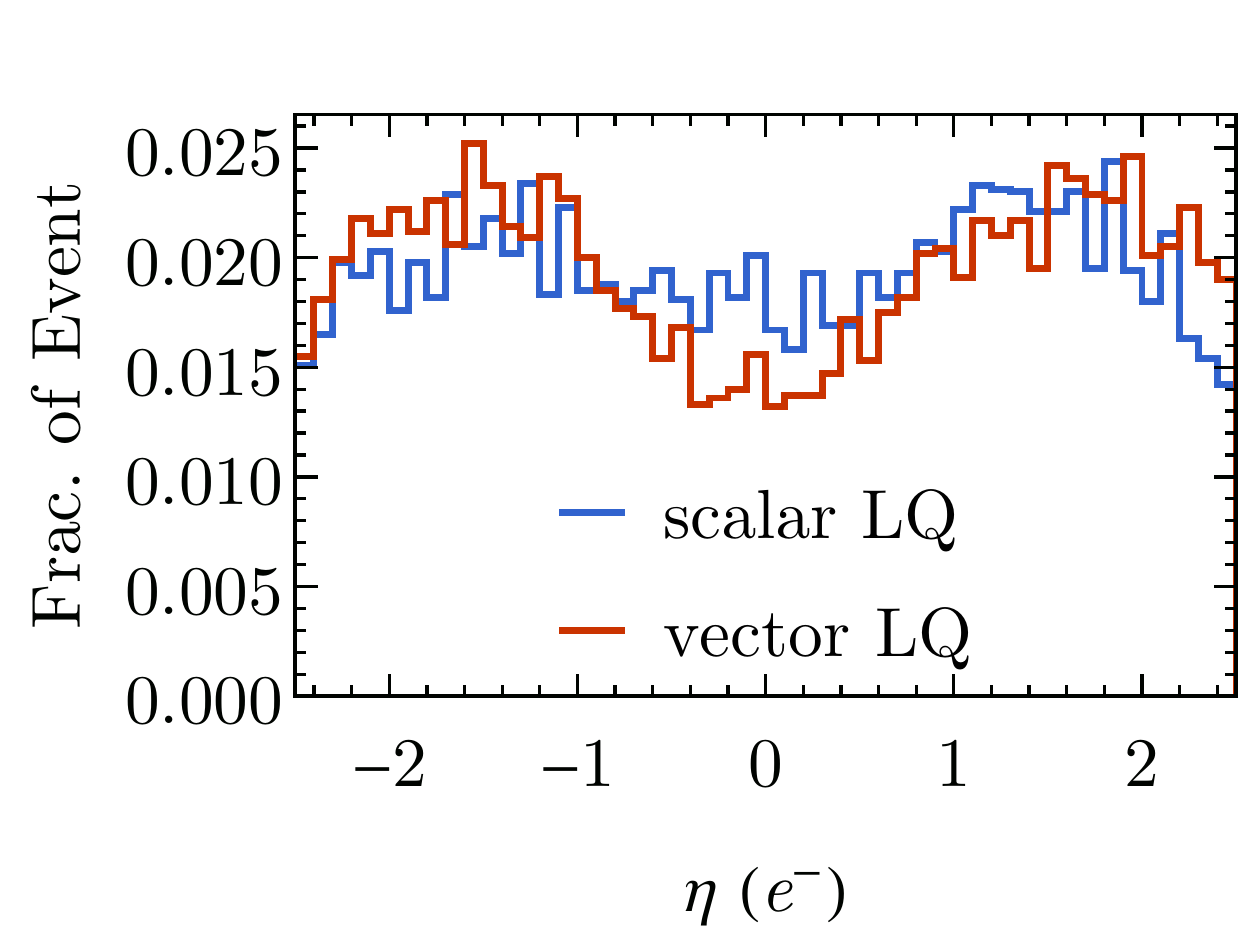}~\includegraphics[width=0.33\textwidth]{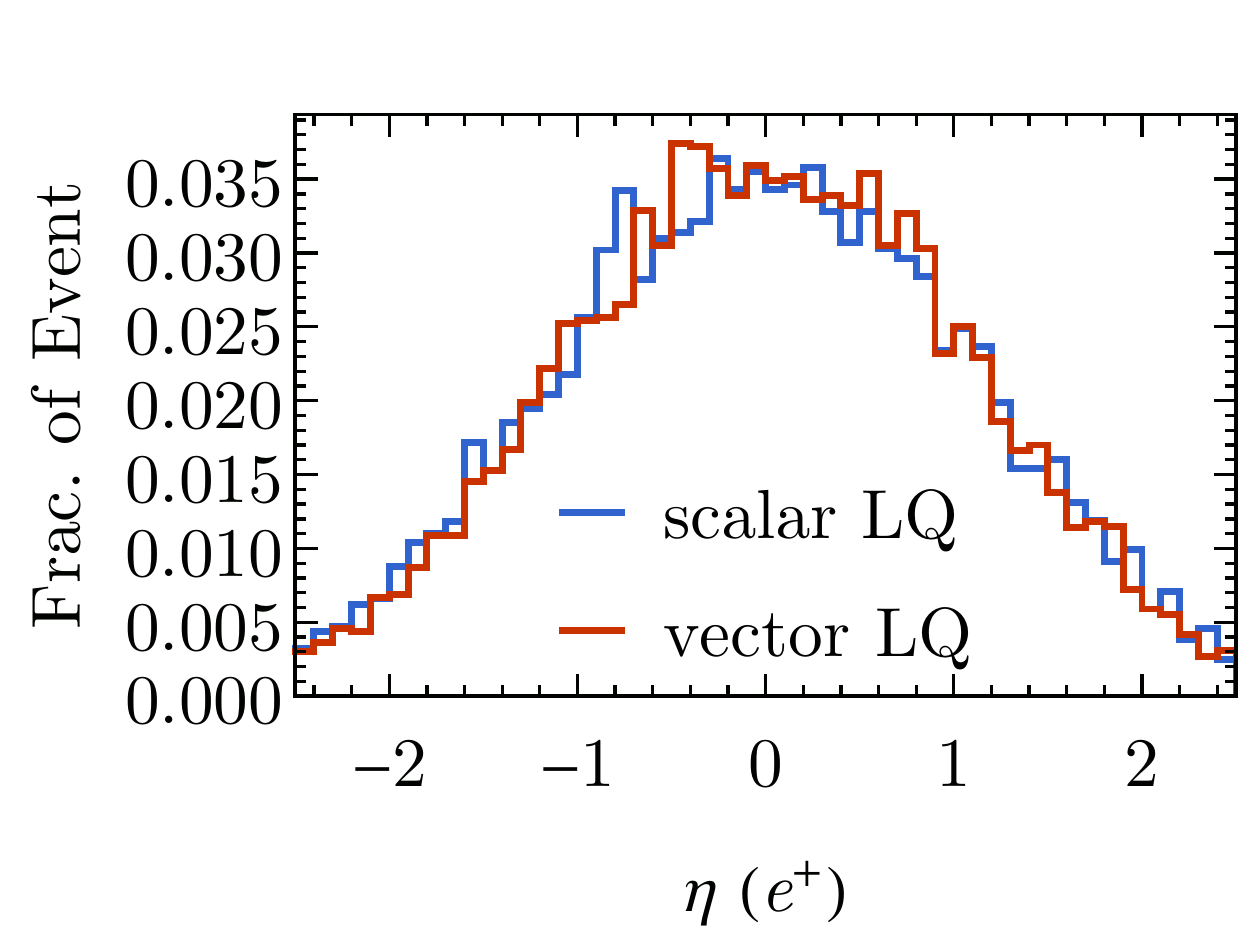}
    % requires the graphicx package
   \caption{$\pt$ and $\eta$ distributions for electrons and positrons in the Drell-Yan process $p p \to e^+ e^-$ with $t$-channel exchange of $\phi_{eu}$. Here we set $m_{\phi_{eu}} = 1.5 \tev$ and $\lambda_{\phi_{eu}} = 1$.}
   \label{fig:kindisDY}
\end{figure}

Here we compare cut efficiencies for the DY-like process $p p  \to e^+ e^-$ for $t$-channel leptoquark exchange. The $t$-channel diagram is non-resonant and it interferes with the SM photon and $Z$ diagrams, giving rise to interesting asymmetries and angular distributions~\cite{Bansal:2018eha}. Here we focus only on the efficiencies, comparing the leptoquarks  
$S_{eu} e_R^T i\sigma^2 u_R$ and $V^*_{\mu\,eu} e_R^\dag \sigma^\mu u_R$. We simulate events using \texttt{MG5} (LO, ME, ${\rm PDF} = {\rm NNPDF2.3LO}$, ${\rm scale} = m_{ee}$). For both models we choose $\lambda_\phi =1$ and scan over leptoquark masses from 500 GeV to 3600 GeV in 100 GeV intervals. As in the case of the single production we expect the leptons to be more aligned with the direction of the initial quarks in the vector case because of spin conservation. Thus the vector case predicts an $e^{-}$ distribution that is boosted towards larger $\eta$ as the $e^-$ goes in the direction of the initial $u$-quark which statistically is expected to have more energy.  This can be observed in the $\eta$-distributions shown in~\figref{kindisDY}.

This peaking of leptons in the forward - large $\eta$ - directions leads to a suppression of the $\eta$ cut efficiency in the vector case. However, for the heavy leptoquark masses and couplings where the DY analysis is interesting the difference in efficiencies is less than 10\%. For lighter leptoquarks this efficiency difference would have been enhanced by the $t$-channel singularity for forward scattering. 

The generator-level and basic DY search cuts, adopted from~\cite{Aaboud:2017buh}, are summarized in~\tabref{sumarryofcutsDYATLAS}.

\begin{table}[htbp]
   \centering
   \topcaption{Summary of basic cuts for the ATLAS DY $ee$ search~\cite{Aaboud:2017buh}.}
   \label{tab:sumarryofcutsDYATLAS}
   \begin{tabular}{@{} lc @{}} % Column formatting, @{} suppresses leading/trailing space
            \hline
            Generator & $p_{\text{T}e} > 15 \gev$, $|\eta_e| < 2.6$, $m_{ee} > 400 \gev$\\
            \hline
      Trigger      & Two electrons $E_\text{T} > 17$ GeV\\
      \hline
      Electron $p_\text{T}$        & $p_\text{T$e_{1, 2}$} > 30$ GeV  \\
      \& $\eta$         & $|\eta_{e_{1, 2}}|< 1.37$ or  $1.52 < |\eta_{e_{1, 2}}|< 2.47$  \\
      Isolation    &  $\Delta R_{e e} > \max\{0.2, 10\gev/p_\text{T$e_{1, 2}$} \}$  \\
      \hline
   \end{tabular}
\end{table}

\section{Leptoquark searches organized by the leptoquark matrices}
\label{sec:searches}

\subsection{Summary of relevant LHC searches}

\begin{figure}[t]
   \centering
   \includegraphics[width=1\textwidth]{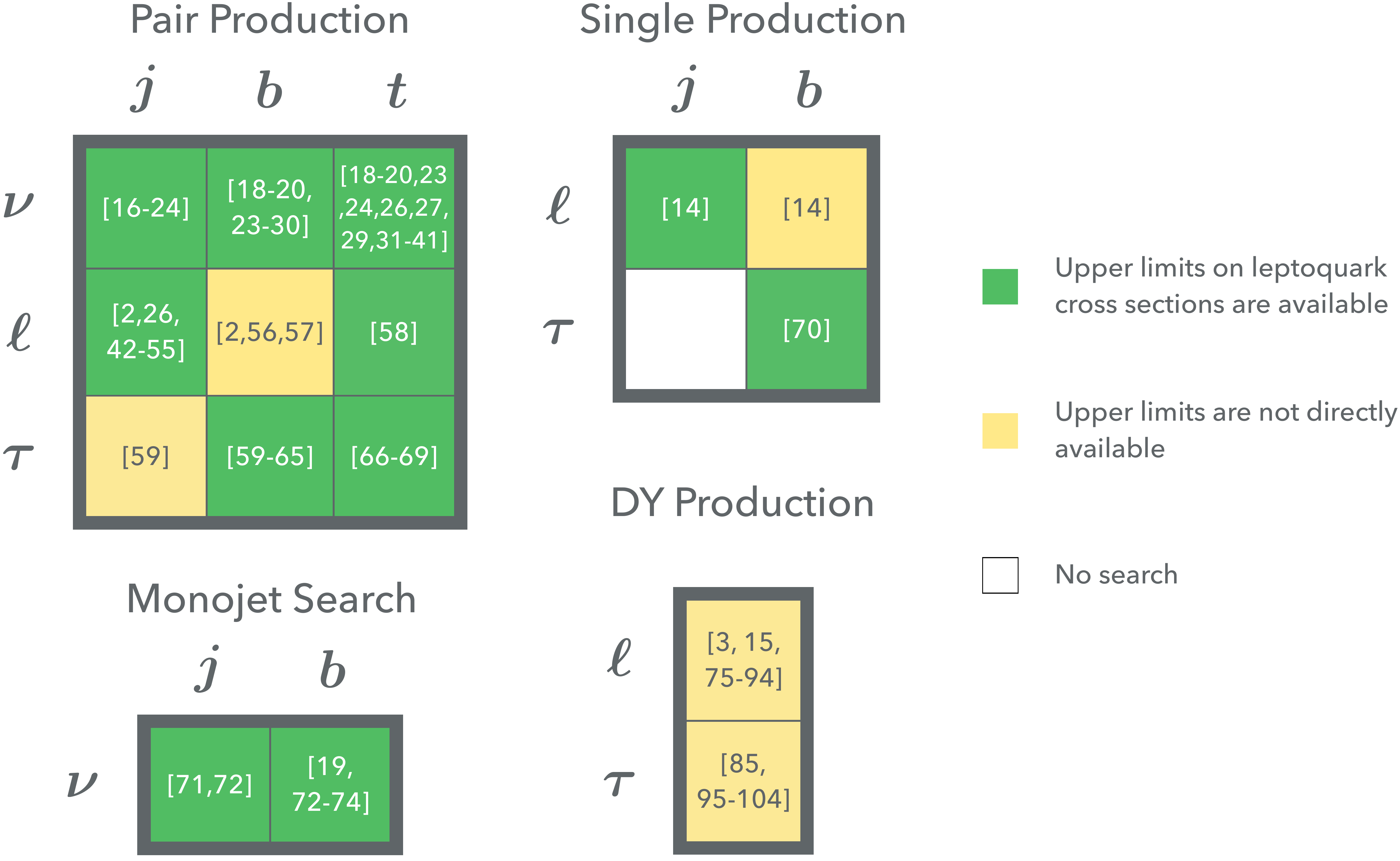}
   \vspace{1mm} % requires the graphicx package
   \caption{Summary of the current status of searches for pair (upper-left), single (upper-right), monojet (lower-left) and DY production (lower-right) of leptoquarks at the LHC after Runs 1 and 2.  Each entry corresponds to a specific final state in an experimental search. We include searches for leptoquarks, dark matter, and supersymmetry, which have identical final states and decay topologies. For such searches one can directly obtain limits on the LQ cross section from the experimental papers, we color such matrix elements in green. Final states for which we obtained limits on the cross section indirectly by reinterpreting a related search are colored in yellow. Final states for which no search has been performed so far are left white. The numbers give the references relevant to the various final states. Pair production matrix:
\hyperref[sec:nujbt]{$({\nu j})$}~\cite{Chatrchyan:2013mys, Khachatryan:2015vra, CMS-PAS-SUS-16-033,CMS-PAS-SUS-16-036,Sirunyan:2017cwe,Aad:2015gna, ATLAS-CONF-2017-022, CMS-PAS-SUS-18-001, Sirunyan:2018kzh}, 
\hyperref[sec:nujbt]{$({\nu b})$} ~\cite{Aad:2013ija, Aad:2015caa, Aad:2015pfx, Aaboud:2016nwl, Sirunyan:2016jpr,Sirunyan:2017cwe,ATLAS-CONF-2017-38,CMS-PAS-SUS-16-033,CMS-PAS-SUS-16-036, CMS-PAS-SUS-18-001,Sirunyan:2018kzh}, 
\hyperref[sec:nujbt]{$({\nu t})$}~\cite{Chatrchyan:2013xna, Aad:2014kra, Aad:2015caa, Aad:2015pfx,  ATLAS-CONF-2016-077, Sirunyan:2016jpr, Khachatryan:2016pxa, Khachatryan:2016oia, Khachatryan:2016pup, ATLAS-CONF-2017-020,Sirunyan:2017cwe,ATLAS-CONF-2017-024,CMS-PAS-SUS-17-001,CMS-PAS-SUS-16-049,CMS-PAS-SUS-16-033,CMS-PAS-SUS-16-036, Sirunyan:2017wif, CMS-PAS-SUS-18-001,Sirunyan:2018kzh}, 
\hyperref[sec:emuj]{$({ej})$} and/or
\hyperref[sec:emuj]{$({\mu j})$}~\cite{Khachatryan:2010mq, Khachatryan:2010mp, Aad:2011ch, Aad:2011uv, Chatrchyan:2011ar, ATLAS:2012aq, Chatrchyan:2012vza, CMS-PAS-EXO-12-041, CMS-PAS-EXO-12-042, Aad:2015caa, Khachatryan:2015vaa, Aaboud:2016qeg,CMS-PAS-EXO-16-007, CMS-PAS-EXO-16-043,CMS-PAS-EXO-17-009,CMS-PAS-EXO-17-003},
\hyperref[sec:emub]{$({eb})$} and 
\hyperref[sec:emub]{$({ \mu b})$}~\cite{ATLAS-CONF-2015-015,ATLAS-CONF-2017-036, CMS-PAS-EXO-17-009},
\hyperref[sec:emut]{$({ \mu t})$}~\cite{Sirunyan:2018ruf},
\hyperref[sec:tauj]{$({\tau j})$}~\cite{CMS-PAS-EXO-17-016},
\hyperref[sec:taub]{$({\tau b})$}~\cite{Chatrchyan:2012sv,Chatrchyan:2012st,ATLAS:2013oea,Khachatryan:2014ura,Khachatryan:2016jqo,Sirunyan:2017yrk,CMS-PAS-EXO-17-016}, and 
\hyperref[sec:taut]{$({\tau t})$}~\cite{CMS-PAS-EXO-12-030,CMS-PAS-EXO-13-010,Khachatryan:2015bsa,Sirunyan:2018nkj}; 
Single production matrix:
\hyperref[sec:emujb]{$({\ell j})$}~\cite{Khachatryan:2015qda},
\hyperref[sec:emujb]{$({\ell b})$}~\cite{Khachatryan:2015qda},
\hyperref[sec:taujb]{$({\tau b})$}~\cite{CMS-PAS-EXO-17-029}; 
Monojet search matrix: \hyperref[sec:nujbsp]{$({\nu j})$}~\cite{Sirunyan:2017jix, Aaboud:2017phn},\hyperref[sec:nujb]{$({\nu b})$}~\cite{CMS-PAS-SUS-16-036, Aaboud:2017phn, ATLAS-CONF-2016-085, Aaboud:2017rzf};
DY production matrix: \hyperref[sec:emuj]{$({\ell j})$ or (${\ell b}$})~\cite{Aad:2011xp, Collaboration:2011tt, Collaboration:2011dca, Aad:2011tq, CMS-PAS-EXO-11-019,  Chatrchyan:2012it, Chatrchyan:2012kc, Chatrchyan:2012oaa, CMS-PAS-EXO-12-015, CMS-PAS-EXO-12-020, Aad:2012cfr, Aad:2012hf, Aad:2012bsa, Aad:2014wca, Khachatryan:2014fba, CMS:2015ooa, Aaboud:2016cth, CMS-PAS-EXO-16-031, Khachatryan:2016zqb, Aaboud:2017buh, Sirunyan:2018exx, Sirunyan:2296689}, 
   \hyperref[sec:emuj]{$({\tau j})$ or $({\tau b})$}~\cite{Aad:2011rv, Chatrchyan:2012hd, Aad:2012gm, Aad:2012cfr, Aad:2014vgg, Khachatryan:2014wca, Aad:2015osa, Khachatryan:2016qkc, Aaboud:2016cre, Aaboud:2017sjh, Sirunyan:2018zut}.}
   \label{fig:review}
\end{figure}

Because of the PDF-dependence of the cross sections bounds on leptoquarks coupling to the different quark-flavors are not simply related. However, from the experimental side, searches for leptoquarks which couple to any of the light quarks involve the same final state (light jets plus leptons), therefore the final states are a useful scheme for classifying searches. In~\figref{review}, we organize leptoquark searches according to their final states into matrices for pair-production, single-production, DY-production, and the monojet-search.  Each distinct matrix element corresponds to a different final state in experimental searches ($j$, $b$, $\nu$, and $\ell$ stand for a light-jet, a $b$-jet, a neutrino, and an electron or a muon, respectively). For example, the entry  in the pair-production matrix $(\ell j)$ denotes searches for the final state $j j\ell^+ \ell^-$.  The corresponding entries in the single-production matrix and the DY-production matrix indicate searches for the final states $j \ell^+ \ell^-$ and the Drell-Yan-like $\ell^+ \ell^-$, respectively. The numbers in the matrices point to references relevant to the various final states. Of course, there is some overlap between ``DY'', ``single'' and ``pair'' production due to cut efficiencies, emission of additional jets at NLO, and fragmentation. However, since jets from the decay of leptoquarks are isolated and very energetic they are not likely to be faked by soft or collinear radiation. Therefore this separation into final states with different numbers of hard jets is useful even in the presence of higher order corrections.\footnote{Note that our simulations are done at the LO ME level. Partial NLO computations for processes involving scalar leptoquarks do exist in the literature, and are incorporated in the ``leptoquark toolbox'' \texttt{MG5} models \cite{Dorsner:2018ynv}.} The monojet-search matrix applies to leptoquarks coupling to neutrinos and light quarks or $b$-quarks. Those searches are simultaneously sensitive to single-production, DY production with initial state radiation, and pair-production because monojet searches usually allow a second hard jet.

\subsection{$\ell u$, $\ell d$, $\ell s$, $\ell c$, and $\ell b$}

\begin{figure}[!htbp]
   \centering
   \includegraphics[width=0.35\textwidth]{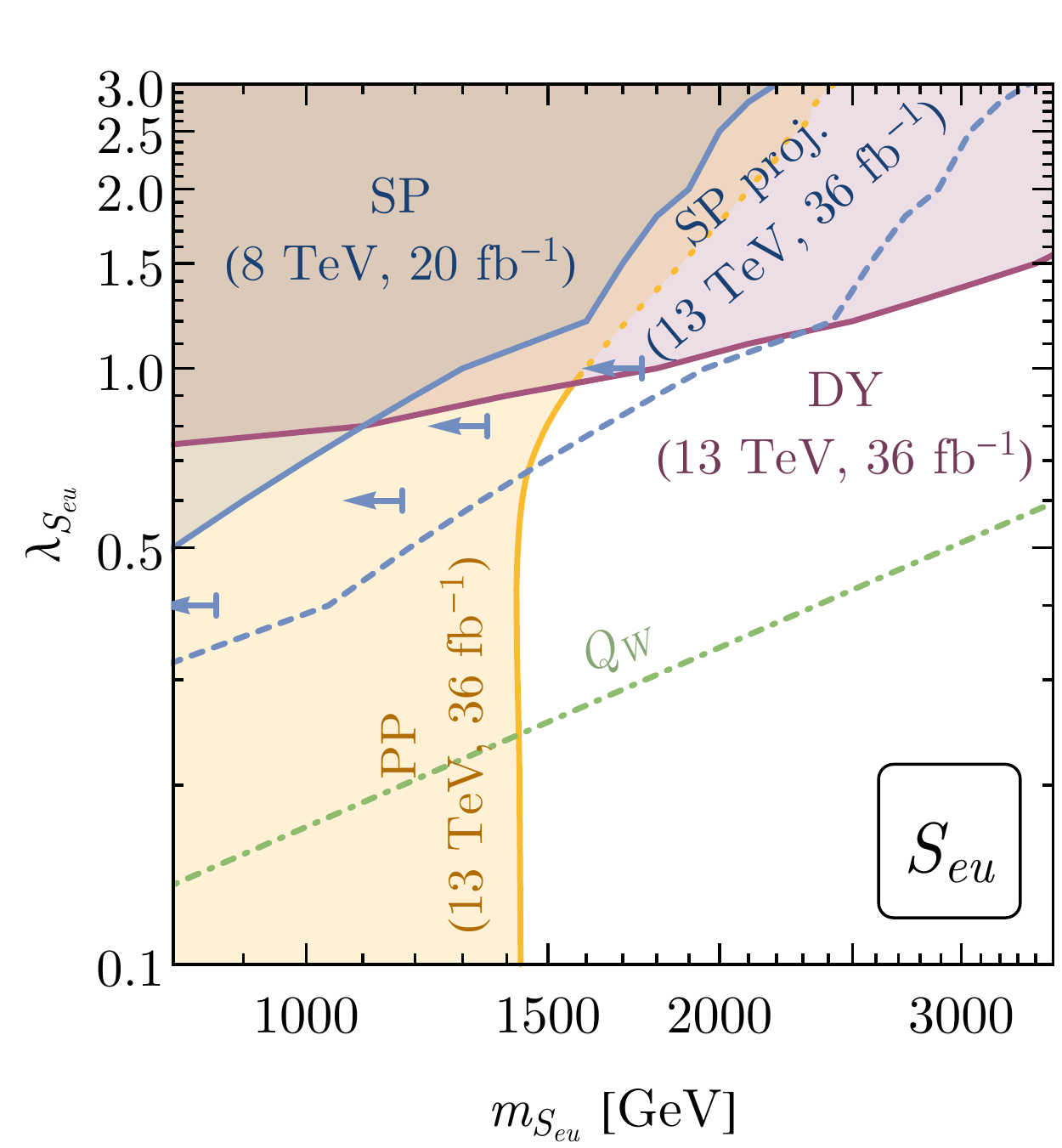}~\includegraphics[width=0.35\textwidth]{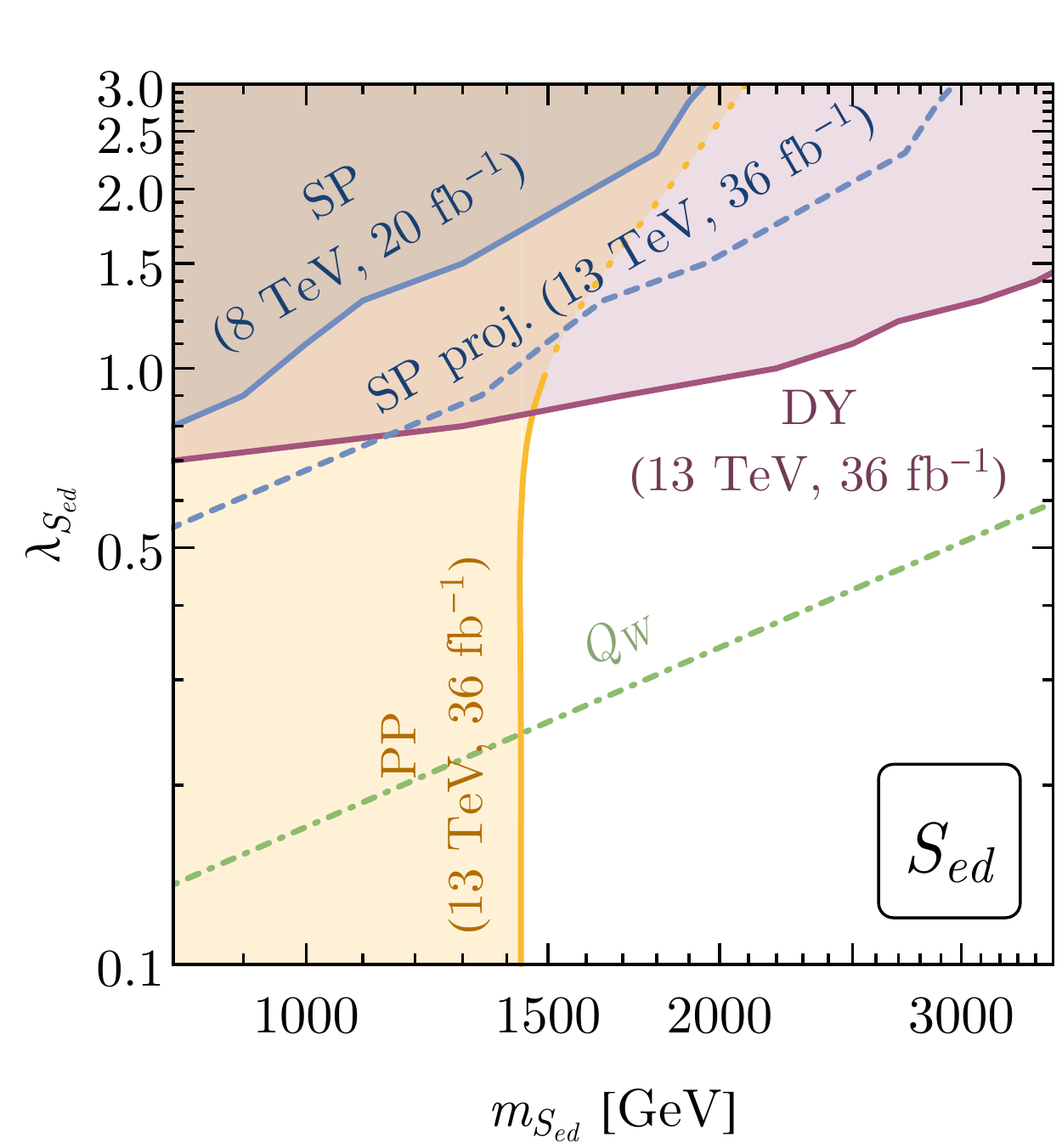}\\
   \includegraphics[width=0.35\textwidth]{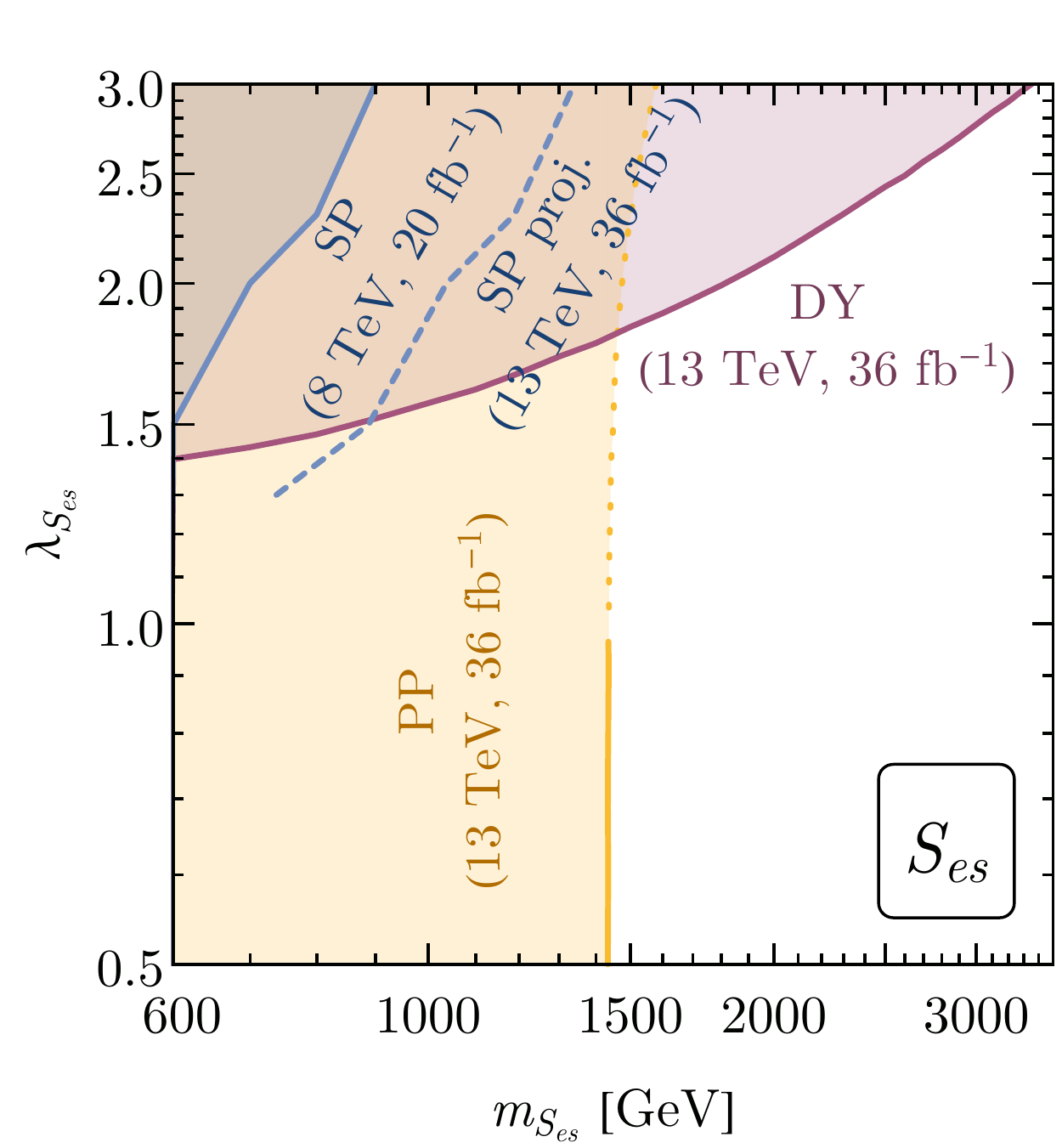}~\includegraphics[width=0.35\textwidth]{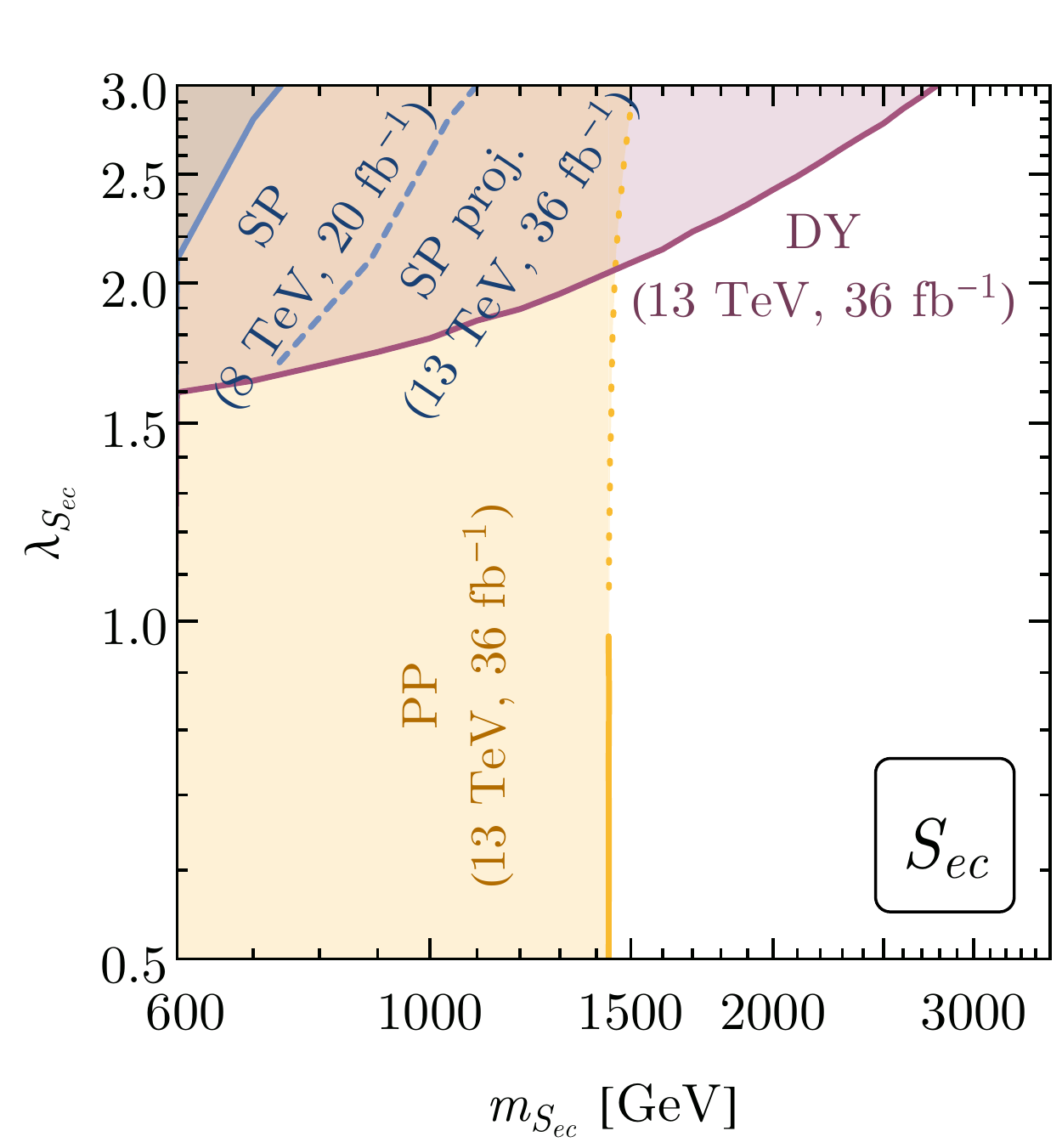}
   \\ \includegraphics[width=0.35\textwidth]{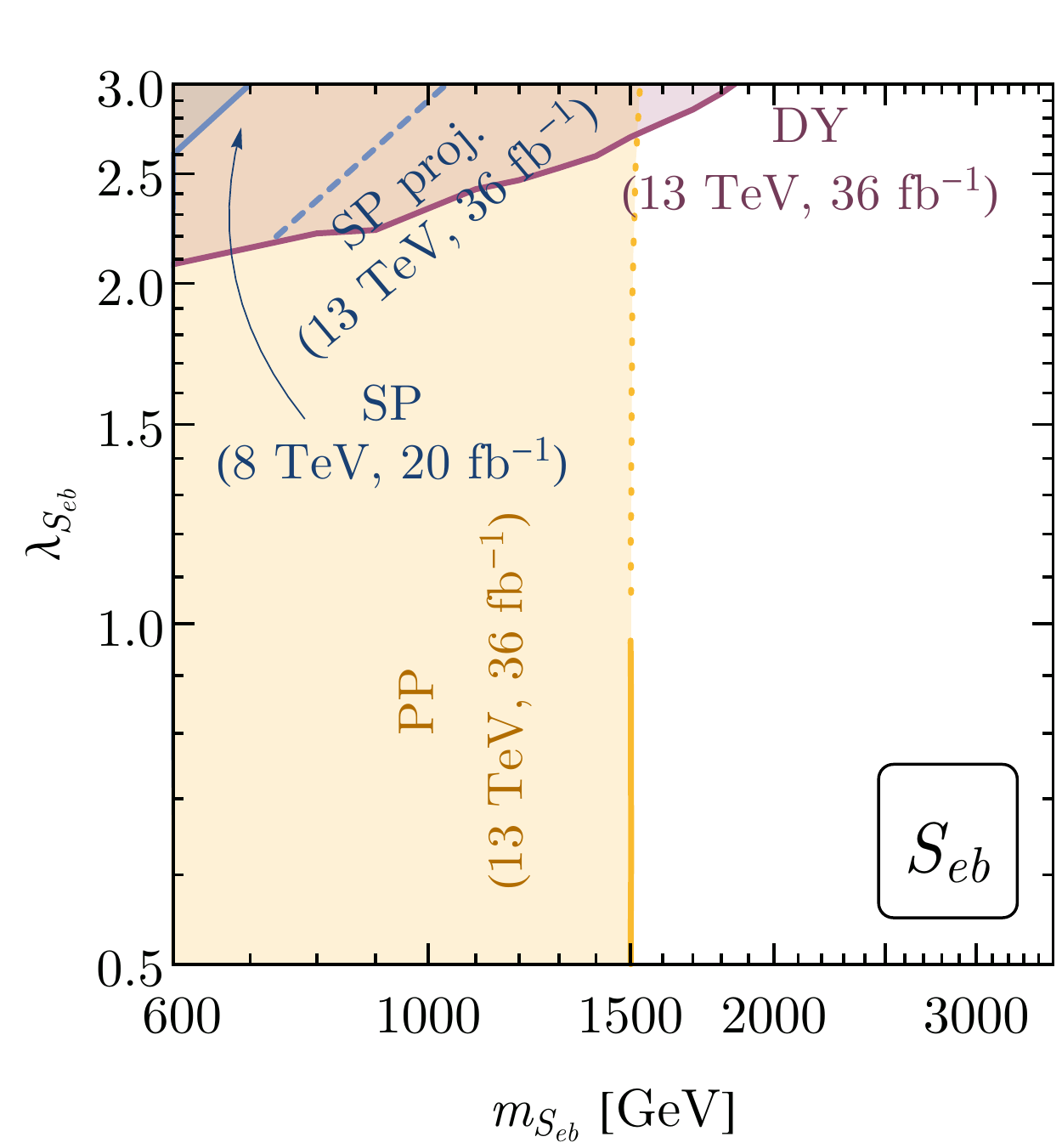}
 \caption{95\% CL limits on the parameter space for scalar MLQ $S_{e q}\,(q=u, d, s, c, b)$. The pair-production (PP) limits are recast from~\cite{CMS-PAS-EXO-17-009}. The single-production (SP) limits are recast from a Run 1 search~\cite{Khachatryan:2015qda} with our projections for the Run 2 reach, and DY limits are based on~\cite{Sirunyan:2018exx}. The green dot-dashed lines show constraints from weak charge measurements  described in \appref{others}. The arrows in the $S_{e u}$ plot show the bounds  directly given by~\cite{Khachatryan:2015qda}.}
   \label{fig:equark}
\end{figure}

\begin{figure}[!htbp]
   \centering
   \includegraphics[width=0.35\textwidth]{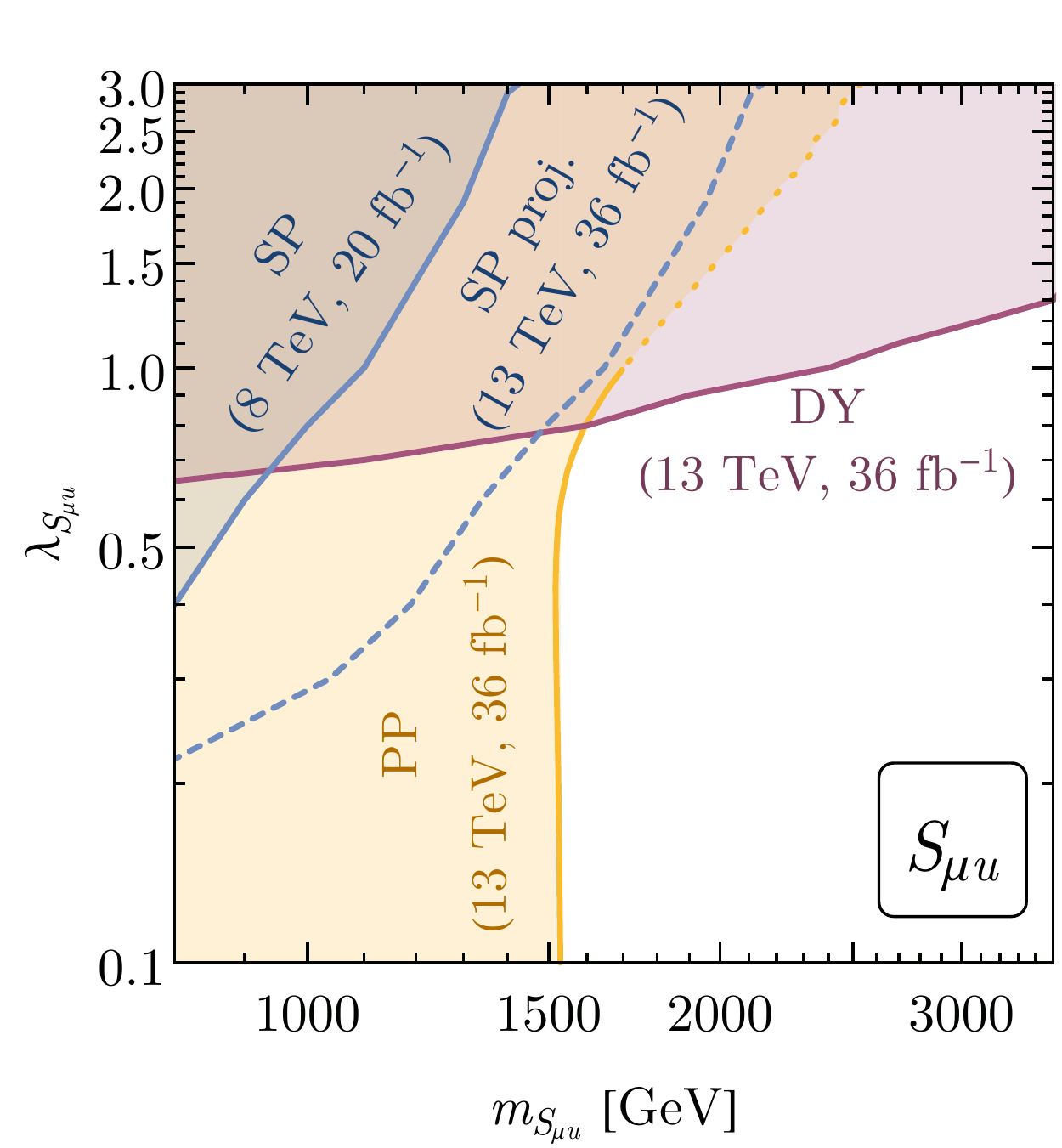}~\includegraphics[width=0.35\textwidth]{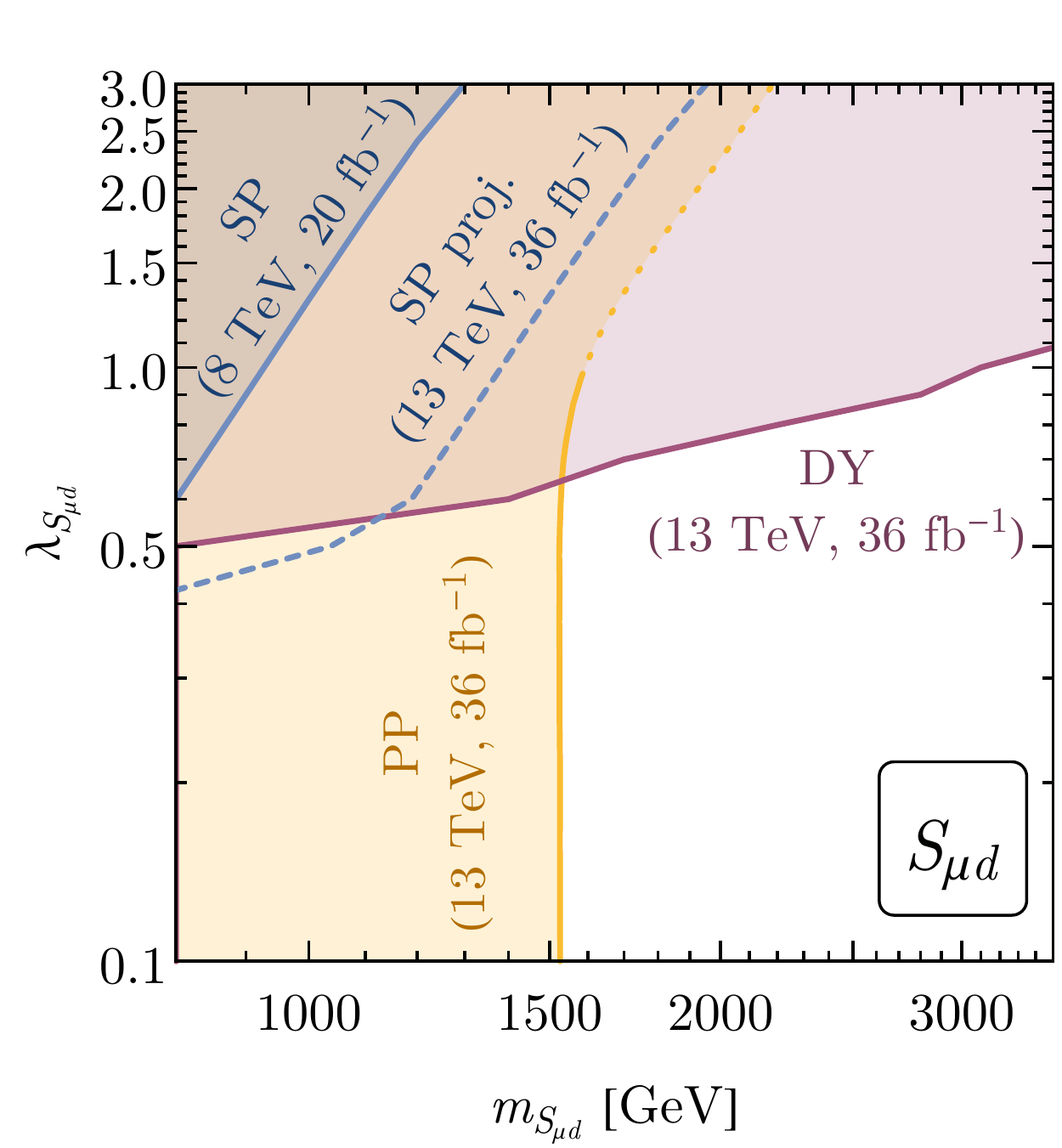}\\
   \includegraphics[width=0.35\textwidth]{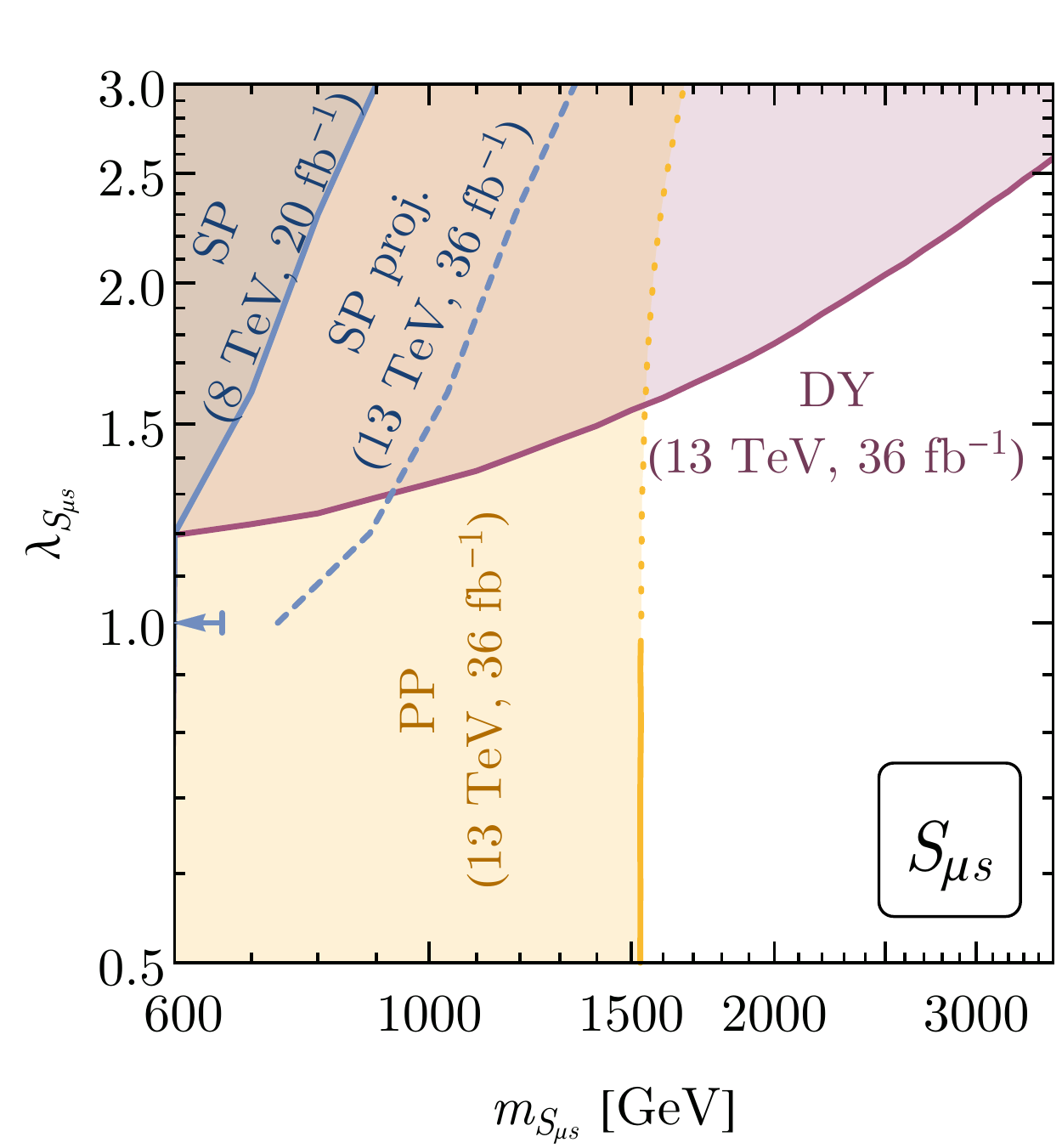}~\includegraphics[width=0.35\textwidth]{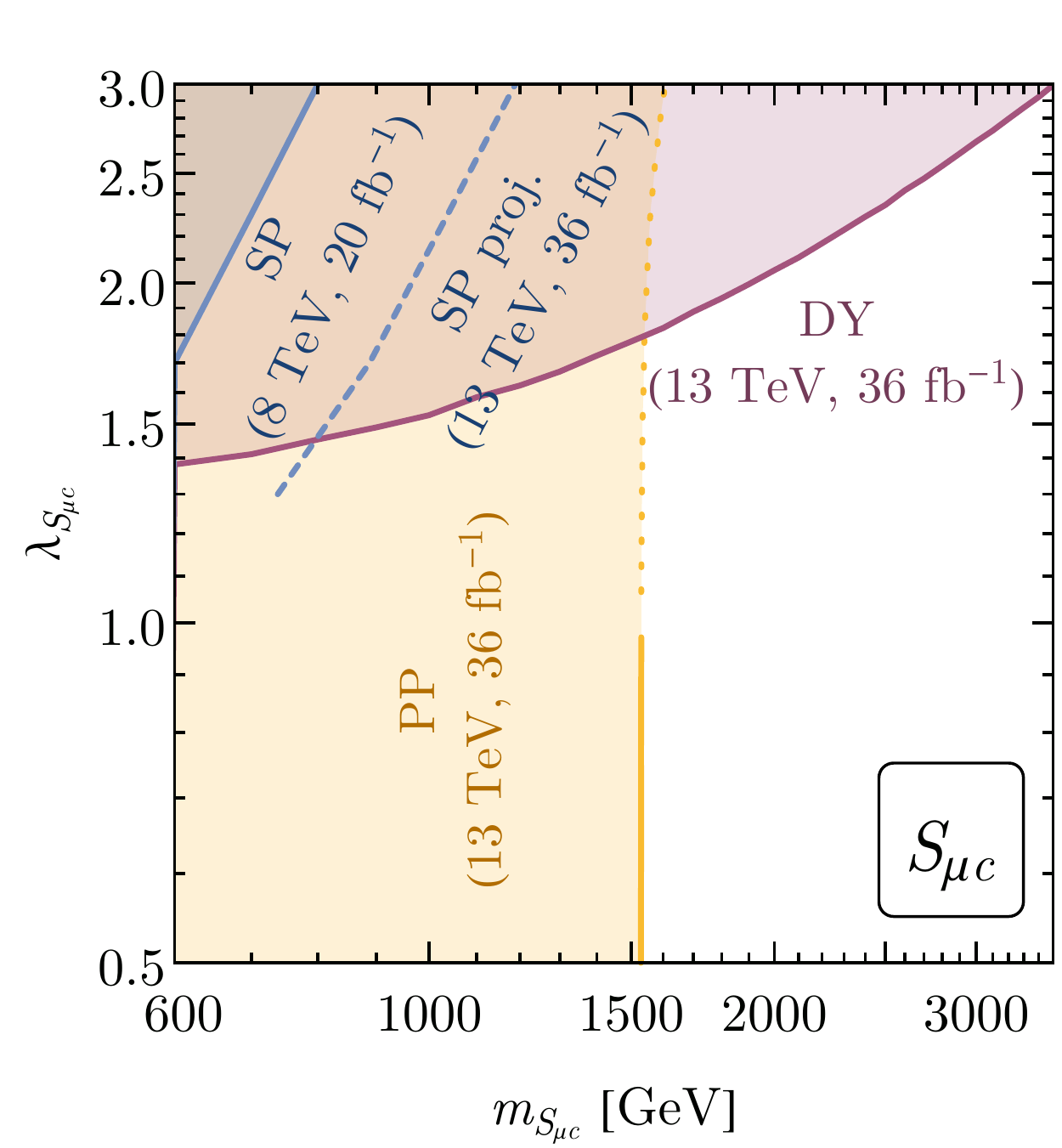}
  \\ \includegraphics[width=0.35\textwidth]{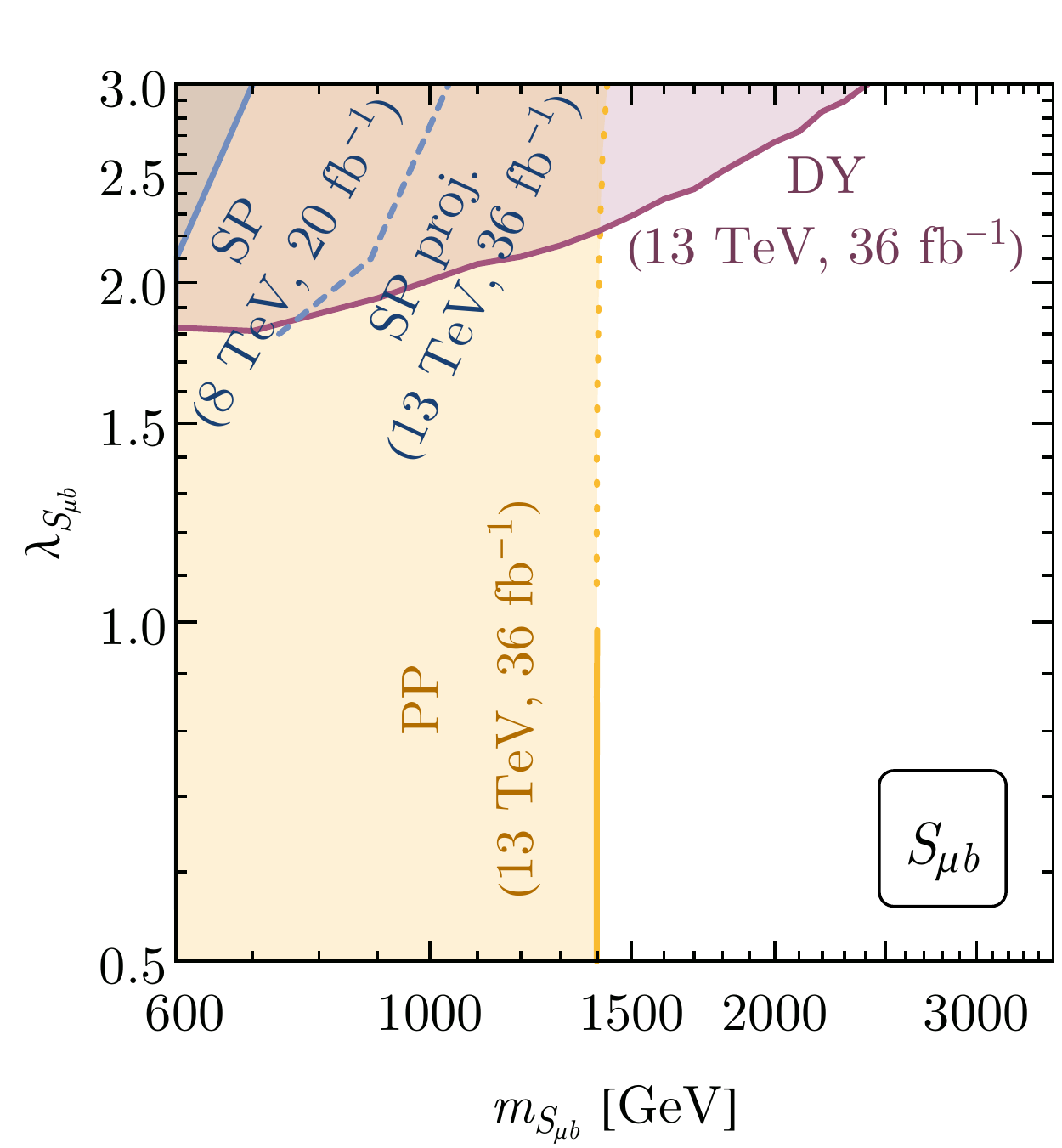}
 \caption{95\% CL limits on the parameter space for scalar MLQ $S_{\mu q}\,(q=u, d, s, c, b)$.  The pair-production (PP) limits are recast from~\cite{CMS-PAS-EXO-17-003}. The single-production (SP) limits are recast from a Run 1 search~\cite{Khachatryan:2015qda} with our projections for the Run 2 reach, and DY limits are based on~\cite{Sirunyan:2018exx}. The arrow in the $S_{\mu s}$ plot shows the bound  directly given by~\cite{Khachatryan:2015qda}.}
   \label{fig:muquark}
\end{figure}

For $\ell u$, $\ell d$, $\ell s$, $\ell c$, and $\ell b$ type leptoquarks, Drell-Yan-like production with the $\ell^+ \ell^-$ final state  puts important constraints in addition to the pair production bounds. In comparison to the pair production, the DY production suffers from quark-PDF suppression but at large $\lambda$ it is enhanced by $\lambda^4$ and does not require as high center-of-mass energies because the leptoquark in this process is not on-shell. Single production provides additional sensitivity to a small region in parameter space with intermediate masses and couplings but is subdominant in most of the parameter space.

To obtain the pair-production leptoquark constraints at large $\lambda$, we extrapolate the result from a CMS Run 2 search, which assumed that the coupling $\lambda$ is negligible in the production cross section, by simply identifying the contour in $\lambda-m$ parameter space which has the same pair-production cross section as the cross section at $\lambda=0$ and mass corresponding to the 95\% confidence-level (CL) limit obtained by CMS. The resulting bounds are shown as the orange shaded regions in~\figref{equark} and~\figref{muquark}.  Note that the $t$-channel lepton exchange diagram (PP-5 of~\figref{mypairdiagrams}) negatively interferes with and partially cancels the $q \bar{q}$-initiated QCD diagrams (PP-1 to PP-4 of~\figref{mypairdiagrams}). 
For which values of $\lambda$ the lepton exchange diagram dominates the total cross section depends on the quark flavor. The curvature of the bounds in the figures indicates that for leptoquarks coupling to valence quarks ($u, d$) the lepton exchange diagram starts to dominate for $\lambda \gtrsim 1$; whereas for leptoquarks coupling to heavier quarks the $\lambda$-dependence does not become important until $\lambda \gg 1$. As a caveat, note that the CMS Run 2 search used a cross section calculation which is NLO in the QCD coupling but not in $\lambda$. Thus the $K$-factor obtained in this limit is only accurate for small $\lambda$. We don't expect this to be a very important effect in most of the parameter region plotted,  but this theoretical uncertainty is the reason for our use of a dotted line for the bound at $\lambda >1$.

For single production, we recast the CMS Run 1 search~\cite{Khachatryan:2015qda}. CMS obtained upper limits on the leptoquark mass for specific choices of the coupling $\lambda = (0.4, 0.6, 0.8, 1.0)$ for $S_{e u}$ and $\lambda = 1$ for $S_{\mu s }$. We indicate these limit with blue arrows in~\figref{equark} and \figref{muquark}. To obtain our generalizations of these bounds we use \texttt{MG5} to simulate signal events (LO, ME, ${\rm PDF}= {\rm CTEQ6L1}$, ${\rm scale} = m_\phi$) and apply the trigger and selection cuts from~\cite{Khachatryan:2015qda} (listed in~\tabref{sumarryofcutsfull}). We find that the on-shell production cross section computed in this way is about a factor of two smaller than the value given in~\cite{Khachatryan:2015qda}. We have not been able to identify the reason for this discrepancy. Ref.~\cite{Raj:2016aky} found a similar resonant cross section to the one computed by us, and also studied and excluded issues such as finite resonance width and scale settings as possible reasons for the discrepancy. To get our bounds, we used our production cross sections and compared them to the observed 95\% CL upper limits on the cross section provided in~\cite{Khachatryan:2015qda}. The resulting LO limit excludes the blue shaded region with solid boundaries and is somewhat weaker than the bounds obtained by~\cite{Khachatryan:2015qda} because of the difference in the predicted cross sections. To illustrate the prospects of single-leptoquark production searches at LHC Run 2, we also estimate the Run 2 reach by scaling the Run 1 mass limit to the Run 2 energy (13 TeV) and luminosity (36 $\fb$) using \texttt{Collider Reach}~\cite{colliderreach}. These projected limits are shown as blue dashed lines in~\figref{equark} and~\figref{muquark}.

\begin{table}[htbp]
   \centering
   \topcaption{Summary of selection cuts for CMS  single-leptoquark production search~\cite{Khachatryan:2015qda}.  Cuts on the sum of transverse momentum $S_\text{T}$ and $m_{\ell j}$ depend on the assumed leptoquark mass. Details can be found in  Tables B.1 and B.2 of ~\cite{Khachatryan:2015qda} where generator level and a higher-level cuts on $m_{\ell j}$ are listed in the 4th and  3rd columns of the tables.}
   \label{tab:sumarryofcutsfull}
   \begin{tabular}{@{} lcc @{}} % Column formatting, @{} suppresses leading/trailing space
   \hline
   Search & $eej$ & $\mu\mu j$ \\
            \hline
      Lepton $p_\text{T}$        & $p_\text{T$e_{1, 2}$} > 45$ GeV & $p_\text{T$\mu_{1, 2}$} > 45$ GeV \\
     \& $\eta$         & $|\eta_{e_{1, 2}}|< 1.442$ or  $1.56 < |\eta_{e_{1, 2}}|< 2.1$  & $|\eta_{\mu_{1, 2}}|< 2.1$\\
      Jet $p_\text{T}$ \& $\eta$         & \multicolumn{2}{c}{$p_\text{T$j$} > 125 \gev$, $|\eta_{j}|<2.4$} \\
      Isolation    &  $\Delta R_{e e}$ > 0.5, $\Delta R_{e j} > 0.3$ & $\Delta R_{\mu \mu}$ > 0.3, $\Delta R_{\mu j} > 0.3$ \\
      $m_{\ell \ell}$ & \multicolumn{2}{c}{$> 110 \gev$}\\
      $S_\text{T}$, $m_{\ell j}$ &  \multicolumn{2}{c}{c.f. Tables B.1 and B.2 of ~\cite{Khachatryan:2015qda}} \\
      \hline
   \end{tabular}

\end{table}

Finally, we recast leptoquark bounds based on a DY search from CMS Run 2~\cite{Sirunyan:2018exx} (comparable bounds can be obtained by recasting the ATLAS search~\cite{Aaboud:2017buh}).  We generated signal events with \texttt{MG5} (LO, ME, ${\rm PDF} = {\rm NNPDF2.3LO}$, $\rm{scale} = m_{\ell \ell }$) for leptoquark masses ranging from 500 GeV to 3600 GeV in 100 GeV steps and couplings from 0.1 to 3.0 with 0.1 steps. For processes involving $S_{\ell u}$ and $S_{\ell d}$, we included interferences with the SM DY diagrams in the signal simulation. The generated events are required to pass through the selection cuts as summarized in~\tabref{sumarryofcutsDYCMS}. We then compared the resulting number of events with dilepton invariant masses $m_{\ell \ell} > 1.8$ TeV to the background predictions provided by~\cite{Sirunyan:2018exx}. Details of how we set these limits are described in~\appref{cutandcount}. The resulting bounds are shown in~\figref{equark} and~\figref{muquark} as purple shaded regions.

\begin{table}[htbp]
   \centering
   \topcaption{Summary of selection cuts for the CMS Drell-Yan  $ee$ and $\mu\mu$ search~\cite{Sirunyan:2018exx}. }
   \label{tab:sumarryofcutsDYCMS}
   \begin{tabular}{@{} lcc @{}} % Column formatting, @{} suppresses leading/trailing space
   \hline
   Search & $ee$ & $\mu\mu$ \\
            \hline
      Lepton $p_\text{T}$        & $p_\text{T$e_{1, 2}$} > 35$ GeV & $p_\text{T$\mu_{1, 2}$} > 53$ GeV \\
      \& $\eta$         & $|\eta_{e_{1, 2}}|< 1.44$  or & $|\eta_{\mu_{1, 2}}|< 2.4$ \\
      &  $|\eta_{e_i}|< 1.44$ and   $1.57 < |\eta_{e_{j}}|< 2.5$ &\\
      Isolation    &  $\Delta R_{e e} > 0.3 $& $\Delta R_{\mu \mu} > 0.3$ \\
 	& & $\Delta \theta_{\mu \mu} < \pi-0.02$\\
      \hline
   \end{tabular}
\end{table}

In~\figsref{equark}{muquark} one sees that the lower leptoquark mass region is mostly bounded by pair production whereas the large coupling region is mostly bounded by the DY search (subject to the caveat of model dependence due to possible dimension 6 operators). Additional parameter space which can be explored with single production searches with Run 2 data exists at intermediate masses and couplings.\footnote{We focused our attention on leptoquarks coupling to right-handed $SU(2)_{weak}$ singlet fermions. A MLQ coupling to left-handed doublet fermions can also contribute to the signal in $p p \rightarrow l \nu$ ``monolepton'' searches. For MLQs coupling to $u$ or $d$ quarks these can be significantly stronger than the dilepton DY bounds~\cite{Bansal:2018eha}.}

\subsection{$\tau u$, $\tau d$, $\tau s$, $\tau c$, and $\tau b$}
\label{sec:tauq}

\begin{figure}[!htp]
   \centering
   \includegraphics[width=0.4\textwidth]{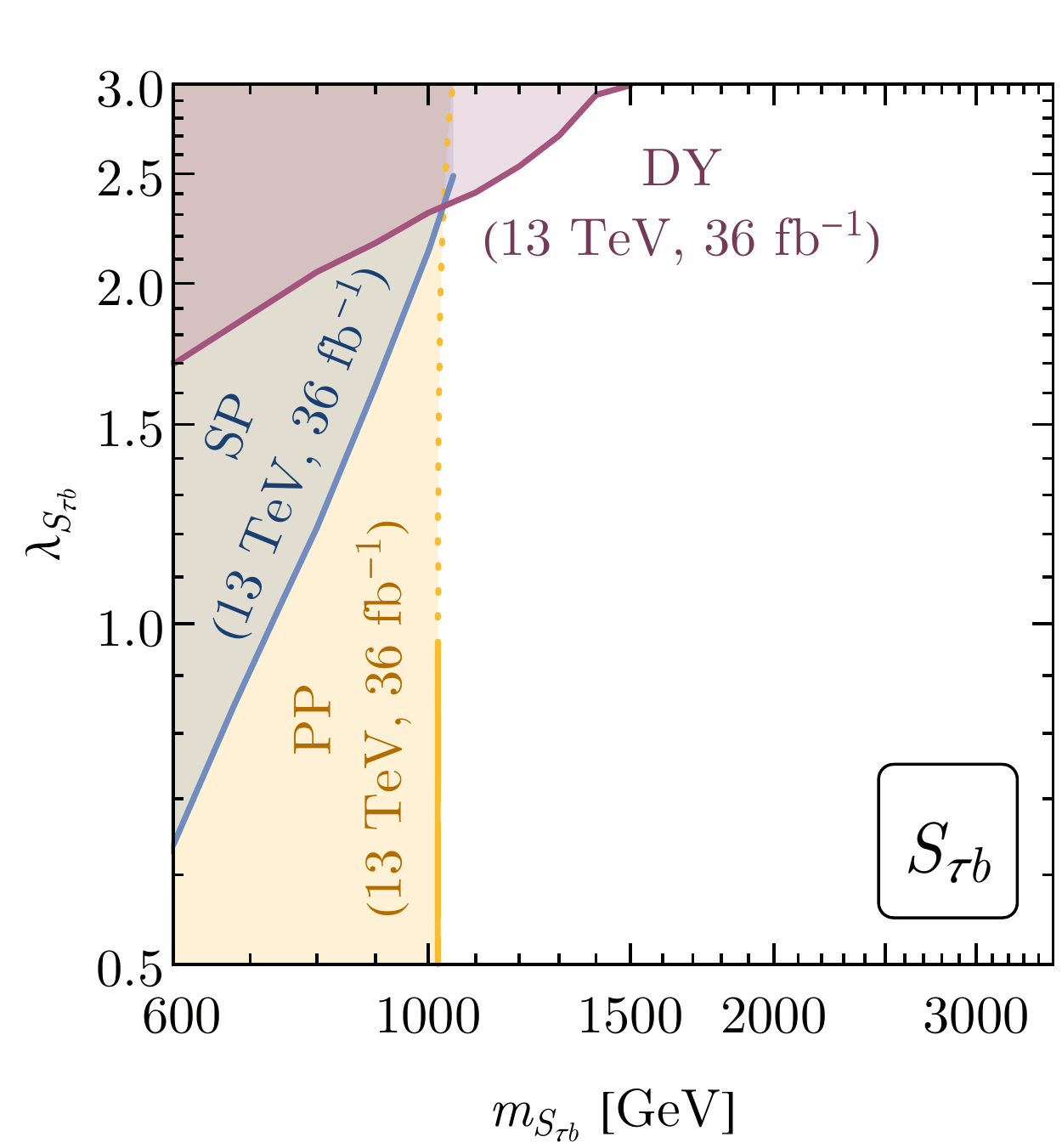}
 \caption{95\% CL limit on the parameter space for scalar MLQ $S_{\tau b}$. The pair-production (PP) limit is recast from~\cite{CMS-PAS-EXO-17-016}. The  single-production (SP) limit is given by~\cite{CMS-PAS-EXO-17-029}, and the DY limit is recast from~\cite{Aaboud:2017sjh}. All searches are from Run 2 with $36 \fb$ of data.}
   \label{fig:tauquark}
\end{figure}

Searches for $\phi_{\tau q}$ ($\phi_{\tau b}$) are similar to $\phi_{\ell q}$  ($\phi_{\ell b}$) searches. Here we focus on $S_{\tau b}$ to provide an example. The CMS collaboration has given the bounds on $S_{\tau b}$ from pair production~\cite{CMS-PAS-EXO-17-016} and single production~\cite{CMS-PAS-EXO-17-029} with $36\fb$ of data at Run 2. ATLAS performed a DY search for the $\tau^+\tau^-$ final state with $36\fb$ Run 2 data~\cite{Aaboud:2017sjh}. Here we simply quote the leptoquark bounds from pair and single production (extending the pair production limit to large couplings), and we recast the DY $\tau\tau$ search. 

Ref.~\cite{Aaboud:2017sjh} which we use for the DY recast included three different $\tau\tau$ final states:  $\tau_h \tau_h$, $\tau_e \tau_h$, and $\tau_\mu \tau_h$,  where the subscript stands for the subsequent decay channel of the $\tau$ to a final state with a hadron ($h$), an electron ($e$), or muon ($\mu$). Their branching ratios are given by
\begin{equation}
\label{eq:tauj}
p p\quad \to\quad   \tau_h  \tau_h \, [42.0\%], \quad  \tau_e  \tau_h \, [23.1\%],\quad \tau_\mu  \tau_h\, [22.5\%], \quad \cdots  
\end{equation}
We select the analysis for the  $\tau_h \tau_h$ channel as our example for the recast because it has the largest branching fraction.  

The selection cuts for the ATLAS $\tau_h\tau_h$ search are summarized in~\tabref{sumarryofcutsDYtautauATLAS}. 
The signal events are simulated with \texttt{MG5} (LO, ${\rm PDF}={\rm NNPDF2.3LO}$, ${\rm scale} = m_{\tau\tau}$) and passed through \texttt{Pythia 8.2}~\cite{Sjostrand:2014zea} and \texttt{Delphes 3.4.1}~\cite{deFavereau:2013fsa} for parton shower and detector simulations. In \texttt{Delphes}, we include $\tau$-tagging rates with ``medium'' (55\% for one-track $\tau_h$, 40\% for three-track $\tau_h$) and ``loose'' (60\% for one-track $\tau_h$, 50\% for three-track $\tau_h$) identification criteria for the leading and sub-leading $\tau_h$ respectively~\cite{Aaboud:2017sjh}. After passing events through the selection cuts, we bin the events according to the total transverse mass,  $$m_\text{T}^\text{tot} \equiv \sqrt{(p_\text{T$\tau_{h1}$}+p_\text{T$\tau_{h2}$}+p_\text{T}^\text{miss})^2-(\vec{p}_\text{T$\tau_{h1}$}+\vec{p}_\text{T$\tau_{h2}$}+\vec{p}^\text{miss}_\text{T})^2},$$ where $\vec p_\text{T$\tau_h$}$ ($\vec p_\text{T}^\text{miss}$) are the  vector momenta of the hadronic taus (missing momentum) projected into the transverse plane,
and compare the resulting histogram to Fig.~3b of the supplemental material of~\cite{Aaboud:2017sjh} where the event selection are $b$-tag inclusive. The analysis procedure is similar to the $\ell q$ analysis except we compare to $m_\text{T}^\text{tot}$ distributions. The resulting constraint is shown as the purple shaded region in~\figref{tauquark}. A stronger bound could be obtained if we combined the search for the  $\tau_\ell \tau_h$ final state with the search for the $\tau_h \tau_h$ final state.

As shown in~\figref{tauquark}, pair-production excludes leptoquark masses less than 1 TeV. DY excludes $\lambda_{S_{\tau b}} > 2.3$ $(>3)$ for $m_{S_{\tau b}} = 1 \tev$ ($1.5 \tev$). Single-production excludes parameter space with $850\gev \leq m_{S_{\tau b}}\leq 1 \tev$ and coupling $1.5\leq \lambda_{S_{\tau b}} \leq 2.5$.

\begin{table}[htbp]
   \centering
   \topcaption{Summary of  cuts for the ATLAS DY $\tau_h \tau_h$ search~\cite{Aaboud:2017sjh}. Note that the $\pt$ cut for the leading $\tau_{h}$ is required to be 5 GeV larger than the trigger cut, where the trigger cuts are $p_\text{T$\tau_{h 1}$} > 80, 125$ or $160$ GeV for three different Run 2 data-taking periods. Since Ref.~\cite{Aaboud:2017sjh} does not specify the integrated luminosities for each data taking period we apply the strongest cut to our signal simulations in order to obtain a conservative estimate. The $\tau$-tagging efficiencies are described in the text.}
   \label{tab:sumarryofcutsDYtautauATLAS}
   \begin{tabular}{@{} lc @{}} % Column formatting, @{} suppresses leading/trailing space
   \hline
  \multicolumn{2}{c} {At least two $\tau_h$'s and no electrons and muons} \\
  \hline
        $\tau_h$ $p_\text{T}$       & $p_\text{T$\tau_{h1}$} > 165 \gev$,  $p_\text{T$\tau_{h2}$} > 65 \gev$\\
      \& $\eta$ &  $|\eta_{\tau_{h1, h2}}|< 1.37$ or $1.52<|\eta_{\tau_{h1, h2}}| < 2.5$\\
       $\tau_h$ charge & opposite charge for $\tau_{h1}$ \& $\tau_{h2}$\\
            $\Delta R (\tau_{h1}, \tau_{h2})$ & $> 0.4$ \\
             $|\Delta \phi (\tau_{h1}, \tau_{h2})|$ &$> 2.7$  \\
      \hline
   \end{tabular}
\end{table}

\subsection{$\nu u$, $\nu d$, $\nu c$, $\nu s$, and $\nu b$}
Here we consider the leptoquark coupling to a neutrino and a light quark or $b$-quark. In principle, the neutrino could be either part of the left-handed $SU(2)_{weak}$ lepton doublet or a sterile neutrino. In the case of the left-handed lepton doublet, the leptoquark necessarily also couples to the charged lepton of the doublet.  LHC searches involving the charged leptons in the final state provide bounds which are at least as strong (stronger for electrons and muons, and comparable for taus) as those from searches for the neutrino final state (missing energy). This is why  we instead focus on the leptoquark coupling to a sterile neutrino that escapes undetected and produces missing energy signatures.

The LHC collaborations performed monojet searches for several different dark matter (DM) models.  In particular, ATLAS~\cite{Aaboud:2017phn} and CMS~\cite{Sirunyan:2017jix} have searched for DM that is produced via the exchange of a colored scalar mediator. 
Our MLQ model is the same as the fermion-portal DM model considered by~\cite{Bai:2013iqa} with couplings to only one of the quark flavors non-zero.\footnote{Very similar and in some cases identical DM models were also considered by other authors: $t$-channel DM~\cite{An:2013xka}, DM with colored mediator~\cite{Papucci:2014iwa}, leptoquark-mediated DM~\cite{Sirunyan:2017jix}, and flavored DM~\cite{Agrawal:2014una}.} CMS~\cite{Sirunyan:2017jix} also took the mediator to couple to only one quark flavor ($u$-quark). Therefore their bound can be directly carried over to the leptoquark $S_{\nu q}$.  ATLAS~\cite{Aaboud:2017phn} chose the mediator to couple to multiple quark flavors at the same time, which makes translating their bound less straightforward.

\begin{figure}[t]
 \centering
 \renewcommand{\thesubfigure}{MJ-SP-1}
             \begin{subfigure}[t]{0.22\textwidth}
        \centering
        \includegraphics[height=0.6\textwidth]{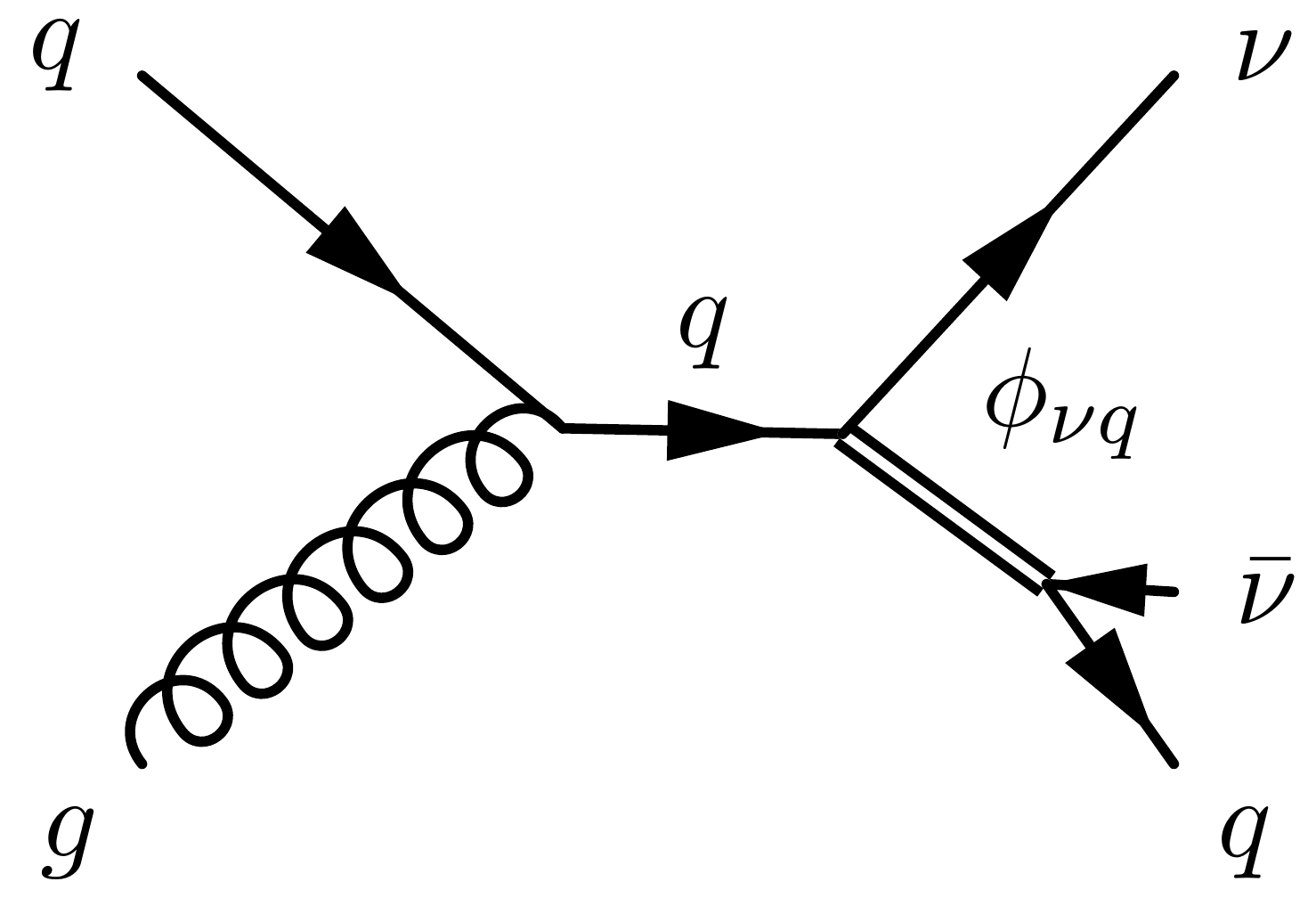}
        \caption{}
         \end{subfigure}
                   \hspace{0.2mm}    
 \renewcommand{\thesubfigure}{MJ-SP-2}
        \begin{subfigure}[t]{0.22\textwidth}
        \centering
        \includegraphics[height=0.6\textwidth]{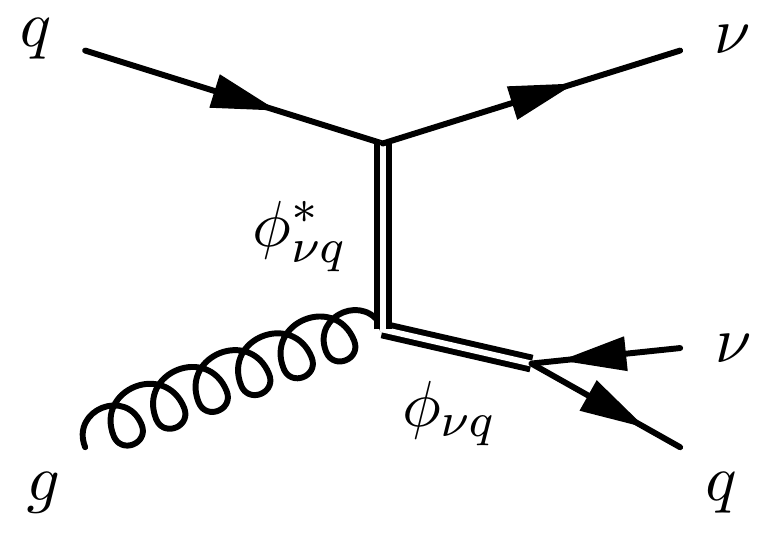}
        \caption{}
         \end{subfigure}%
\hspace{0.00mm}
  \renewcommand{\thesubfigure}{MJ-DY-1}
        \begin{subfigure}[t]{0.22\textwidth}
        \centering
        \includegraphics[height=0.6\textwidth]{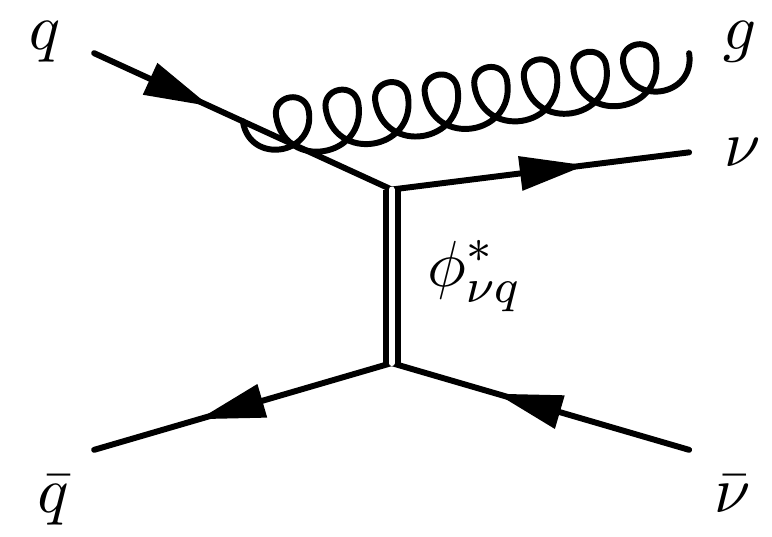}
        \caption{}
         \end{subfigure} 
          \hspace{0.2mm}
          \\    
          \vspace{5mm}
\renewcommand{\thesubfigure}{MJ-DY-2}
        \begin{subfigure}[t]{0.22\textwidth}
        \centering
        \includegraphics[height=0.6\textwidth]{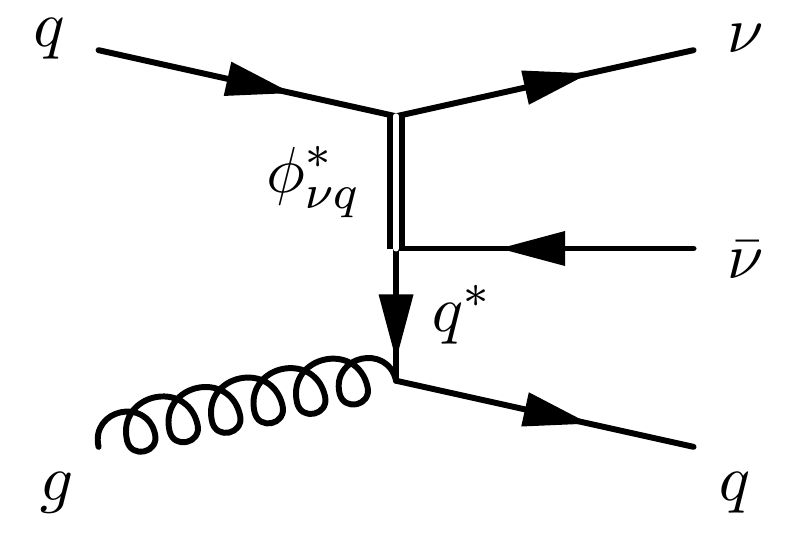}
        \caption{}
         \end{subfigure} 
\renewcommand{\thesubfigure}{MJ-DY-3}
        \begin{subfigure}[t]{0.22\textwidth}
        \centering
        \includegraphics[height=0.6\textwidth]{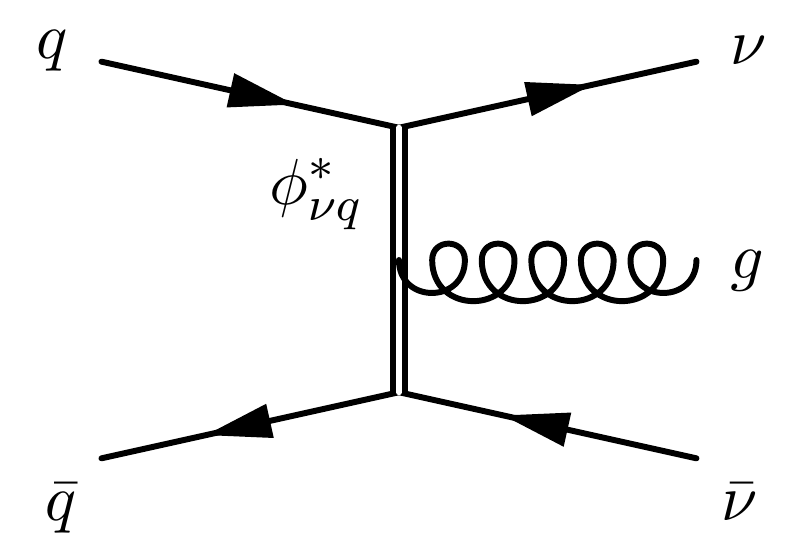}
        \caption{}
         \end{subfigure} 
\renewcommand{\thesubfigure}{MJ-PP-1}
        \begin{subfigure}[t]{0.22\textwidth}
        \centering
        \includegraphics[height=0.6\textwidth]{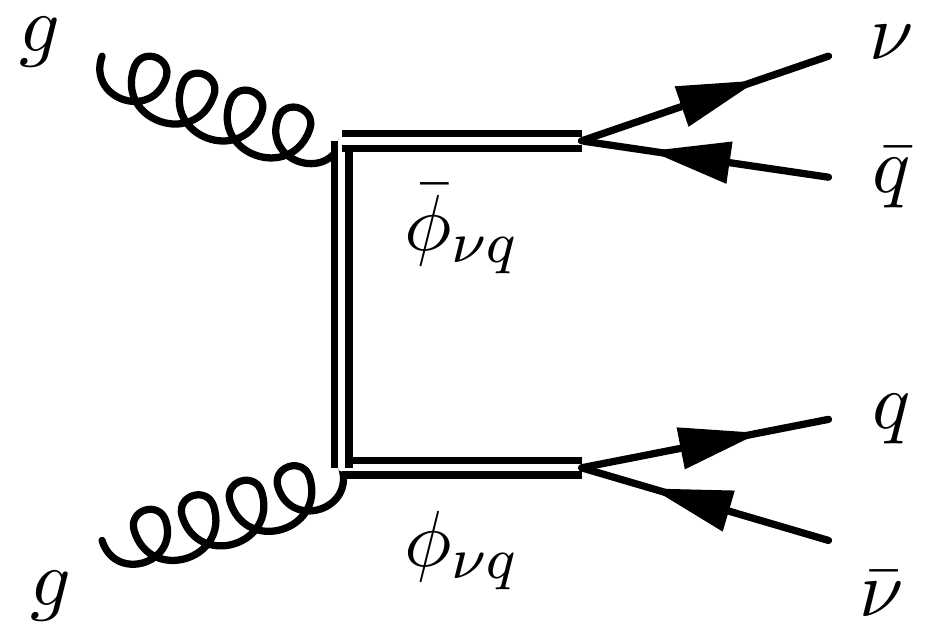}
        \caption{}
         \end{subfigure} 
         \\
         \vspace{5mm}
\renewcommand{\thesubfigure}{MJ-PP-2}
        \begin{subfigure}[t]{0.22\textwidth}
        \centering
        \includegraphics[height=0.6\textwidth]{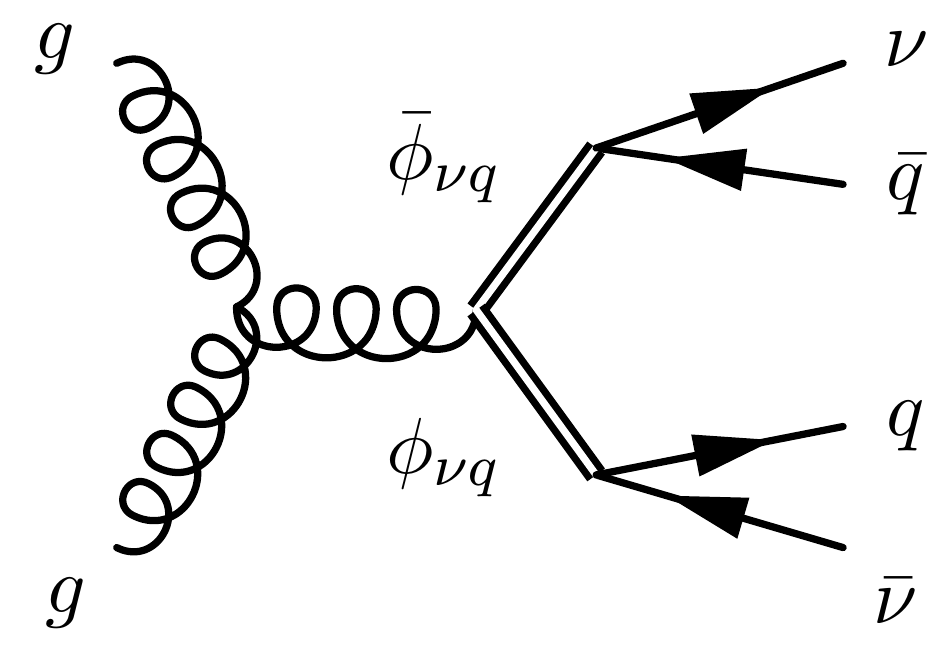}
        \caption{}
         \end{subfigure}         
\renewcommand{\thesubfigure}{MJ-PP-3}
        \begin{subfigure}[t]{0.22\textwidth}
        \centering
        \includegraphics[height=0.6\textwidth]{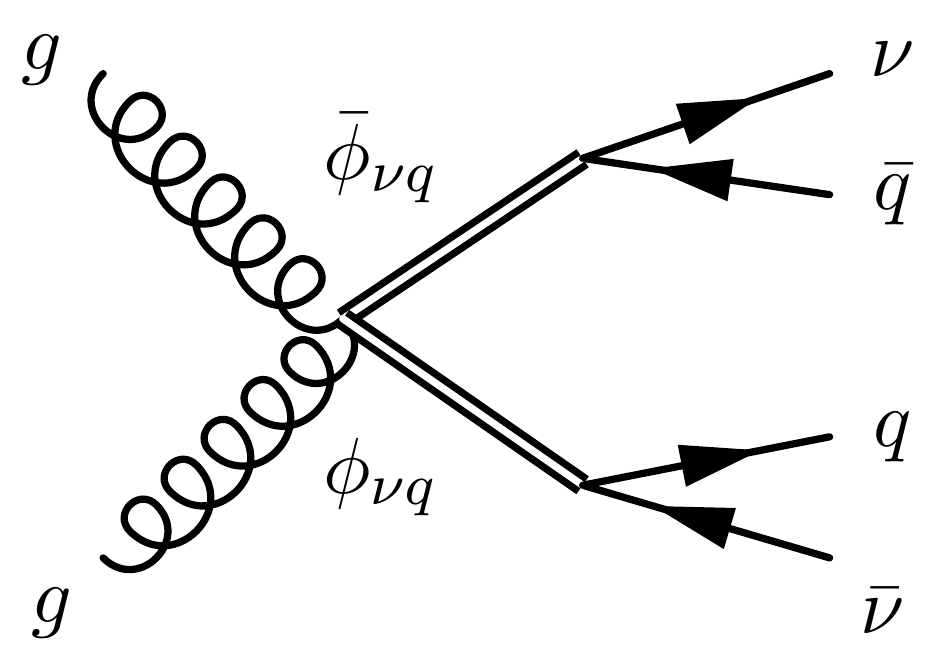}
        \caption{}
         \end{subfigure}  
\renewcommand{\thesubfigure}{MJ-PP-4}
        \begin{subfigure}[t]{0.22\textwidth}
        \centering
        \includegraphics[height=0.6\textwidth]{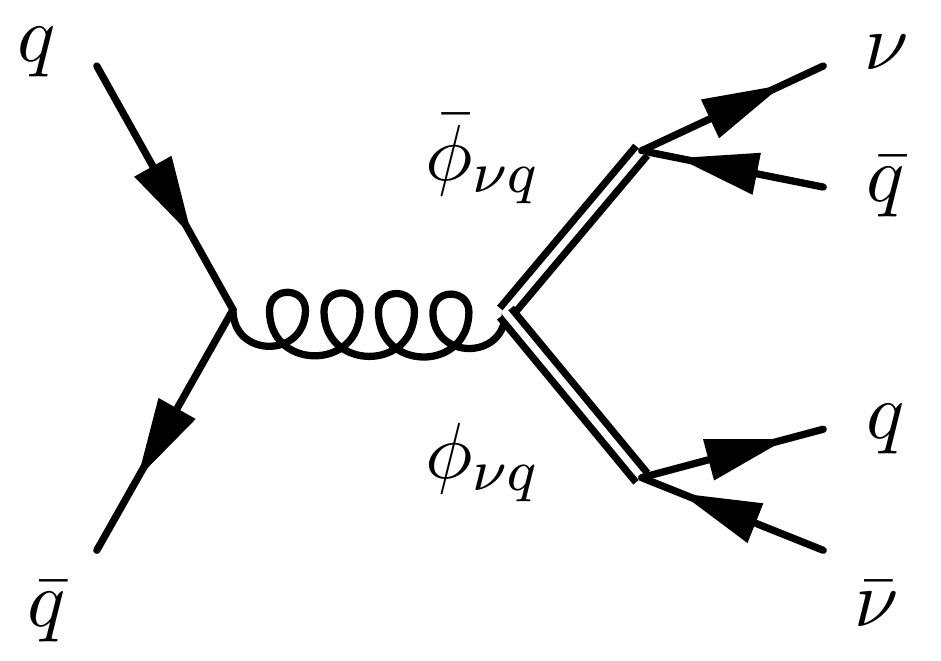}
        \caption{}
         \end{subfigure} 
\renewcommand{\thesubfigure}{MJ-PP-5}
        \begin{subfigure}[t]{0.22\textwidth}
        \centering
        \includegraphics[height=0.6\textwidth]{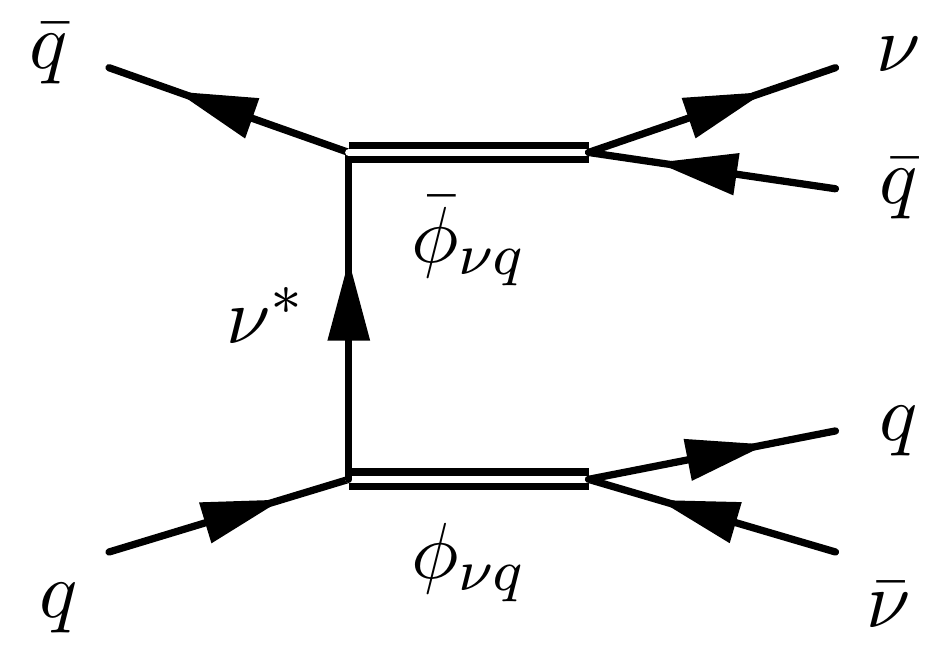}
        \caption{}
         \end{subfigure}                                                            

   \caption{Leading order diagrams contributing to the monojet searches for  $\phi_{\nu q}$. Searches for $\phi_{\nu q}$ are identical to searches with flavored-dark-matter for vanishing dark matter mass. Diagrams can be classified as MJ-SP (here the monojet is the decay product of a singly produced leptoquark),  MJ-DY (here the monojet stems from NLO radiation emanating from a process which is DY-like at leading order). In addition, we include $2j+\met$ from leptoquark pair production MJ-PP because monojet searches usually allow a second hard jet and hence are also sensitive to the $2j+\met$ signal. Other $2j + \met$ diagrams with one or zero leptoquarks on shell are less significant in the parameter space that we are interested in. Hence we do not show them here.}
   \label{fig:leptoquarkvsDM}
\end{figure}

\begin{table}[htbp]
   \centering
   \topcaption{Summary of selection cuts for the CMS mono-jet search~\cite{Sirunyan:2017jix}. Here $H_\text{T}^{20}$ is computed as the sum of the magnitudes of the vector $p_{\rm T}$ of jets with $p_\text{T$j$} > 20 \gev$ and $|\eta_j| < 5.0$.}
   \label{tab:DMcuts}
   \begin{tabular}{@{} lc @{}} % Column formatting, @{} suppresses leading/trailing space
      \hline
      Trigger     & $p_\text{T}^\text{miss} > 110$ GeV, $H_\text{T}^{20}> 110 \gev$ \\
      \hline
      $p_\text{T}^\text{miss}$       & $> 250 \gev$  \\
      Jet $p_\text{T}$ \& $\eta$       & $p_\text{T$j$} > 100 \gev$, $|\eta_{j}|<2.4$ \\
	 $\Delta \phi (\vec{p}_\text{T}^\text{miss}, \vec{p}_\text{T$j$})$  &  $> 0.5$\\
      \hline
   \end{tabular}

\end{table} 

\begin{figure}[htbp]
   \centering
   \includegraphics[width=0.35\textwidth]{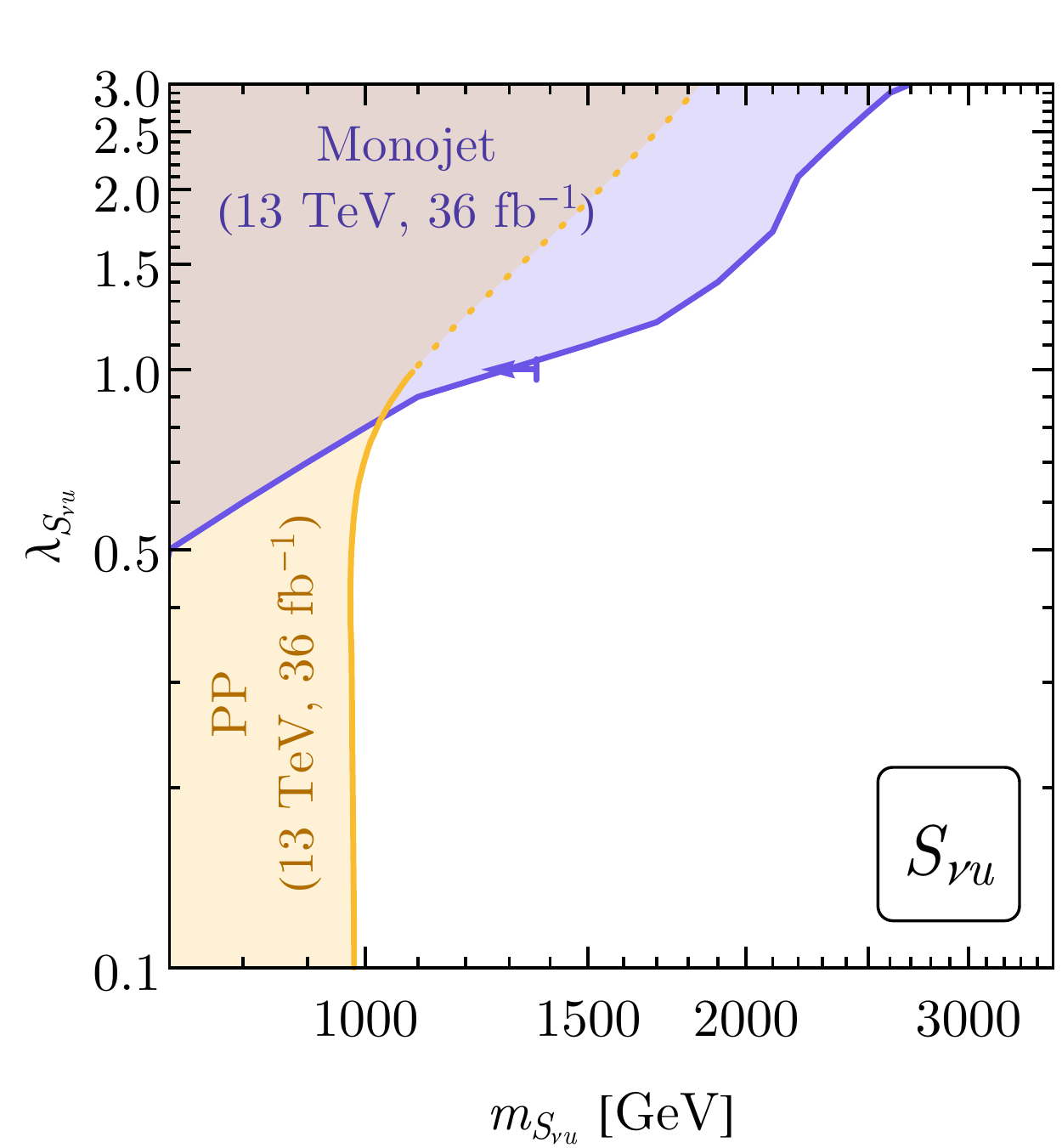}~\includegraphics[width=0.35\textwidth]{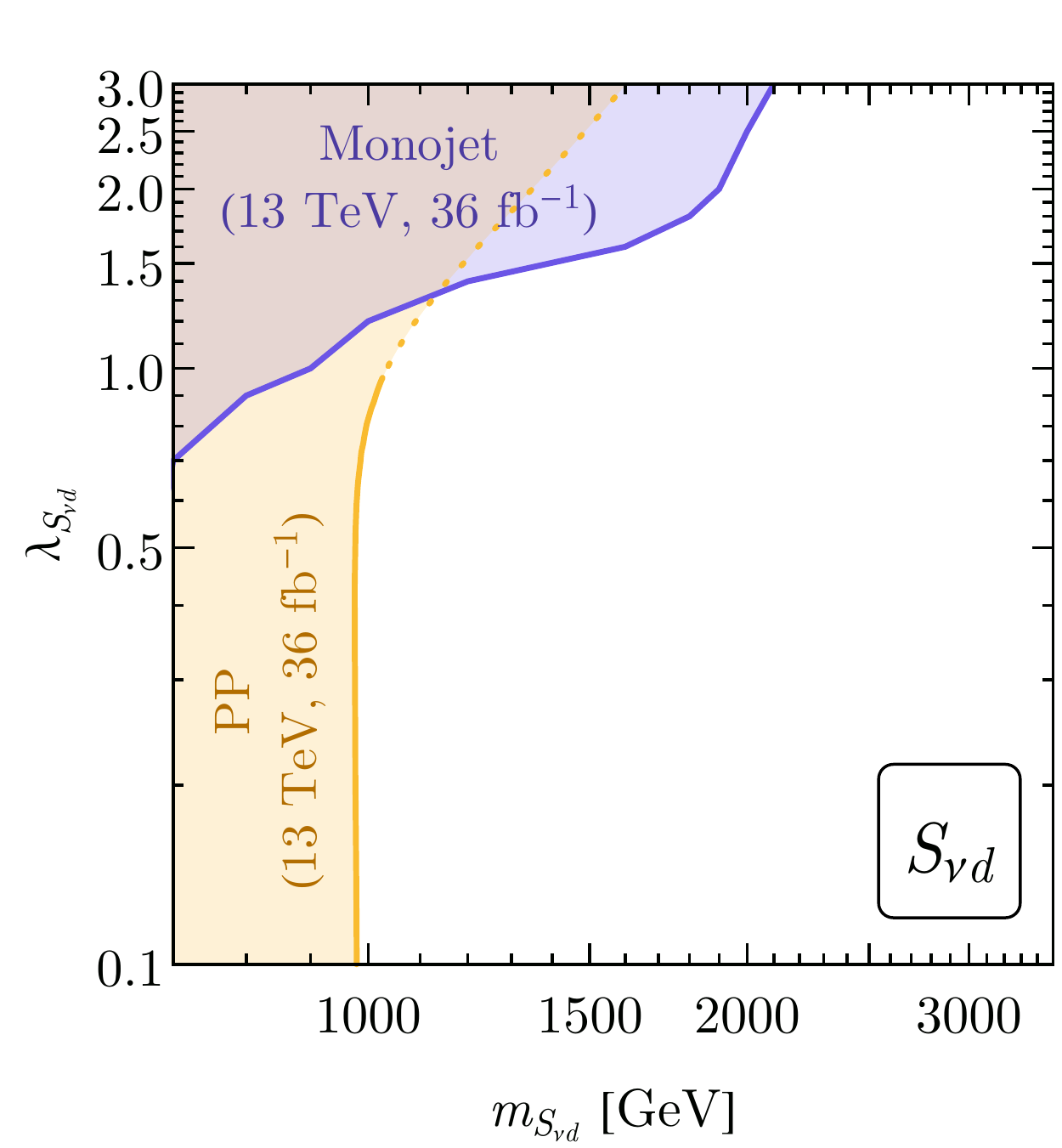}\\
   \includegraphics[width=0.35\textwidth]{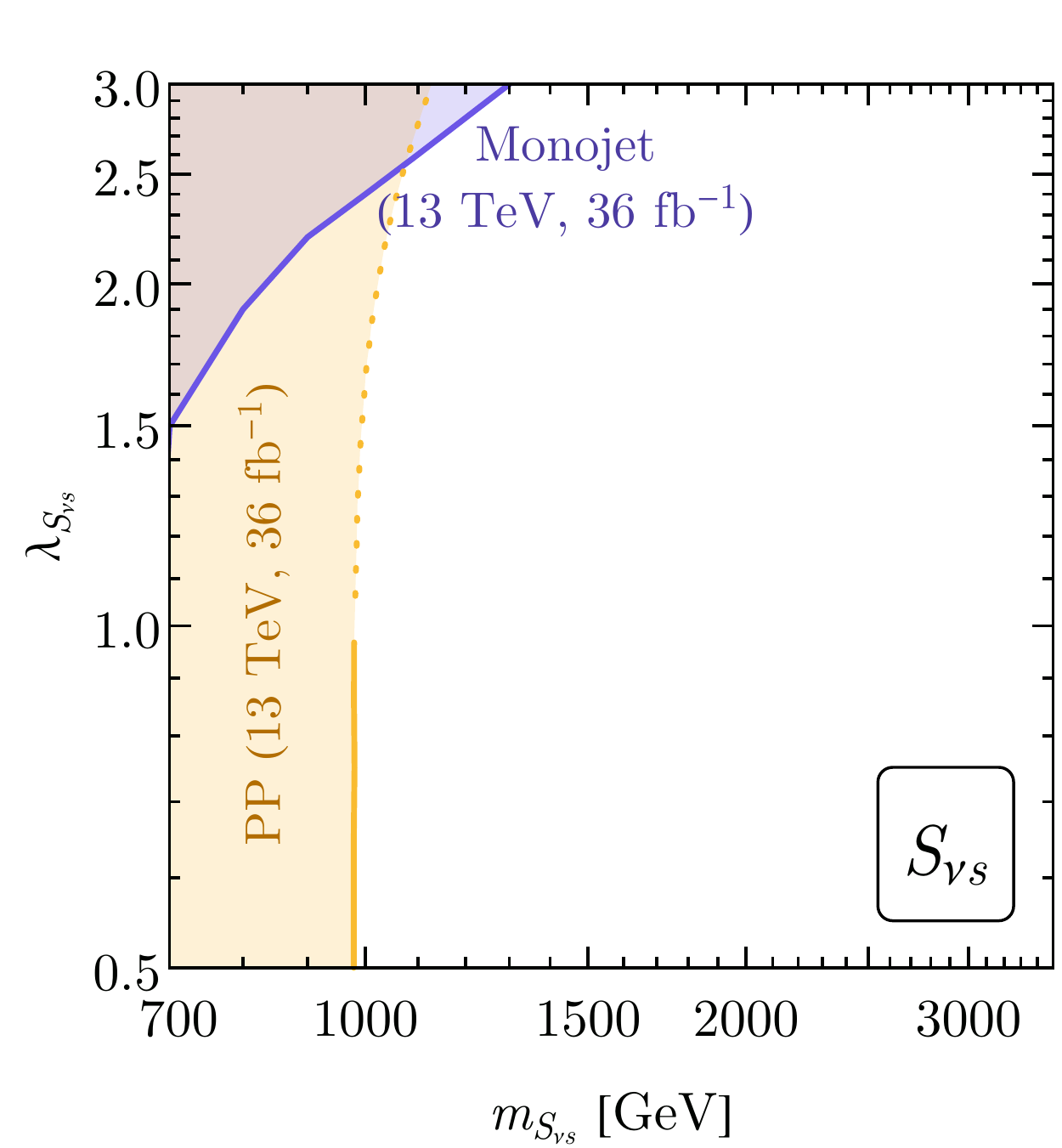}~\includegraphics[width=0.35\textwidth]{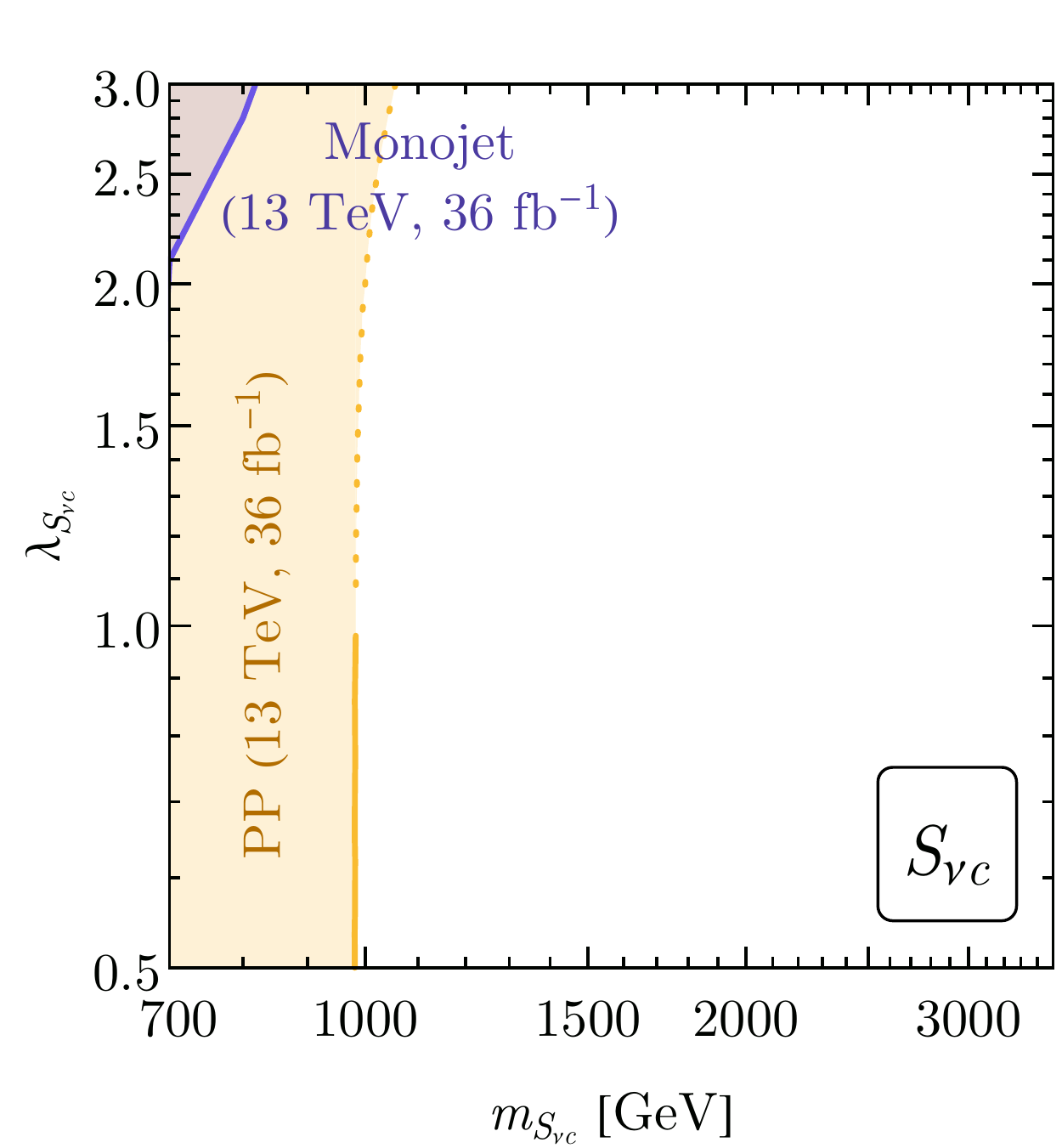}\\
    \includegraphics[width=0.35\textwidth]{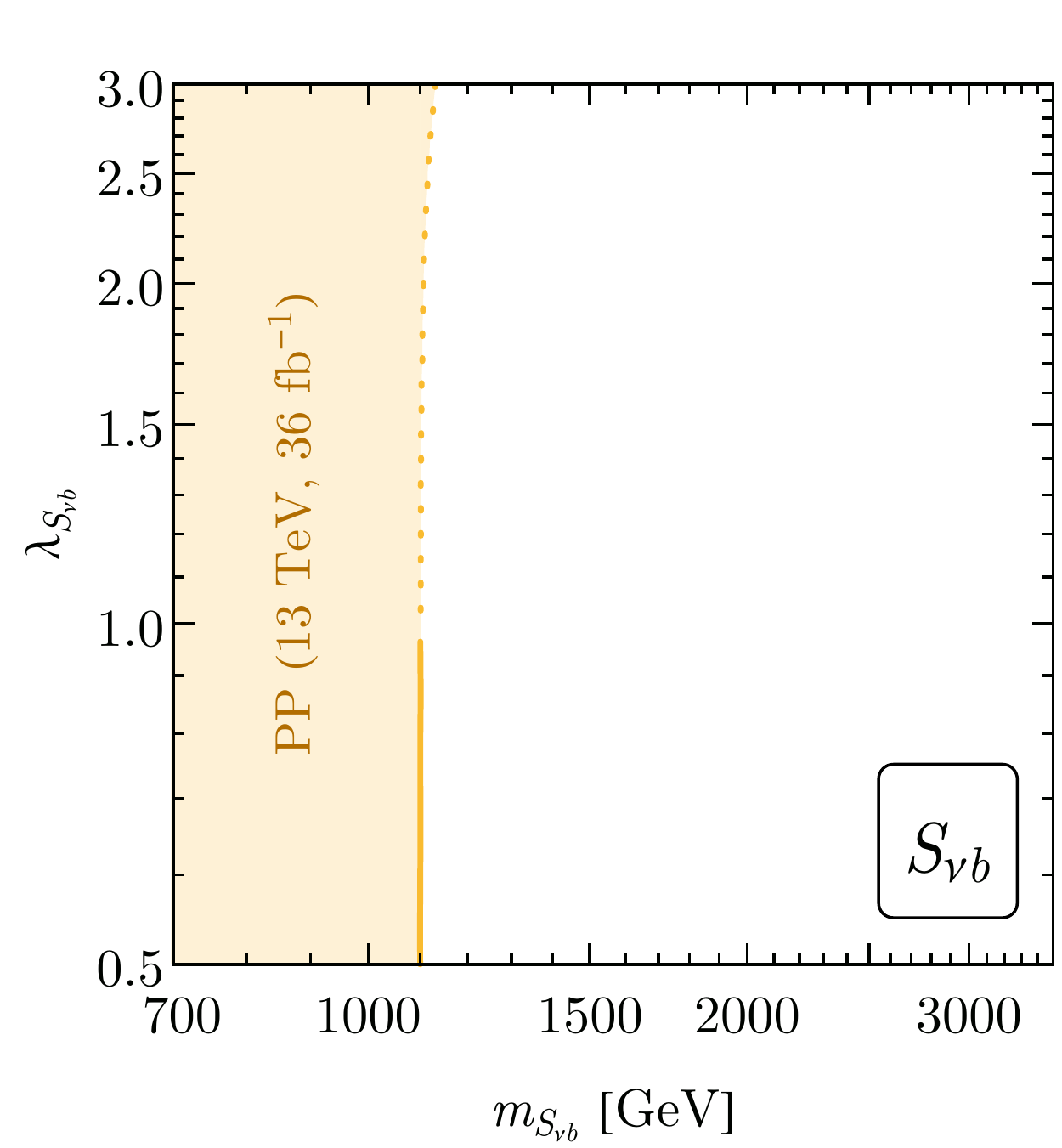}
 \caption{95\% CL limit on the parameter space for the scalar MLQ $S_{\nu q}\,(q=u, d, s, c, b)$ that decays into the $\nu j$ final state. The pair production and the monojet search limits are recast from~\cite{Sirunyan:2018kzh} and~\cite{Sirunyan:2017jix}, respectively. The arrow in the $S_{\nu u}$ plot indicates the bound on $m_{S_{\nu u}}\leq1365 \gev$ for $\lambda_{S_{\nu u}}=1$ from~\cite{Sirunyan:2017jix}.}
   \label{fig:nuudcsb}
\end{figure}

For simplicity, for our discussion we choose the sterile neutrino mass to be negligible.\footnote{Refs.~\cite{Aaboud:2017phn, Sirunyan:2017jix,Aad:2014vea, Aaboud:2017rzf} cover $S_{\chi q}$ masses as low as $m_\chi = 1\gev$ in their parameter scan. This is sufficiently close to massless when compared with the energy scale of a typical event which is set by the leptoquark mass.} In~\cite{Sirunyan:2017jix} CMS performed a search with only one value for the coupling, fixing $\lambda_{S_{\nu u}} =1$, and obtained the 95\% CL mass limit $m_{S_{\nu u}}\geq 1365 \gev$ which we show as the blue arrow in the $S_{\nu u}$ panel of~\figref{nuudcsb}. To place a constraint on $S_{\nu q}$ coupling to light quarks ($q = u, d, s, c$) with other values of the coupling, we recast the analysis from~\cite{Sirunyan:2017jix} as follows. We simulate signal events in \texttt{MG5} (LO, ME, ${\rm PDF} = {\rm NNPDF3.0}$, ${\rm scale} = m_\phi$) with masses from 500 GeV to 3200 GeV in 100 GeV steps and couplings from 0.1 to 3.0 in 0.1 steps. We then pass the events through the trigger and selection cuts given in~\cite{Sirunyan:2017jix} that are summarized in~\tabref{DMcuts}. Finally we bin the selected events according to missing-transverse-momentum $\pt^\text{miss}$ and compare the total number of events above the $\pt^\text{miss}$ cut to the number provided in Tab. 4 of~\cite{Sirunyan:2017jix}. Details of the comparison procedure are described in~\appref{cutandcount}, and the resulting limits are shown in ~\figref{nuudcsb} (blue shaded region) together with the pair production limits (orange shaded region). Note that the pair production limits are largely quark-flavor-independent, and for $s, c, b, t$ quarks they are much stronger than the monojet search limits. 

To suppress the top-quark background, Ref.~\cite{Sirunyan:2017jix} explicitly vetoed $b$-jets in the final state and hence it is not ideal to recast their search to put bounds on $S_{\nu b}$. On the other hand, ATLAS performed a $b$-flavored DM search in~\cite{Aad:2014vea, Aaboud:2017rzf} which has diagrams that are identical to $S_{\nu b}$ when the DM is massless. In Fig. 7 of the supplement material~\cite{ATLASbnu} of Ref.~\cite{Aaboud:2017rzf}, the 95\% CL upper limit on the production cross section is given for $m_{S_{\nu b}}= (800, 1000, 1200, 1400, 1600) \gev$ with $m_\chi = 1 \gev$. In order to determine the sensitivity of this search to $S_{\nu b}$ single production, we simulated signal events $p p \to \nu \bar{\nu} b$, together with $p p \to \nu \bar{\nu} \bar b$ and $p p \to \nu \bar{\nu} b \bar b$, via \texttt{MG5} (LO, ME, PDF=NNPDF2.3LO, scale = default, no generator level cuts)\footnote{We thank Marie-H\'el\`ene Genest for providing details of the simulation.} for a range of leptoquark masses and couplings. We then compared the resulting production cross sections with the upper limits from~\cite{ATLASbnu}. We find that for the mass values investigated by~\cite{Aaboud:2017rzf} the bound on $\lambda_{S_{\nu b}}$ is always greater than 3. Hence the limit does not reach the parameter space shown in the $S_{\nu b}$ panel of~\figref{nuudcsb}.

\subsection{$\nu t$, $\ell t$, and $\tau t$}
Since the top-quark PDFs in the proton are negligibly small, the production of $\phi_{l t}$ proceeds only via its QCD coupling from the $gg$ and $q \bar q$ initial states (see diagrams PP-1 to PP-4 in~\figref{mypairdiagrams}). Therefore we expect that meaningful bounds only result from pair production searches at the LHC, and that the bounds are largely independent of the leptoquark coupling.\footnote{For very large couplings the leptoquark width is also large and resonance reconstruction becomes more challenging, potentially degrading the bounds. Thus bounds from resonance searches only apply for $\lambda \lesssim 3$.} Pair production bounds are simply lower bounds on the mass of $\phi_{lt}$.

CMS has searched for pair production of $\phi_{\nu t}$~\cite{Sirunyan:2018kzh}, $\phi_{\mu t}$~\cite{Sirunyan:2018ruf}, and $\phi_{\tau t}$~\cite{Sirunyan:2018nkj} with 36 $\fb$ of data at Run 2. The resulting lower bounds on the leptoquark masses are summarized in~\tabref{lt} for the complete list of MLQs. Note that Ref.~\cite{Sirunyan:2018ruf}  performed a dedicated search for $\phi_{\mu t}$ with reconstruction of the leptoquark resonances. The dedicated search obtained a very impressive bound on the mass of the fiducial MLQ of $m_{\phi_{\mu t}} \geq 1.4 \tev$. This is much stronger than the previous best available bound of $m_{\phi_{\mu t}} \geq 800 \gev$~\cite{Diaz:2017lit} which we obtained by recasting a cut and count multi-signal region search for R-parity violating supersymmetry~\cite{CMS-PAS-SUS-16-041}. We expect that a similar dramatic improvement of the bound can be obtained with a dedicated search for $\phi_{et}$.

\begin{table}[!htbp] 
   \topcaption{ 95\% CL lower limits on the masses of MLQs which decay into ($l t$)($l t$) final states with $l = \nu, e, \mu, \tau$ for scalar leptoquarks (\emph{left}) and vector leptoquarks (\emph{right}). The recasted bounds (bounds without references) are obtained using the $\mu$-factor scaling described in~\cite{Diaz:2017lit}.}
   \resizebox{\columnwidth}{!}{%      
   \begin{tabular}{@{} |c | c|  @{}} % Column formatting, @{} suppresses leading/trailing space
      \hline
           &  $U^{c} L, Q N^c, U^{c} N^c\!, (Q L$ triplet)\\
           \hline
           $S_{\nu t}$ & $m_{S_{\nu t}} \geq 1 \tev$~\cite{Sirunyan:2018kzh}\\
       \hline
   \end{tabular}
   \quad
   \begin{tabular}{@{} | c| c| c| @{}} % Column formatting, @{} suppresses leading/trailing space
      \hline
            & \quad $U^{c\dag} L, Q^\dag N^c, U^{c\dag} N^c$\, &   $Q L$ triplet, $Q L$ singlet\\
           \hline
           $V_{\nu t}$ & $m_{V_{\nu t}}\geq 1.8 \tev$~\cite{Sirunyan:2018kzh} &  $m_{V_{\nu t}} \geq 1.5 \tev$~\cite{Sirunyan:2018kzh}\\
       \hline
   \end{tabular}
  } 
  \resizebox{\columnwidth}{!}{% 
   \begin{tabular}{@{} |c |c| c|  @{}} % Column formatting, @{} suppresses leading/trailing space
      \hline
           & $U^c L, U^cE^c, QE^c$ & ($QL$ triplet),  ($QL$ singlet) \\
           \hline
           
           $S_{e t}$ & $m_{S_{e t}}\geq 900 \gev$~\cite{Diaz:2017lit} & $m_{S_{e t}}\geq 600 \gev$ \\
           \hline
            $S_{\mu t}$ & $m_{S_{\mu t}}\geq 1.4 \tev$~\cite{Sirunyan:2018ruf} & $ m_{S_{\mu t}} \geq 1.1 \tev$ \\
            \hline
            $S_{\tau t}$ & $m_{S_{\tau t}}\geq 900 \gev$~\cite{Sirunyan:2018nkj} & $m_{S_{\tau t}} \geq  560 \gev$\\
  
              \hline
   \end{tabular}
   \quad
     \begin{tabular}{@{}| c| c|   @{}} % Column formatting, @{} suppresses leading/trailing space
       \hline
           &        {$U^{c\dag} L, U^{c\dag} E^c, Q^\dag E^c, (Q^\dag L$ triplet)}  \\
                      \hline
                      
                       $V_{e t}$ & $m_{V_{e t}}\geq 1.5 \tev$ \\
                       \hline
                        $V_{\mu t}$ & $m_{V_{\mu t}}\geq 2 \tev$\\
                        \hline
                        $V_{\tau t}$ & $m_{V_{\tau t}}\geq 1.45 \tev$\\

      \hline
   \end{tabular}
}   

\label{tab:lt}
\end{table}

\section{A vector leptoquark case study: $U_1$ in the 4321 model}
\label{sec:4321}

The $R_{D^{(*)}}$ and $R_{K^{(*)}}$ anomalies motivated a lot of model building with leptoquarks which ultimately settled on the vector leptoquark, $U_1$, with the SM quantum numbers $(\bar 3, 1)_{-2/3}$ as the most promising explanation~\cite{Hiller:2014yaa,Hiller:2014ula,Calibbi:2015kma,Buttazzo:2017ixm}. Refs.~\cite{DiLuzio:2017vat,DiLuzio:2018zxy} proposed a specific implementation of the $U_1$ in a 4321 model~\cite{Diaz:2017lit} to satisfy existing flavor constraints. At the benchmark point provided by~\cite{DiLuzio:2018zxy}, $U_1$ only couples to the second and third generation leptons and quarks in the $SU(2)_{weak}$ singlet combination bilinear $Q^\dag L$. The interaction is given by
\begin{align}
\mathcal L \supset  U_{1, \mu} &\left(\lambda_{33} \bar{q}_3 \gamma^ \mu P_L l_3 +\lambda_{32} \bar{q}_3 \gamma^\mu P_L l_2 + \lambda_{23} \bar{q}_2 \gamma^\mu P_L l_3 + \lambda_{22} \bar{q}_2 \gamma^\mu P_L l_2 \right) +h.c.
\label{eq:U1new}
\end{align}
where
\beq
q_i =\left (
  \begin{tabular}{c}
  $V^*_{ji} u_j$ \\
   $d_i$
  \end{tabular}
\right ),\quad 
l_i =\left (
  \begin{tabular}{c}
  $\nu_i$ \\
   $e_i$
  \end{tabular}
\right )\quad 
(i= 2, 3;\,j = 1, 2, 3)
\eeq
and $V_{ij}$ is the CKM matrix. At the benchmark point, the model parameters are
\beq
m_{U_1} = 2\tev,\quad \lambda_{33} = 1.12,\quad \lambda_{23} = 0.42,\quad \lambda_{32} = -0.10, \quad \lambda_{22} = 0.039
\eeq
with more details shown in~\appref{4321}. Given $\lambda_{33}, \lambda_{23} \gg \lambda_{32, 22}$, the most relevant searches are those with taus rather than muons in the final state. And given $\lambda_{33}/\lambda_{23}\approx 0.4$, an on-shell $U_1$ is more likely to decay to the $b \tau$ and $t \met$ final states than to $j \tau$ or $j \met$.\footnote{Note that $j\met$ can also occur from $U_{1, \mu} \lambda_{33} \bar{q}_3 \gamma^ \mu P_L l_3$ because of CKM mixing but is suppressed by a small CKM angle.} Therefore, we conclude that the most relevant search channel is leptoquark pair production 
\beq
gg \to U_1 \bar{U}_1 \to \frac{1}{4}\left[(\tau^- \bar{b})(\tau^+ b) + t \bar{t} \met + (\tau^- \bar{b}) t \met + (\tau^+ b) \bar{t}\met\right]
\eeq
for small leptoquark couplings\footnote{At small $\lambda$, $b\bar{b}, s\bar{s}, b \bar{s}, s\bar{b}$-initiated processes are less significant than $gg$-initiated processes.} and the DY production,\beq
b\bar{b}, s\bar{s}, b \bar{s}, s\bar{b} \to \tau^+ \tau^-
\eeq
for large couplings.
Since we found that limits from single production are generically weaker than the combination of  pair production and DY searches we will not investigate single production further.

\begin{figure}[t]
   \centering
   \includegraphics[width=0.48\textwidth]{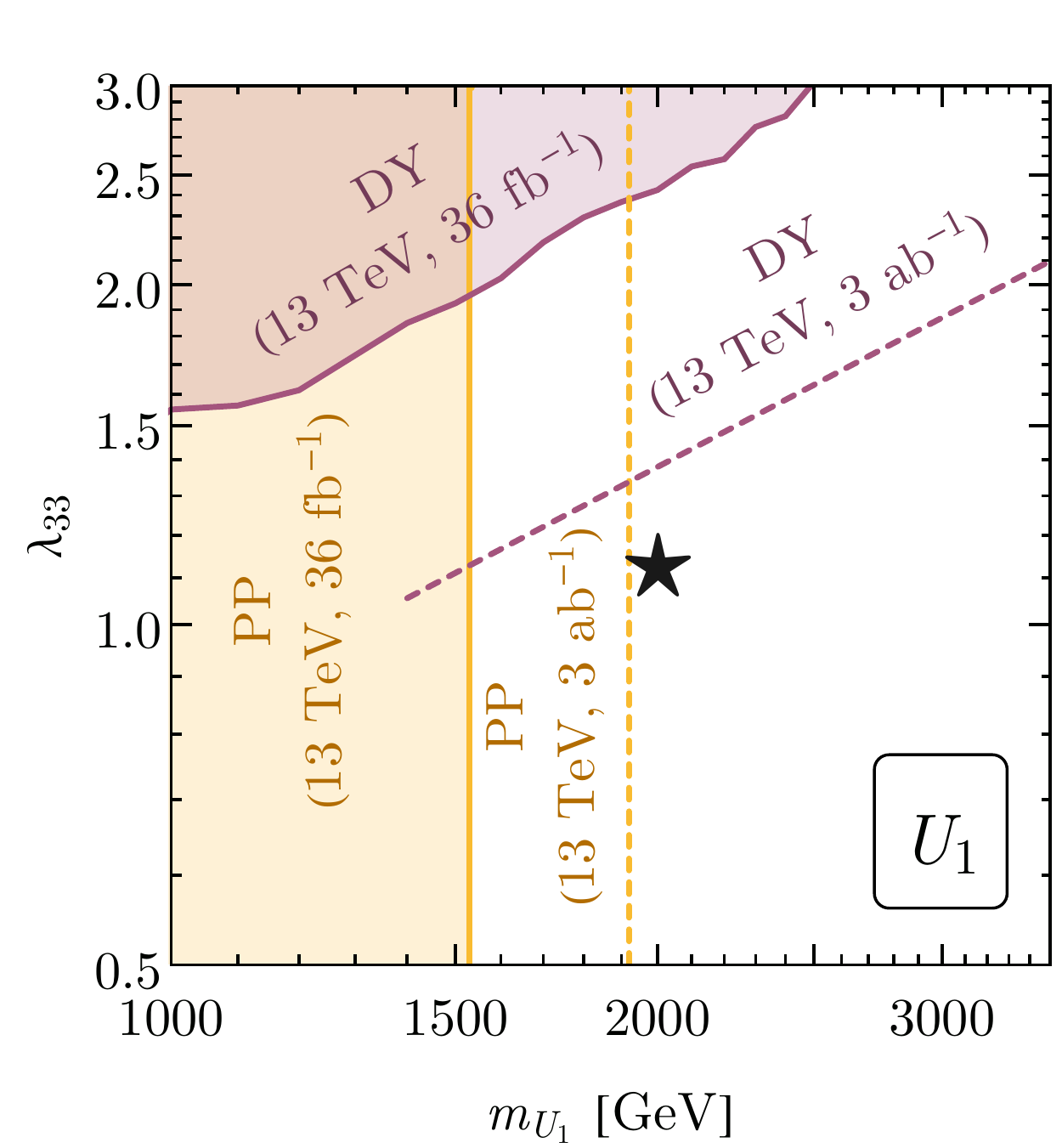} % requires the graphicx package
   \caption{Current (solid) and projected (dashed) LHC constraints on the vector leptoquark $U_1$ mass and its coupling to the third generation quark and lepton doublets, $\lambda_{33}$. The pair production and DY bounds are recast from~\cite{Sirunyan:2018kzh} and~\cite{Aaboud:2017sjh}, respectively. The star marks the benchmark point ($m_{U_1} = 2 \tev$ and $\lambda_{33} = 1.12$) from the 4321 model that explains the $R_{D^{(*)}}$ and $R_{K^{(*)}}$ anomalies~\cite{DiLuzio:2018zxy}. See text for more details.}
   \label{fig:U1}
\end{figure}

Ref.~\cite{Sirunyan:2018kzh} performed an updated $pp \to VV \to (t\nu)(\bar{t}\nu)$ search and obtained the 95\% CL lower limit $m_{U_1} \geq 1530$ GeV (see lower panel of Fig.~3 in the reference). 
Reinterpreting the result from the $pp \to SS \to (b\tau)(\bar{b}\tau)$ search~\cite{CMS-PAS-EXO-17-016}, we obtain a slight weaker lower limit on $m_{U_1}$ of 1400 GeV.\footnote{This limit is mostly $\lambda$-independent until $\lambda$ reaches $3$. At $\lambda =3$,  the  95\% CL lower limit grows to $m_{U_1}\geq 1430 \gev$ for the current data ($13\tev$, $36\fb$) and $m_{U_1}\geq1790 \gev$ for the HL-LHC ($13\tev$, $3\ab$).} To get the DY limit, we simulate the DY production of  $U_1$ via \texttt{MG5+Pythia+Delphes} and perform the cut and count analysis for the highest bin of the $m_\text{T}^\text{tot}$ distribution given by~\cite{Aaboud:2017sjh}. Details of the simulation and analysis are identical to the ones described in~\secref{tauq}. We find a strong limit on $\lambda_{33}$ which equals  $\lambda_{33}\geq 2.5$ for the leptoquark mass  $m_{U_1} = 2 \tev$ corresponding to the benchmark point, see~\figref{U1}. Note that our limits are slightly weaker than those in~\cite{Angelescu:2018tyl} because we prefer a more conservative analysis of the statistical and systematical uncertainties (see~\appref{cutandcount} for details).

Also shown in~\figref{U1} are our projections for the reach of pair production and DY searches at the high-luminosity LHC (HL-LHC) at 13 TeV with 3 $\ab$ of data. For the pair production projection, we assume that the uncertainty is dominated by statistics and scale the expected 95\% upper limit on the production cross section from~\cite{Sirunyan:2018kzh} down by a factor of $\sqrt{3000/36}$. 
By comparing this projected upper limit on the cross section with the tree-level prediction for the production cross section, $\sigma(pp \to U_1 \bar{U}_1 \to (t\nu)(\bar{t}\nu))$, as a function of $m_{U_1}$ and $\lambda_{33}$ we obtain our projected 95\% CL lower limit on $m_{U_1}$ of 1920 GeV.\footnote{To cross check the validity of our scaling treatment, we used the same method to project Run 2 LHC bounds on stop ($\t{t}$) masses with 300 $\fb$ and 3 $\ab$. We obtained bounds on $m_{\t t}$ of 1.2 TeV and 1.5 TeV respectively, which is in good agreement with~\cite{ATL-PHYS-PUB-2013-011}.} Performing the projection procedure for $pp \to U_1 \bar{U}_1 \to (b\tau)(\bar{b}\tau)$ search based on~\cite{CMS-PAS-EXO-17-016}, we project a 95\% CL lower limit on $m_{U_1}$ of 1760 GeV for HL-LHC ($13 \tev$, $3 \ab$). In~\figref{U1} we show the stronger reach from the $(t\nu)(\bar{t}\nu)$ search.

In order to project the DY-production limit, we assume that both the statistical and the systematical uncertainties can be reduced as the integrated luminosity increases. To be more specific, we take the number of observed and expected background events from the highest bin of the $m^\text{tot}_\text{T}$ distribution from~\cite{Aaboud:2017sjh} and scale both of them by a factor of $3000/36$. Meanwhile, we optimistically (and perhaps unrealistically) assume that the systematical uncertainties are negligible. We then apply the cut and count analysis described in~\appref{cutandcount} and obtain a projected 95\% CL limit in the $m_{U_1}-\lambda_{33}$ parameter space.

Note that our projected sensitivities for both the pair-production and DY searches at HL-LHC are tantalizingly close to excluding the benchmark point or discovering the $U_1$ in the 4321 model.

\section{Distinguishing leptoquarks with different weak quantum numbers}
\label{sec:distinguish}

In this paper we have emphasized that the flavor of final state particles are the best way to organize leptoquark searches and summarize bounds on leptoquark cross sections. The $SU(2)_{weak}$ quantum numbers and whether leptoquarks couple to left- or right-handed quarks and leptons do not play an important role for the experimental analysis. Of course, they do play an important role in predicting cross sections and branching fractions.    

Imagine that a leptoquark has been discovered in some channels. From the final state products we then know which lepton it couples to and whether the quark is a light quark (light-jet), a $b$-quark or a top-quark. Next we would like to know the $SU(2)_{weak}$ quantum numbers, charge, spin, and whether the coupling is to left- or right-handed quarks. In the following we summarize a few points to indicate how that could be done by combining information from pair production, single production and DY.
  \begin{figure}[t]
   \centering
   \includegraphics[width=0.8\textwidth]{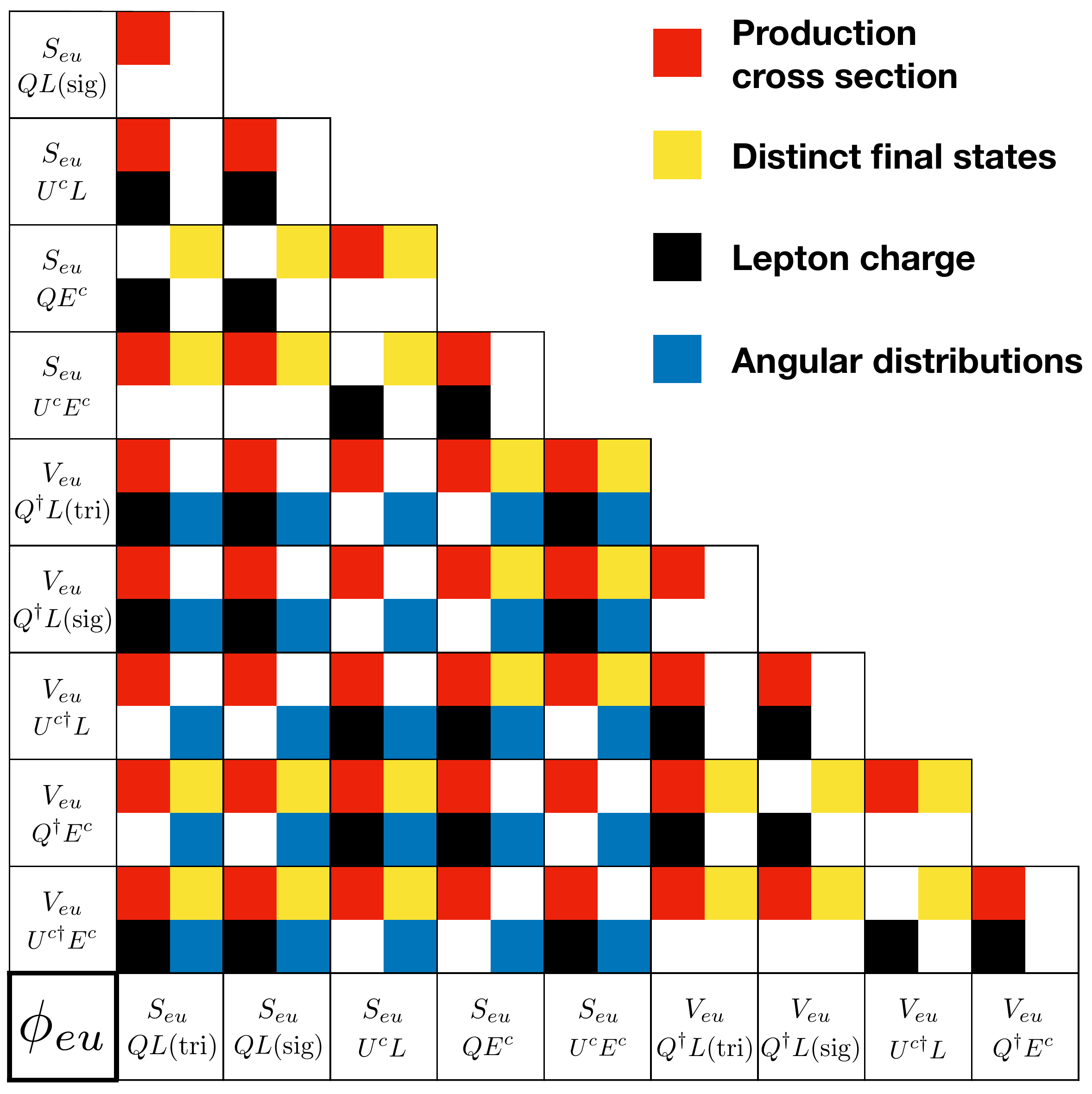} % requires the graphicx package
   \caption{Signal features which distinguish leptoquarks of the same flavor (coupling to up quarks and electrons or positrons) with different spin or electroweak quantum numbers. Here we use colors to represent the different methods that are discussed in~\secref{distinguish} for distinguishing between leptoquarks: the production cross section (red), the final states (yellow), the lepton charge (black), and the angular distributions (blue). If two different leptoquarks can be distinguished by any of above features, we filled their cross cell with the corresponding color.}
   \label{fig:checkers}
\end{figure}

\begin{itemize}
\item{\bf Production cross section.}
For small couplings, $\lambda \lesssim 1$, the pair production cross section of leptoquarks is completely determined by the QCD coupling and the leptoquark mass. However, for the same mass the vector leptoquark pair production cross section is almost an order of magnitude larger than the scalar one.  In addition, there can be multiplicity factors if the leptoquark is a non-trivial $SU(2)_{weak}$ multiplet. Thus a precision measurement of the pair production cross section can determine both spin and $SU(2)_{weak}$ quantum numbers. Leptoquarks with significant couplings to the first generation quarks can also be singly produced. However since the leptoquark coupling is an adjustable parameter the single production cross section cannot be used to distinguish between models.

\item{\bf Distinct final states.} 
If the leptoquark couples to left-handed quarks or leptons then it decays to multiple distinct final states with simply related branching fractions. This may be because the leptoquark itself is an $SU(2)_{weak}$ doublet or triplet and the different members of the multiplet couple and decay to different quarks and leptons, or in the case the that leptoquark is a singlet it  couples to a singlet combination of SM fermion doublets, for example to the $SU(2)_{weak}$ singlet combination $u_L \nu_L - d_L e_L$.

\item{\bf Lepton charge.}
Single production of leptoquarks allows distinction between a ``lepto-quark'' and an ``antilepto-quark'' where by ``antilepto-quark'' we mean a particle which decays to an antilepton and a quark. In pair-production, this measurement is difficult because it requires determining whether a final state jet is initiated from a quark or an anti-quark. This is impossible for light quarks, very difficult for $c$ or $b$ quarks and easy for top-quarks. A leptoquark which is singly produced is likely to have come from a collision of a gluon and a valence quark. This determines that the final state jet is also more likely a quark jet (as opposed to an antiquark jet) and therefore the charge of the lepton that combines with this jet distinguishes between the lepto- or antilepto-quark possibility.    

\item{\bf Angular distributions.}
Finally, the angular distributions of leptons which do not come from resonance decay in the single production and the DY production, carry information about the spin of the fermions involved and can be used to further distinguish between leptoquark couplings to left- and right-handed quarks and leptons. Angular distributions of the leptoquark decay products are not as useful as they are mostly just back-to-back because of the large mass of the leptoquarks. Possible spin correlations of the leptoquarks do not lead to significant differences in angular distributions \cite{Diaz:2017lit}.
\end{itemize}

A detailed strategy of how to distinguish between the different possibilities will depend on which production channels have been observed with which precision and will likely combine several of the points listed above. In \figref{checkers} we imagine as an example that a leptoquark coupling to up quarks and electrons or positrons has been observed. In the matrix we indicate with which features of the searches one can distinguish any pair of leptoquark models.

\acknowledgments
We thank Zeynep Demiragli, Ilja Dor\v{s}ner, Darius Faroughy, Marie-H\'el\`ene Genest, Admir Greljo, Marcus Morgenstern, Vojtech Pleskot, Ruth Pottgen, Nirmal Raj, and Francesco Romeo for helpful discussions. We thank the Aspen Center for Physics, which is supported by NSF grant PHY-1607761, for hospitality during work on this project. YZ also thanks the KITP at UCSB , which is supported by the NSF grant PHY1748958, for hospitality during work on this project. The work of MS and YZ is supported by DOE grant DE-SC0015845.

\appendix

\section{Cut and count analyses}
\label{sec:cutandcount}

In DY searches or monojet searches, no resonance can be reconstructed in the distributions of the final state particle kinematics ($m_{\ell \ell}$ or $m_\text{T}^\text{tot}$ distributions for DY and $\pt^\text{miss}$ distributions for monojets). Often the contributions from leptoquarks present as distortions of the tails of distributions. To constrain leptoquark models with such searches, we  perform cut and count analyses.

For the DY searches with $\ell^+\ell^-$ ($\tau^+ \tau^-$) final states, we focus on the highest bin of the $m_{\ell \ell}$ ($m_\text{T}^\text{tot}$) distribution where the SM backgrounds are relatively small. Given $N$ observed events and $B$ expected background events
we compute 95\% CL upper limit on the number of the signal events, $S_\text{limit}$, using the ${\rm CL}_{\rm s}$ method for Poisson distributions~\cite{read2000modified, Olive:2016xmw, statistics}. More precisely, and in order to obtain a conservative bound, we use a conservative background expectation, $B'\equiv B-2 \delta B$, in the ${\rm CL}_{\rm s}$ method. Here $B$ and $\delta B$ are the SM background and its systematic uncertainty determined by the experimentalists.  
Once  $S_{\rm limit}$ is obtained in this way, we compare it to the theoretical prediction for the number of leptoquark signal events (including leptoquark-SM interference) as a function of the parameters $m_\phi$ and $\lambda_\phi$. We calculate these predictions using  \texttt{MG5} and impose the experimental efficiencies, acceptance, and selection cuts. By sampling over the parameter space, we find the region in the $m_\phi-\lambda_\phi$ plane in which the predicted number of signal events is below the 95\% CL upper limit $S_{\rm limit}$.

For the monojet search, we perform a variation of above procedure. Here we combine the last several bins in the tail of the $\pt^\text{miss}$ distributions into a single signal bin to achieve a better sensitivity to the signal. For this combined bin, the number of observed events $N$, expected background events $B$, and systematic uncertainty on the background events $\delta B$ are
\beq
N = \sum_{i=1}^p N_i,\quad B =\sum_{i=1}^p B_i, \quad \delta B = \sum_{i,j=1}^p  \delta B_i \rho_{ij} \delta B_j,
\eeq 
where $\rho_{ij}$ is the correlation matrix for the uncertainties of the expected background events determined by the experimentalists (e.g. Fig. 20 of~\cite{Sirunyan:2017jix}). Here $i$ and $j$ run over the $p$ highest bins in the $\pt^\text{miss}$ distribution. In order to optimize the sensitivity of the search to the new physics we perform a scan over all possible choices of $p$, the number of combined bins, at any given point in $m_\phi - \lambda_\phi$ parameter space. For each choice of $p$ we perform the analysis for setting limits described in the previous paragraph, except that we use the number of expected background events $B$ in place of the observed data $N$. Of these choices we then select the one which gives the strongest bound on the model in this background-only hypothesis. Once the optimal combined signal bin has been determined in this way we then use the observed data to determine if the parameter space point is excluded at the 95\% confidence level. By scanning over parameter space and repeating the above procedure for each point we find the 95\% CL allowed region in the $m_\phi - \lambda_\phi$ plane.

\section{Constraints from the weak charge measurements and other probes}
\label{sec:others}

The weak charge of protons and nuclei is measured in experiments of parity-violating electron scattering  (PVES) and atomic parity violation (APV). Given the effective Lagrangian between electrons and up and down quarks
\beq
\mathcal L = \f{1}{2 v^2} \bar{e}\gamma^\mu \gamma^5 e \left(C_{1u} \bar{u}\gamma_{\mu} u + C_{1d}\bar{d}\gamma_\mu d \right),
\eeq
where $C_u$ and $C_d$ stands for the couplings and the vacuum expectation value $v=246 \gev$, the weak charge of an atom (or proton) is given by
\beq
Q_W= -2 \left(Z(2 C_{1u}+ C_{1d}) + N (C_{1u}+ 2C_{1d})\right)
\eeq
where $Z$ and $N$ are the number of protons and neutrons of an atom respectively. ~\tabref{weakcharge} lists results from PVES and APV measurements and their SM predictions.
% Requires the booktabs if the memoir class is not being used
\begin{table}[htbp]
   \centering
      \topcaption{Summary of experimental measurements of weak charges of the proton, cesium, and thallium.}

   %\topcaption{Table captions are better up top} % requires the topcapt package
   \begin{tabular}{@{} cccccc @{}} % Column formatting, @{} suppresses leading/trailing space
      \hline
       $Q_W$   & $Z$ & $N$ & Experimental value &  Standard Model value & Reference\\
      \hline
     $p$ (PVES fit) & 1 & 0 & $0.0719\pm 0.045$ & $0.0708 \pm 0.0003$ &~\cite{Androic:2018kni}\\
       Cs & 55 & 78 & $-72.62 \pm 0.43$ & $-73.25 \pm 0.01$ &~\cite{Olive:2016xmw}\\
       Tl & 81 & 124 & $-116.4 \pm 3.6$ & $-116.90 \pm 0.02$ & ~\cite{Olive:2016xmw}\\
      \hline
   \end{tabular}
   \label{tab:weakcharge}
\end{table}

Extra contributions to the weak charge from leptoquark exchange between electrons and up and down quarks can be expressed as
\beq
\delta Q_W =  -2 \left(Z(2 \delta C_{1u}+ \delta C_{1d}) + N (\delta C_{1u}+ 2 \delta C_{1d})\right)
\eeq
where the couplings $\delta C_{u,d}$ are given by
\beq
\delta C_{1u} = \kappa_u \f{v^2 \lambda^2_{eu}}{4 m^2_{\phi_{eu}}},\quad \delta C_{1d} = \kappa_d \f{ v^2 \lambda^2_{ed}}{4 m^2_{\phi_{ed}}}.
\label{eq:lqapv}
\eeq
 Here $\kappa_{u,d}$ represent the flavor-dependent pre-factors listed in~\tabref{APV}. We collect them from Tab.~4 of~\cite{Dorsner:2016wpm}. To constrain the ratio of $\lambda_\phi/m_\phi$ for various leptoquarks listed in~\tabref{APV}, we construct a $\chi^2$ function
 \beq
 \chi^2 = \sum_{i = p, \text{Cs},\text{Tl}} \frac{\left(Q_{W, \text{SM}}(i) + \delta Q_W - Q_{W,\text{exp}}(i)\right)^2}{\sigma_\text{exp}^2(i) + \sigma_\text{SM}^2(i)}
 \label{eq:chisq}
 \eeq
to fit the experimental measurements listed in~\tabref{weakcharge}. Note that $\sigma_\text{exp}$ and $\sigma_\text{SM}$ in~\eqref{chisq} stand for the $1\sigma$ experimental and SM theoretical systematic uncertainties, respectively. The resulting $95\%$ CL upper limits on $\lambda_\phi/m_\phi$ are shown in the last column of~\tabref{APV}. For example, the 95\% CL bound on $S_{eu} (U^c E^c)$ is given by $\lambda_{S_{eu}} < 0.17 (m_{S_{eu}} / 1 \tev)$ and is shown as the green dot-dashed line in the $S_{eu}$ plot of~\figref{equark}.

\begin{table}[!htbp]
\centering
   \topcaption{The $\kappa$ coefficients for \eqref{lqapv} and the 95\% CL upper limits on $\lambda_\phi/m_\phi$ for MLQ $\phi_{eu}$ and $\phi_{ed}$ with different SM gauge quantum numbers.} % requires the topcapt package
   
   \begin{tabular}{@{} llccr @{}} % Column formatting, @{} suppresses leading/trailing space
      \hline
      Scalar LQ  & $(SU(3),SU(2))_Y$  & $\kappa_u$ & $\kappa_d$ & 95\% CL limit\\
      \hline
      $QL$ triplet & $(3,3)_{-1/3}$  & $-2$ & $-2$ & $\lambda_S\leq 0.21 \left(\f{m_S}{\text{TeV}}\right)$\\
      \\[-1em]
      $QL$ singlet & $(3,1)_{-1/3}$  & $-1$ & $\times$ & $\lambda_S\leq 0.40 \left(\f{m_S}{\text{TeV}}\right)$\\
      \\[-1em]
      $U^c L$ & $(\bar 3,2)_{-7/6}$  & $+1$  & $\times$ & $\lambda_S\leq 0.17 \left(\f{m_S}{\text{TeV}}\right)$ \\
      \\[-1em]
      $D^c L$ & $(\bar 3,2)_{-1/6}$ &   $\times$ & $+1$ &$\lambda_S\leq0.17 \left(\f{m_S}{\text{TeV}}\right)$\\
      \\[-1em]
      $Q E^c$ &$(3,2)_{7/6}$  & $-1$ &  $-1$ & $\lambda_S\leq0.30 \left(\f{m_S}{\text{TeV}}\right)$ \\
      \\[-1em]
      $U^c E^c$ & $(\bar 3,1)_{1/3}$ &  $+1$  & $\times$ & $\lambda_S\leq0.17 \left(\f{m_S}{\text{TeV}}\right)$ \\
      \\[-1em]
      $D^c E^c$ & $(\bar 3,1)_{4/3}$  & $\times$ & $+1$ & $\lambda_S\leq 0.17 \left(\f{m_S}{\text{TeV}}\right)$\\
      \hline
      \hline
      Vector LQ     & $(SU(3),SU(2))_Y$    & $\kappa_u$ & $\kappa_d$ & 95\% CL limit \\
      \hline
      $Q^\dag L$ triplet  & $(\bar 3,3)_{-2/3}$  & $+4$ &$+2$ & $\lambda_V\leq0.069 \left(\f{m_V}{\text{TeV}}\right)$\\
      \\[-1em]
      $Q^\dag L$ singlet & $(\bar 3,1)_{-2/3}$   & $\times$ & $+2$ & $\lambda_V\leq0.12 \left(\f{m_V}{\text{TeV}}\right)$\\
      \\[-1em]
      $U^{c \dag} L$ & $( 3,2)_{1/6}$   & $-2$ & $\times$ & $\lambda_V\leq0.28 \left(\f{m_V}{\text{TeV}}\right)$ \\
      \\[-1em]
      $D^{c\dag} L$ & $(3,2)_{-5/6}$ & $\times$ & $-2$ & $\lambda_V\leq0.32 \left(\f{m_V}{\text{TeV}}\right)$\\
      \\[-1em]
      $Q^\dag E^c$ &$(\bar 3,2)_{5/6}$  & $+2$ & $+2$ & $\lambda_V\leq0.085 \left(\f{m_V}{\text{TeV}}\right)$ \\
      \\[-1em]
      $U^{c\dag} E^c$ & $( 3,1)_{5/3}$ &  $-2$ & $\times$ & $\lambda_V\leq0.28 \left(\f{m_V}{\text{TeV}}\right)$  \\
      \\[-1em]
      $D^{c\dag} E^c$ & $( 3,1)_{2/3}$  & $\times$ & $-2$ & $\lambda_V\leq0.32 \left(\f{m_V}{\text{TeV}}\right)$\\
      \hline
   \end{tabular}
   \label{tab:APV}
\end{table}

As can be seen in~\tabref{weakcharge}, the current measured values for the weak charges of the proton, Cs, and Tl are all larger than the ones predicted by the SM. As a result, leptoquarks with positive $\kappa$'s, which contribute negatively to $Q_\text{w}^{p}$ and $Q_\text{w}^\text{Cs}$, are more strongly bounded than leptoquarks with negative $\kappa$'s. In the fits for leptoquarks with positive $\kappa$'s, pulls are dominated by  the Cs measurement. On the other hand, for leptoquarks with negative $\kappa$'s, pulls from PVES are competitive with those from Cs and even become dominant for leptoquarks with $\kappa_u < 0$ and $\kappa_d =0$.

Other probes for MLQs which are sensitive to leptoquarks with large couplings, such as rare lepton decays, rare $Z$-boson decays, anomalous magnetic moments of electrons or muons (see e.g. review by~\cite{Dorsner:2016wpm}) or Non-Standard Neutrino Interactions at IceCube~\cite{Dutta:2015dka,Becirevic:2018uab,Dey:2017ede},  NuTeV, and Super-K~\cite{Wise:2014oea}, yield weaker bounds than current LHC searches. We therefore do not discuss them further. A notable exception would be leptoquarks which couple to both left- and right-handed leptons. Loops with such leptoquarks contribute to electric and magnetic dipole moments and could lead to competitive constraints (see e.g.~\cite{Fuyuto:2018scm, Andreev:2018ayy, Cesarotti:2018huy}). Such contributions are automatically suppressed by the small lepton masses when the leptoquark couplings are chiral as in our MLQ models.

\section{Benchmarks for the 4321 model to explain the $R_{D^{(*)}}$ and $R_{K^{(*)}}$ anomalies}
\label{sec:4321app}

The 4321 model was proposed in Ref.~\cite{DiLuzio:2018zxy} as a simultaneous solution to both the $R_{D^{(*)}}$ and $R_{K^{(*)}}$ problems. The model introduces a vector leptoquark, $U_1$, with SM quantum numbers $({3},{1})_{2/3}$ and couplings to the SM fermions
\begin{align}
\mathcal L \supset \f{g_4}{\sqrt{2}} U_{1,\mu} &\left[c_{\theta_{LQ}} s_{l_3} s_{q_3} \bar{b}_L\gamma^ \mu \tau_L- s_{\theta_{LQ}}s_{l_2} s_{q_3} \bar{b}_L \gamma^\mu \mu_L \right.\nonumber\\
&\left.   + s_{\theta_{LQ}} s_{l_3}  s_{q_2} \bar{s}_L \gamma^\mu\tau_L +c_{\theta_{LQ}}  s_{l_2} s_{q_2} \bar{s}_L\gamma^\mu \mu_L +h.c.\right] 
\label{eq:U1}
\end{align}
To solve the $R_{D^{*}}$ anomalies, the contributions from the leptoquark should satisfy
\beq
\Delta R_{D^{(*)}} =  2\f{ g_4^2 v^2}{4 m_{U_1}^2}\left(  s_{\theta_{LQ}} c_{\theta_{LQ}} s_{l_3}^2 s_{q_2} s_{q_3}\f{V_{cs}}{V_{cb}} + c_{\theta_{LQ}}^2 s_{l_3}^2 s_{q_3}^2\right) = 0.217 \pm 0.053.
\label{eq:RD}
\eeq
where the CKM matrix elements are $V_{cs} = 0.97344$ and $V_{cb} = 0.0412$. The best fit in~\eqref{RD}  (and the value in~\eqref{RK}) is quoted from~\cite{Capdevila:2017bsm}.

To solve the $R_{K^{(*)}}$ anomalies, the contributions from the leptoquark (assuming tree-level only) should satisfy
\beq
\Delta C_9 (U_1) = -\Delta C_{10} (U_1) = -\f{2\pi}{\alpha V_{tb} V_{ts}}\f{ g_4^2 v^2}{4 m_{U_1}^2} s_{\theta_{LQ}} c_{\theta_{LQ}}  s_{l_2}^2 s_{q_2} s_{q_3} = -0.66 \pm 0.18.
\label{eq:RK}
\eeq
where $\alpha = 137.04^{-1}$ is the fine structure constant and the CKM matrix elements are $V_{tb} = 1.009$ and $V_{ts} = 40.0\times 10^{-3}$. At the benchmark point proposed by~\cite{DiLuzio:2018zxy}
\beq
m_{U_1} = 2 \tev, \quad g_4 = 3.5,\quad s_{l_3} = 0.8,\quad s_{q_2} = 0.3,\quad s_{q_3} = 0.8,\quad \theta_{LQ} = \pi/4\ ,
\eeq
one obtains $\Delta R_{D^{(*)}} = 0.187$ which is within $1\sigma$ of the best fit value for $R_{D^{(*)}}$. Substituting the benchmark values into~\eqref{RK} yields
\beq
s_{l_2}=0.075
\eeq 
to get the best fit value of $R_{K^{(*)}}$.  Plugging these values back into~\eqref{U1}, and comparing the result to the Lagrangian~\eqref{U1new}, i.e.,  $\lambda_{ij} U_{1, \mu} q_i \gamma^\mu P_L l_j$, where $i , j$ label the generations of quark and lepton doublets, we find
\beq
\lambda_{33} = 1.12,\quad \lambda_{23} = 0.42,\quad \lambda_{32} = -0.10, \quad \lambda_{22} = 0.039 \ .
\eeq
Thus the $U_1$ couplings to ${b\tau}$ or ${s\tau}$ are much larger than those to ${b\mu}$ and ${s\mu}$. This implies that DY searches for $\tau \tau$ final states are more sensitive than those with $\tau \mu$ or $\mu \mu$ final states. The pair production searches with $(b\tau) (b\tau)$ or ($t \met$)($\bar{t} \met$) final states give the best bounds on $m_{U_1}$ for small $\lambda$'s.

\section{Leptoquarks with alternative electroweak quantum numbers}
\label{sec:alternativecouplings}

Throughout the paper we have focused on the simplest possible case with only a single leptoquark coupling to a single lepton-quark
bilinear. Here we briefly review the other MLQ models defined in~\cite{Diaz:2017lit} in which the fermions
involved in the leptoquark coupling and the leptoquark itself can have non-trivial $SU(2)_{weak}$ representations.  For simplicity we focus only on
first generation leptons and first generation quarks. Similar models exist for general $i-j$ generation leptoquarks.

Consider first scalar leptoquarks. If the quark involved in the coupling is an $SU(2)_{weak}$ doublet $Q=(u_L,d_L)$ with left chirality and the lepton is a singlet $E^c$ with right chirality then the leptoquark must also be a doublet $S_{eq} =(S_{eu}, S_{ed})$. We obtain the coupling
\beq
{\mathcal L} \supset \lambda_{eq} S_{eq} (E^c)^T i\sigma^2 Q
=  \lambda_{eq} ( S_{eu} e_R^\dag u_L + S_{ed} e_R^\dag d_L ) \ .
\eeq 

Alternatively, the lepton can be a doublet $L=(\nu_L,e_L)$ and the quark a singlet. Then we have
 \beq
{\mathcal L} \supset \lambda_{lu} S_{lu} (L_L^T i \sigma^2 u^c)^*
=  -\lambda_{lu} ( S_{\nu u} \nu_L^\dag u_R + S_{eu} e_L^\dag u_R ) \ ,
\eeq 
and similarly for a leptoquark coupling to down quarks.

Another possibility is that the lepton and the quark are both left-chirality doublets $Q$ and 
$L$. In this case, the leptoquark can either be an $SU(2)_{weak}$ singlet $S$ or a triplet $\vec S=(S^{\frac23},S^{-\frac13},S^{-\frac43})$
 \bea
   \lambda_{lq} S^*_{lq}\, L_{L}^T i\sigma^2 Q_L 
&= \lambda_{lq}& S^*_{lq}\, \frac{e_{L}^T i\sigma^2 u_L - \nu_{L}^T i\sigma^2 d_L}{\sqrt{2}}  \quad \quad \quad ({singlet}) \\
\lambda_{lq} \vec S^*_{lq}\, \cdot L_{L}^T i\sigma^2 \vec \tau Q_L 
&=  \lambda_{lq}& \Big( (S^{\frac23}_{lq})^*\, \nu_{L}^T i\sigma^2 u_L + (S^{-\frac43}_{ql})^*\, e_{L}^T i\sigma^2 d_L \nonumber \\
&& \ \ +\,  (S^{-\frac13}_{lq})^*\,  \frac{e_{L}^T i\sigma^2 u_L + \nu_{L}^T i\sigma^2 d_L}{\sqrt{2}}\Big) \quad ({triplet}) 
\eea
where $\vec \tau$ are the Pauli matrices in $SU(2)_{weak}$ space and the normalization factor of $1/\sqrt{2}$ ensures that the width formula in \eqref{scalarwidth} also applies to the singlet and triplet leptoquarks.\footnote{LHC and future hadron collider bounds on $\vec{S}$ are discussed in~\cite{Hiller:2018wbv}.}

Leptoquarks can also be spin-1 vectors. In that case the fermion bi-linear must also be a vector. We already defined a vector coupling to the lepton and quark right-handed singlets in Section \ref{sec:normalization}. If either the lepton or quark is an $SU(2)_{weak}$ doublet then the leptoquark is a doublet and we have
\beq
{\mathcal L} \supset
 \lambda_{lu} V^*_{\mu\,lu}\, (U^c)^\dag \bar \sigma^\mu L 
= - \lambda_{lu}( V^*_{\mu\,eu}\, e_L^T i\sigma^2 \sigma^\mu u_R + V^*_{\mu\,\nu u}\, \nu_L^T i\sigma^2 \sigma^\mu u_R)\ ,
\eeq 
or
\beq
{\mathcal L} \supset 
  \lambda_{eq} V_{\mu\,eq}\, (E^c)^\dag \bar\sigma^\mu Q 
=  -\lambda_{eq}( V_{\mu\,eu}\, e_R^T i\sigma^2\bar\sigma^\mu u_L + V_{\mu\,ed}\, e_R^T i\sigma^2\bar\sigma^\mu d_L)\ .
\eeq 
If both lepton and quark are doublets then they couple to either a singlet $V_\mu$ or triplet $\vec V_\mu$
 \bea
   \lambda_{lq} V_{\mu\,lq}\, L^\dag \bar\sigma^\mu Q 
&= \lambda_{lq}& V_{\mu\,lq}\, \frac{\nu_{L}^\dag \bar\sigma^\mu u_L + e_{L}^\dag \bar\sigma^\mu d_L}{\sqrt{2}}  \quad \quad \quad ({singlet}) \\
\lambda_{lq} \vec V_{\mu\,lq} \cdot L^\dag \bar\sigma^\mu \vec \tau\, Q 
&=  \lambda_{lq}& \Big( V^{-\frac53}_{\mu\,lq}\, e_{L}^\dag \bar\sigma^\mu u_L + V^{\frac13}_{\mu\,lq}\, \nu_{L}^\dag \bar\sigma^\mu d_L \nonumber \\
&& \ \ \ +\ V^{-\frac23}_{\mu\, lq}\,  \frac{\nu_{L}^\dag \bar\sigma^\mu u_L - e_{L}^\dag \bar\sigma^\mu d_L}{\sqrt{2}}\Big) \quad ({triplet}) 
\eea

The $SU(2)_{weak}$-singlet vector leptoquark with couplings primarily to third generation leptons $L_3$ and quarks $Q_3$ and subdominant couplings to $Q_2$ and $L_2$ has recently generated a lot of interested because it can be used to explain the famous anomalies in $B$-meson decays, $R_{K^{(*)}}$ and $R_{D^{(*)}}$. 

In this paper we focused on the different possible flavor quantum numbers of the leptoquark and limited our attention mostly to searches for scalar leptoquarks with couplings to singlet quarks and leptons. We also demonstrated that efficiencies and acceptances do not strongly depend on spin and the $SU(2)_{weak}$ quantum  numbers of the leptoquark. Thus it is fairly straightforward to reinterpret the bounds for other cases. Techniques for distinguishing between the different $SU(2)_{weak}$ representations are discussed in~\secref{distinguish}.

\section{Parton-level differential cross sections}
\label{sec:partonlevel}

\subsection{Single-leptoquark production}

\begin{figure}[htbp]
   \centering
   \includegraphics[width=0.48\textwidth]{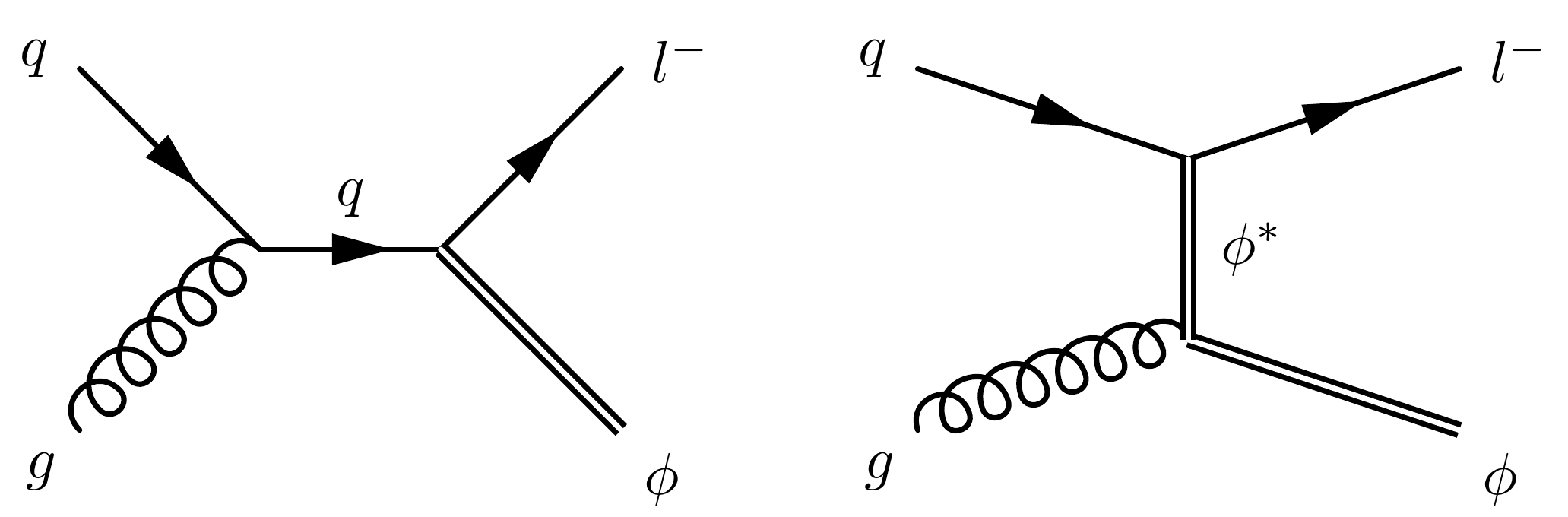} % requires the graphicx package
   \caption{The two leading diagrams for single leptoquark production.}
   \label{fig:twoonshell}
\end{figure}

Here we re-derive the analytic expression for the differential cross section of single scalar leptoquark production. The relevant diagrams are shown in \figref{twoonshell}. 

For simplicity, we assume that the leptoquark decays to massless leptons and quarks ($m_q = m_l \simeq 0$). Motivated by the appearance of the second Feynman diagram in~\figref{twoonshell} we define $\theta^*$ to be the scattering angle between the incoming quark and the outgoing lepton in the center-of-mass frame. The Mandelstam variables are then given by
\beq
\hat t \equiv (p_l - p_q)^2 = (m_S^2-\hat s)\f{1-\cos \theta^*}{2}\quad \hat u \equiv (p_S-p_q)^2 = ( m_S^2-\hat s)\f{1+\cos \theta^*}{2}
\label{eq:mymand}
\eeq
The averaged amplitude square  is
\beq
\overline{|M|^2} = \f{\pi\alpha_s \lambda^2}{3}\left[ \f{\hat s-m_S^2}{\hat s}+  \f{2 \hat t^2}{(\hat t-m_S^2)^2}+\f{\hat t(2 \hat s - \hat t - m_S^2)}{\hat s(m_S^2-\hat t)}\right]
\label{eq:singlemsquare}
\eeq

The first two terms in the square bracket of \eqref{singlemsquare} correspond to the amplitude squared of the diagrams SP-1 and SP-2  respectively.  The last term shows the interference between the two. Since $\hat s > m_S^2$ and $\hat t <0$, the interference is always \emph{destructive}. Using \eqref{singlemsquare}, we obtain an analytic expression for the differential cross section 
\begin{align}
\f{\d \sigma_{S_{eu}}}{\d\hat t} = \f{2}{\hat s - m_S^2}\f{\d \sigma}{\d \cos \theta^*} ={}&\f{\alpha_s \lambda^2}{48  \hat s^2}\left[   \f{\hat s-m_S^2}{\hat s}+\f{2 \hat t^2}{(\hat t-m_S^2)^2}+ \f{\hat t(2 \hat s - \hat t - m_S^2)}{\hat s(m_S^2-\hat t)}\right]\\
={}&\f{\alpha_s \lambda^2}{48  \hat s^2}\left[\f{\hat s +\hat t -m_S^2}{\hat s} + \f{\hat t (\hat t +m_S^2)}{(\hat t -m_S^2)^2} + \f{\hat t (\hat s - 2 m_S^2)}{\hat s (m_S^2-\hat t)}\right]\\
={}&\f{\alpha_s \lambda^2}{48  \hat s^2}\f{-\hat u (\hat t^2 +m_S^4)}{\hat s (\hat t -m_S^2)^2}
\label{eq:singlediff}
\end{align}

This result is consistent with~\cite{Eboli:1987vb,Djouadi:1989md} and the implementation of the leptoquark model in \texttt{Pythia 6} and the most recent version of \texttt{Pythia 8} (\texttt{v.235}).
For single vector leptoquark production the differential cross section is given by
\beq
\f{\d \sigma_{V_{eu}}}{\d \hat t} = \frac{\alpha_s \lambda^2}{24 \hat s^2} \f{-\hat u \left[(\hat s - m_V^2)^2 + (\hat u - m_V^2)^2\right]}{\hat s (\hat t- m_V^2)^2}
\eeq 
where the definition of the Mandelstam variables (and the scattering angle) is identical to~\eqref{mymand}.
 
\subsection{Drell-Yan production} 
Here we compare the tree-level Drell-Yan process $u \bar{u} \to e^+ e^-$ mediated by scalar leptoquark ($S_{eu}$ coupling to $U^c E^c$) exchange to the same process mediated by vector leptoquark ($V_{eu}$ coupling to $U^{c\dag} E^c$) exchange. The differential cross section for the scalar leptoquark only process is given by
\beq
\f{\d\sigma_{S_{eu}}}{\d \hat t} =\f{1}{16\pi \hat s^2}\overline{|\mathcal M_{S_{eu}}|^2} =\f{1}{16\pi \hat s^2} \frac{ 3\lambda^4 \hat t^2}{4(m_S^2-\hat t)^2}
\eeq
where \beq
\hat t = (p_u -p_{e^+})^2 = -\f{\hat s}{2} (1-\cos \theta^*), \quad
\hat u = (p_{\bar u} -p_{e^+})^2 = -\f{\hat s}{2} (1+\cos \theta^*)
\label{eq:mymand2}
\eeq and $\theta^*$ is the angle between the outgoing $e^+$ and incoming $u$ in the center-of-mass frame. Adding $Z$-boson exchange, we find that the interference term in the matrix element squared is given by
\beq
\f{|\mathcal M_{S_{eu}-Z}|^2}{|\mathcal M_{S_{eu}}|^2} = -\f{32 \pi \alpha t_w^2}{3 \lambda^2} \f{m_{S}^2 -\hat t}{\hat s - m_Z^2} \approx \frac{0.07}{\lambda^2} \f{m_{S}^2 }{\hat s}
\eeq
where $t^2_w \equiv \tan^2 \theta_w$ is the tangent of the weak mixing angle and $\alpha$ is the electromagnetic fine structure constant. The negative sign indicates destructive interference between $S_{eu}$-exchange and the $Z$-boson mediated diagram. 

The situation is different for vector leptoquark's. The leptoquark only cross section and the interference amplitude squared are respectively given by
\beq
\f{\d\sigma_{V_{eu}}}{\d \hat t}  =\f{1}{16\pi \hat s^2} \f{3 \lambda^4 \hat t^2}{ (m_V^2-\hat u)^2}, 
\eeq
and
\beq
\f{|\mathcal M_{V_{eu}-Z}|^2}{|\mathcal M_{V_{eu}}|^2} =
\f{16\pi \alpha t_W^2}{3\lambda^2}\f{m_V^2 -\hat u}{\hat s - m_Z^2}\approx \f{0.04}{\lambda^2}\f{m_V^2}{\hat s}\ ,
\eeq
where our definition of the Mandelstam variables (and the scattering angle) is as in~\eqref{mymand2}.
Note that $V_{eu}$ constructively interferes with the $Z$-boson. The sign difference between $S_{eu}$ and $V_{eu}$ can be understood from the Fierz rearrangements
\beq
 (u_L^\dag e_R) (e_R^\dag u_L) = -\f{1}{2} (u_L^\dag \bar \sigma^\mu u_L) (e_R^\dag \sigma_\mu e_R) = -\frac{1}{2} (\bar{u} \gamma^\mu P_L u) (\bar{e} \gamma_\mu P_R e) 
\eeq
\beq
 (u_R^\dag \sigma^\mu e_R) (e_R^\dag \sigma_\mu u_R) =  (u_R^\dag  \sigma^\mu u_R) (e_R^\dag \sigma_\mu e_R) \ .
\eeq
The factor of $1/2$ explains why in the higher dimensional operator limit $m_S^2 \gg \hat t, \hat u$ one has $\f{\d\sigma_{V_{eu}}}{\d \hat t} \simeq 4 \f{\d\sigma_{S_{eu}}}{\d \hat t}$.

\bibliographystyle{JHEP} 
\bibliography{leptoquark.bib}

\end{document}